\documentclass[
prx,
lengthcheck,
nofootinbib,
 amsmath,amssymb,
 aps,
 superscriptaddress,
 floatfix,
]{revtex4-2}

\usepackage[dvipsnames]{xcolor}
\usepackage{graphicx} % Include figure files
\usepackage[caption=false]{subfig}
\usepackage{bm} % bold math
\usepackage[colorlinks,allcolors=blue]{hyperref}
\usepackage{microtype}
\usepackage{physics}
\usepackage{enumerate}

\def\equationautorefname#1#2\null{Eq.#1(#2\null)}

\newcommand{\appref}[1]{\hyperref[#1]{\appendixautorefname~\ref*{#1}}}

\renewcommand{\Re}{\operatorname{Re}}
\renewcommand{\Im}{\operatorname{Im}}

\newcommand{\las}{\mathrm{L}}
\newcommand{\wlas}{\omega_{\las}}

\allowdisplaybreaks{}

\begin{document}

\title{Spatially resolved photon statistics of general nanophotonic systems}

\author{Maksim Lednev}
    \email{maksim.lednev@uam.es}
    \affiliation{Departamento de Física Teórica de la Materia Condensada, Universidad Autónoma de Madrid, E-28049 Madrid, Spain}
    \affiliation{Condensed Matter Physics Center (IFIMAC), Universidad Autónoma de Madrid, E-28049 Madrid, Spain}
\author{Diego Fernández de la Pradilla}
    \email{diego.fernandez@uam.es}
    \affiliation{Departamento de Física Teórica de la Materia Condensada, Universidad Autónoma de Madrid, E-28049 Madrid, Spain}
    \affiliation{Condensed Matter Physics Center (IFIMAC), Universidad Autónoma de Madrid, E-28049 Madrid, Spain}
\author{Frieder Lindel}
    \email{frieder.lindel@uam.es}
    \affiliation{Departamento de Física Teórica de la Materia Condensada, Universidad Autónoma de Madrid, E-28049 Madrid, Spain}
    \affiliation{Condensed Matter Physics Center (IFIMAC), Universidad Autónoma de Madrid, E-28049 Madrid, Spain}
\author{Esteban Moreno}
    \email{esteban.moreno@uam.es}
    \affiliation{Departamento de Física Teórica de la Materia Condensada, Universidad Autónoma de Madrid, E-28049 Madrid, Spain}
    \affiliation{Condensed Matter Physics Center (IFIMAC), Universidad Autónoma de Madrid, E-28049 Madrid, Spain}
\author{Francisco J. García-Vidal}
    \email{fj.garcia@uam.es}
    \affiliation{Departamento de Física Teórica de la Materia Condensada, Universidad Autónoma de Madrid, E-28049 Madrid, Spain}
    \affiliation{Condensed Matter Physics Center (IFIMAC), Universidad Autónoma de Madrid, E-28049 Madrid, Spain}
\author{Johannes Feist}
    \email{johannes.feist@uam.es}
    \affiliation{Departamento de Física Teórica de la Materia Condensada, Universidad Autónoma de Madrid, E-28049 Madrid, Spain}
    \affiliation{Condensed Matter Physics Center (IFIMAC), Universidad Autónoma de Madrid, E-28049 Madrid, Spain}
\date{\today}

\begin{abstract}
While experimental measurements of photon correlations have become routine in laboratories, theoretical access to these quantities for the light generated in complex nanophotonic devices remains a major challenge. Current methods are limited to specific simplified cases and lack generality. Here we present a novel method that provides access to photon statistics resolved in space and frequency in arbitrary electromagnetic environments.
Within the macroscopic QED framework, we develop a practical tool to compute electric field correlations for complex quantum systems by including lossy two-level systems that act as field detectors within the system. To make the implementation feasible, we use a recently developed multi-emitter few-mode quantization method to correctly account for fully retarded light propagation to the detectors. 
We demonstrate the effectiveness and robustness of the proposed technique by studying the photon correlations of one and two emitters in close proximity to a plasmonic nanoparticle. The simulations show that even in these relatively simple configurations, the light statistics exhibit a strong angular dependence. These results highlight the importance of going beyond conventional quantum-optical approaches to 
fully capture the analyzed physical effects and enable the study of the quantum light generation in realistic nanophotonic devices.
\end{abstract}

\maketitle

\section{Introduction}
\label{sec.introduction}

Quantum optics is the branch of modern science which focuses on the quantum features of light emitted from quantum sources. This field drives strategic technological advances in quantum information processing~\cite{OBrien2009}. Notable examples include the generation and observation of entangled photons in various platforms~\cite{Yao2012,Hiesmayr2016,Huber2017,Prasad2020,Schimpf2021,Wein2022,Zhang2022,Vento2023}, and the development and characterization of single- and two-photon sources, essential for future quantum communication protocols~\cite{Darquie2005,Lounis2005,Reindl2017,Gao2023,Olejniczak2024,Khalid2024,Esmann2024}. In this context, photon correlations have become an essential tool in characterizing the properties of light-emitting devices. 
For example,
the first-order correlation function, $G^{(1)}$, provides information regarding the intensity of emitted light, and the second-order correlation function, $G^{(2)}$, offers insights into the statistical properties of light and is used to distinguish coherent emission from single-photon or thermal light~\cite{Ben-Asher2023,Shan2023,Groiseau2024,Putintsev2024}.

The advancement in the field of quantum technologies hinges on the development of efficient light-matter interfaces, which facilitate the coupling of quantum emitters to electromagnetic (EM) fields. It is therefore important to control and maximize this interaction in order to effectively manipulate the properties of light emitted by quantum sources. Given the inherently weak interactions in free-space, special attention has been paid to optical (wavelength-scale) cavities which act as amplifiers of the light-matter coupling~\cite{Faraon2008,Chang2014Quantum,Loredo2019,Tomm2021,Lukin2023}. In such setups, the interaction can be well described by the simplified picture developed within the field of cavity quantum electrodynamics (cavity QED), where emitters are coupled to confined photonic modes. This effective framework then grants access to the photon correlations~\cite{Ridolfo2010,Auffeves2011,delValle2012,Saez-Blazquez2017,Saez-Blazquez2018Photon,Peyskens2018,Rousseaux2020}.

The coupling strength can be enhanced even more by using subwavelength confinement, achievable in nanophotonic devices such as plasmonic or dielectric nanocavities~\cite{Chikkaraddy2016,Benz2016,Franke2019,Heintz2021}. Interactions enabled by such nanostructures allow to achieve the strong-coupling regime with a few emitters, promising development of a new generation of quantum technologies~\cite{Gonzalez-Tudela2024}. Nevertheless, the EM fields supported by these structures are rather complex and generally cannot be mapped to the simplified models extensively used within cavity QED\@. The first major challenge lies in the quantization of the EM field in the presence of dispersive and absorbing media. To this end, macroscopic quantum electrodynamics (MQED) provides a systematic approach towards quantizing light in complex environments described through their constitutive relations~\cite{Huttner1992,Scheel2008,Buhmann2012I,Buhmann2012II}. However, the power and generality of MQED comes at the cost of a four-dimensional continuum of field operators in space and frequency, substantially restricting its straightforward applicability. Over the last years, several approaches to simplify the full MQED Hamiltonian to a quantum-optics like description in terms of a (small) number of discrete modes have been developed. One is the quantization of quasinormal modes~\cite{Hughes2018,Franke2019}, while another is a few-mode model in which the EM continua are mapped to an equivalent set of discrete, coupled and lossy modes via a fitting procedure~\cite{Medina2021,Sanchez-Barquilla2021,Sanchez-Barquilla2022,Lednev2024}.

Nevertheless, an important limitation of these approaches is that information about the emitted EM field is not straightforwardly available. While quasinormal mode expansions could formally provide this, convergence of this expansion requires the inclusion of an untenable number of modes~\cite{Kristensen2020}. Alternatively, the field operators and correlations can be obtained by relating them to emitter correlation functions, as done in Ref.~\cite{Medina2021}. However, this requires the evaluation of nested integrals of $2N$-time correlation functions of the emitters to obtain even equal-time field correlators $G^{(N)}$, which quickly becomes numerically intractable for $N\geq2$. This unfavorable scaling with $N$ makes it expedient to find a different way to retrieve the information more readily accessible in experiments, that is, photodetection signals and their correlations resolved in space, time, and frequency.

\begin{figure}[tb]
    \includegraphics[width=\linewidth]{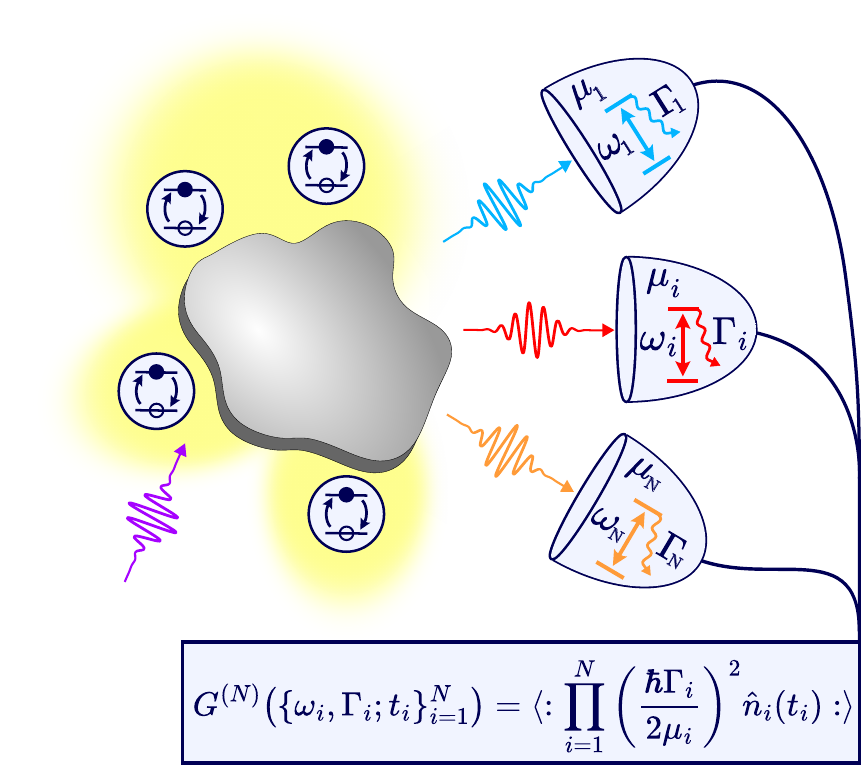}
    \caption{Sketch of the method. The light coming from a complex nanophotonic setup is probed by auxiliary lossy TLSs. Field correlation functions can then be computed as correlations between the operators of the inserted TLSs.}
    \label{fig:sketch}
\end{figure}

We here demonstrate a straightforward approach to resolve this problem, based on the simple idea of directly mirroring experiments by explicitly describing the detectors, modeled as lossy two-level systems (TLSs), as part of the system (see the sketch in \autoref{fig:sketch}). Simplified models of such detectors and the signals produced by them are indeed how the electric field operators of interest are typically derived in quantum optics~\cite{Loudon2000,Grynberg2010}. In this sense, the method proposed here simply corresponds to treating the photodetectors explicitly instead of calculating the field correlation functions that correspond to the signals they produce. A similar idea has been successfully applied in cavity QED models to access frequency-resolved correlations of a single mode~\cite{delValle2012,Holdaway2018}, which otherwise also require the computation of nested integrals of high-order temporal correlators~\cite{Nienhuis1993,Joosten2000}. 

In order to be able to represent general nanophotonic setups, we exploit the recently developed multiemitter few-mode model~\cite{Sanchez-Barquilla2022} to efficiently treat both the emitters and detectors and the light propagation between them, fully including retardation and propagation effects. In doing so, we develop, to the best of our knowledge, the first theoretical tool that provides photon correlations in arbitrary nanophotonic setups resolved in polarization, space, time, and frequency. We also note that the method is not limited to the weak-coupling limit or the approximation of emitters as two-level systems, but can be applied to arbitrary multi-level emitters and coupling regimes, including in the ultrastrong-coupling limit~\cite{Lednev2024}.

We showcase the effectiveness and power of our approach through the examination of photon correlations in the emission of one and two quantum emitters in close proximity to a spherical plasmonic nanoparticle. Our results reveal that, even for this simple geometry, the quantum correlations of the emitted light exhibit a strong dependence on the position and frequency of the detectors and the timing of the detection. The contents of the article are organized as follows: First, we outline the theoretical and methodological aspects of our approach in \autoref{sec.theory}, and then highlight some practical aspects of its implementation in \autoref{sec.implementation}. In \autoref{sec.test}, we study the interaction of a TLS with a quasi-single-mode environment that allows us to compare our method with other approaches widely used in quantum optics. We then analyze spatially resolved correlations in a more complex, multimode system in detail in \autoref{sec.multimode}. Last, \autoref{sec.conclusions} summarizes the main results and conclusions of our study.

\section{Theoretical framework}
\label{sec.theory}
In this section, we provide the theoretical foundation of our method. First, using the MQED framework we show that the field correlation functions can be directly related to the operators of auxiliary TLSs weakly coupled to the EM field.

Second, we will show that the numerical complexity of this problem can be made manageable by using the few-mode quantization method~\cite{Medina2021,Sanchez-Barquilla2021,Sanchez-Barquilla2022,Lednev2024}. This combination then provides a computationally feasible way to calculate electric field correlations. 

\subsection{Two-level detectors in MQED}
\label{sec.aux_tls}

We start by considering $N_\mathrm{e}$ quantum emitters interacting with the modes of an general EM environment. Within MQED, the Hamiltonian of the system in dipole approximation and within the Power-Zienau-Woolley picture reads
\begin{align}
	\hat{H} &= \sum_{\alpha=e,m} \int \mathrm{d}^3r \int_0^\infty \mathrm{d}\omega\ \hbar\omega \hat{\mathbf{f}}_\alpha^\dagger(\mathbf{r},\omega) \cdot \hat{\mathbf{f}}_\alpha(\mathbf{r}, \omega)\nonumber\\
	&+\sum^{N_\mathrm{e}}_{i=1} \hat{H}^\mathrm{e}_i - \sum^{N_\mathrm{e}}_{i=1} \hat{\boldsymbol{\mu}}^{\mathrm{e}}_i \cdot\hat{\mathbf{E}}(\mathbf{r}_i).
	\label{H_N}
\end{align}
Here, $\hat{H}^\mathrm{e}_i$ and $\hat{\boldsymbol{\mu}}^{\mathrm{e}}_i$ represent the bare Hamiltonian and electric dipole operator of the $i$-th emitter, respectively. Additionally, $\hat{\mathbf{f}}_\alpha(\mathbf{r}, \omega)$ and $\hat{\mathbf{f}}_\alpha^\dagger(\mathbf{r},\omega)$ are bosonic annihilation and creation operators for the macroscopic (medium-assisted) EM field, fulfilling $[\hat{\mathbf{f}}_\alpha(\mathbf{r}, \omega), \hat{\mathbf{f}}_\beta^\dagger(\mathbf{r}', \omega')] = \delta_{\alpha\beta} \boldsymbol{\delta}(\mathbf{r}-\mathbf{r}') \delta(\omega-\omega')$, where the label $\alpha=\{e,m\}$ denotes electric and magnetic contributions, respectively. The electric field operator $\hat{\mathbf{E}}(\mathbf{r})$ is given in terms of $\hat{\mathbf{f}}_\alpha(\mathbf{r}, \omega)$ and $\hat{\mathbf{f}}_\alpha^\dagger(\mathbf{r},\omega)$ and contains information about the nanophotonic structure through the EM Green's tensor (see \appref{appendix.Efield_op} for details).

One of the main pillars of this article is that the response of realistic photodetectors can be modeled with auxiliary TLSs~\cite{delValle2012}, with the ensuing computational simplification in the calculation of field correlations. Let us, therefore, include into our system $N_\mathrm{d}$ auxiliary two-level systems with vanishingly small dipole moments $\mu^{\mathrm{d}}_i$, oriented along $\mathbf{n}_i$.

The corresponding Hamiltonian reads:
\begin{multline}
    \hat{H}_\mathrm{d} = \hat{H} + \sum_{i=1}^{N_\mathrm{d}}\hbar\widetilde{\omega}_i\hat{d}_i^\dagger \hat{d}_i\\
	-\sum^{N_\mathrm{d}}_{i=1} \mu^{\mathrm{d}}_i \mathbf{n}_i \cdot\left[\hat{\mathbf{E}}^+(\mathbf{r}_i)\hat{d}_i^\dagger + \hat{\mathbf{E}}^-(\mathbf{r}_i)\hat{d}_i\right],
\label{eq:H_d}  
\end{multline}
where $\hat{H}$ is given by \autoref{H_N}, $\hat{\mathbf{E}}^{\pm}$ represent the positive and negative frequency components of the electric field operator, $\hat{d}_i$ ($\hat{d}^\dagger_i$) are the lowering (raising) operators of the auxiliary TLSs, and $\widetilde{\omega}_i = \omega_i - \frac{i}{2}\Gamma_i$ are their complex frequencies, whose imaginary parts account for the detector bandwidths\footnote{For simplicity, we here directly write the effective non-Hermitian Hamiltonian that describes the coherent evolution of the system within a Lindblad master equation, as the additional ``refilling'' or ``quantum jump'' term does not affect the Heisenberg equations of motion in the low-excitation limit for the detectors.}.

It is important for our purposes that the coupling between the auxiliary TLSs and the electric field must be perturbative, such that the dynamics of $\hat{\mathbf{E}}^{\pm}$ is not affected by the auxiliary TLSs. In other words, the TLSs only probe the electric field without modifying it. Accordingly, we applied the rotating-wave approximation in \autoref{eq:H_d}. Furthermore, the small coupling allows us to derive a simple relation between $d_i$ and $\mathbf{E}^+$ through the Heisenberg equations of motion:
\begin{equation} \label{eq:dipole_to_E}
	\hat{d}_i(t) = i\frac{\mu^{\mathrm{d}}_i}{\hbar}\int_0^t \mathrm{d}\tau\ \mathbf{n}_i\cdot\hat{\mathbf{E}}^+(\mathbf{r}_i,\tau)\cdot e^{-i(\omega_i - \frac{i}{2} \Gamma_i)(t-\tau)}.
\end{equation}
To obtain this expression, we assumed the TLSs to initially be in their ground state, and treated the TLSs as harmonic oscillators with $[d_i, d_j^\dagger]=\delta_{ij}$, which is valid in the small coupling limit.

The dipole operator in \autoref{eq:dipole_to_E} is proportional to the electric field operator probed by a realistic detector, including a finite detection bandwidth~\cite{Eberly1977, Knoll1984, Arnoldus1984} (see \appref{appendix.Efield_op} for details). This allows us to write
\begin{equation}
    \hat{E}_{\mathrm{d}_i}^+(t) = -i\frac{\hbar\Gamma_i}{2\mu^{\mathrm{d}}_i}\hat{d}_i(t),
    \label{eq:equivalence1}
\end{equation}
where $\hat{E}^{+}_{\mathrm{d}_i}$ is the positive-frequency component of the frequency-filtered electric field operator polarized along $\mathbf{n}_i$ at the position of detector $i$. We provide an explicit numerical validation of this expression in \autoref{fig:comparison} in \appref{appendix.Efield_op}. The obtained result provides a recipe to access the correlations of the electric field via correlations between the auxiliary TLSs:
\begin{subequations}
    \begin{align}
        G^{(N)}\big(\{\omega_i,\Gamma_i; t_i&\}_{i=1}^N\big) \equiv \langle :\prod_{i=1}^N \hat{E}^-_{\mathrm{d}_i}(t_i)\hat{E}^+_{\mathrm{d}_i}(t_i): \rangle \label{eq:equivalence2.0}
        \\ &= \langle :\prod_{i=1}^N \left(\frac{\hbar\Gamma_i}{2\mu^{\mathrm{d}}_i}\right)^{2}\hat{d}^\dagger_i(t_i)\hat{d}_i(t_i): \rangle,
        \label{eq:equivalence2}
    \end{align}
\end{subequations}
where $\hat{E}^{-}_{\mathrm{d}} = (\hat{E}^{+}_{\mathrm{d}})^\dagger$, and $:\!\hat{O}\!:$ denotes normal ordering of $\hat{O}$. This equivalence demonstrates that the auxiliary weakly-coupled TLSs indeed act as detectors of the EM field. In \appref{appendix.class}, we show explicitly that this expression remains valid when the field is split into a quantum and a classical part.

We emphasize that \autoref{eq:equivalence2} makes photon correlations much more accessible than directly evaluating \autoref{eq:equivalence2.0}, as the simple temporal evolution of the system leads to an automatic evaluation of the nested temporal integrals inherent to \autoref{eq:equivalence2.0}. The result is similar to the one obtained in Ref.~\cite{delValle2012}, frequently used to compute frequency-filtered correlations of a \emph{single} field mode in cavity QED setups. In contrast, the expression in \autoref{eq:equivalence2} relates detector observables to the full \emph{multimode} electric field correlations sampled at arbitrary points in space and with arbitrary polarization, and also including frequency filtering.

\subsection{Few-mode quantization}
\label{sec.fm_quant}
While expression \autoref{eq:equivalence2} provides a formal expression to compute field correlations that eliminates the need to explicitly perform temporal convolutions, a straightforward implementation of this expression within MQED is computationally extremely demanding due to the presence of a four-dimensional continuum of macroscopic EM field modes. To mitigate this issue, we use the few-mode quantization method developed in~\cite{Medina2021, Sanchez-Barquilla2022}, which allows us to replace the EM environment by a collection of discrete, lossy and interacting bosonic modes with annihilation operators $\hat{a}_\alpha$. The dynamics of the system consisting of the $N_\mathrm{m}$ modes interacting with $N_\mathrm{e}$ emitters and $N_\mathrm{d}$ detectors is given by:
\begin{subequations}\label{eq:Hmodel}
    \begin{align}
        &\hat{H}^\mathrm{mod} = \sum_{k} \hat{H}^\mathrm{e}_k + \sum_{l} \hbar\omega_l \hat{d}_l^\dagger \hat{d}_l + \sum_{\alpha,\beta} \hbar\omega_{\alpha\beta} \hat{a}_\alpha^{\dagger} \hat{a}_\beta + \nonumber\\
        &\sum_{k,\alpha} \hbar\lambda_{\mathrm{e},k\alpha}(\hat{a}_\alpha^{\dagger} + \hat{a}_\alpha)\frac{\hat{\mu}^{\mathrm{e}}_k}{\mu^{\mathrm{e}}_k} + \sum_{l,\alpha} \hbar\lambda_{\mathrm{d},l\alpha}(\hat{a}_\alpha^{\dagger}\hat{d}_l + \hat{a}_\alpha \hat{d}_l^\dagger),\\
        &\dot{\hat{\rho}} = -\frac{i}{\hbar}[\hat{H}^\mathrm{mod},\hat\rho] + \sum_{\alpha=1}^{N_\mathrm{m}} \kappa_\alpha L_{\hat{a}_\alpha}[\hat{\rho}] + \sum_{i=1}^{N_\mathrm{d}} \Gamma_i L_{\hat{d}_i}[\hat{\rho}],
    \end{align}
\end{subequations}
with $L_{\hat{O}}[\hat{\rho}] = \hat{O}\hat{\rho}\hat{O}^\dagger - \frac{1}{2}\{\hat{O}^\dagger\hat{O},\hat{\rho}\}$. The indices take the values $\alpha,\beta=1,\dots,N_\mathrm{m}$; $k=1,\dots,N_\mathrm{e}$; and $l=1,\dots,N_\mathrm{d}$. The modes are characterized by their frequencies and couplings, $\omega_{\alpha\beta}$, and by their losses $\kappa_\alpha$. Their coupling to the emitters (detectors) is given by the matrices $\boldsymbol{\lambda}_\mathrm{e}$ ($\boldsymbol{\lambda}_\mathrm{d}$), where for simplicity of notation we have assumed that only one dipole direction per emitter is relevant. We have also written the emitter dipole operators normalized to arbitrary reference dipole moments $\mu^{\mathrm{e}}_k$ to ensure that $\boldsymbol{\lambda}_\mathrm{e}$ and $\boldsymbol{\lambda}_\mathrm{d}$ have the same units.

This model is exactly equivalent to the Hamiltonian in \autoref{eq:H_d} when the EM spectral density for the real system and the few-mode model match~\cite{Breuer2007,Tamascelli2018,Menczel2024}. These spectral densities are given, respectively, by $(N_\mathrm{e}+N_\mathrm{d})\times(N_\mathrm{e}+N_\mathrm{d})$ matrix-valued functions:
\begin{subequations}
    \begin{align}
	   J_{ij}(\omega)&=\frac{\omega^2}{\hbar\pi\epsilon_0 c^2}\mathrm{Im}\left[\boldsymbol{\mu}_i\cdot\boldsymbol{\mathcal{G}}(\mathbf{r}_i,\mathbf{r}_j,\omega)\cdot\boldsymbol{\mu}_j\right], \label{eq:Jij}\\
        J_{ij}^\mathrm{mod}(\omega)&=\frac{1}{\pi}\mathrm{Im}\left[\boldsymbol{\Lambda}\cdot \frac{1}{\widetilde{\mathbf{H}}-\omega} \cdot\boldsymbol{\Lambda}^T\right]_{ij},
	\end{align}
\end{subequations}
where $\boldsymbol{\mathcal{G}}$ is the dyadic Green's tensor, $\boldsymbol{\Lambda} = \begin{pmatrix}\boldsymbol{\lambda}_\mathrm{e}\\ \boldsymbol{\lambda}_\mathrm{d}\end{pmatrix}$, $\widetilde{H}_{\alpha\beta}=\omega_{\alpha\beta}-\frac{i}{2}\delta_{\alpha\beta}\kappa_\alpha$, and the indices $i,j=1,\dots,N_\mathrm{e}+N_\mathrm{d}$ span both the emitters and the detectors. The equivalence of both approaches [\autoref{eq:H_d} and \autoref{eq:Hmodel}] can then be achieved by fitting the matrix function $\mathbf{J}^\mathrm{mod}(\omega)$ to the physical spectral density $\mathbf{J}(\omega)$ by way of the real parameters $\omega_{\alpha\beta}$, $\kappa_\alpha$, $\lambda_{\mathrm{e}, k\alpha}$ and $\lambda_{\mathrm{d}, l\alpha}$. Once the parameter values are determined, the dynamics of the system can be found by solving the affordable Lindblad equation~\eqref{eq:Hmodel}, as opposed to propagating the dynamics with a four-dimensional continuum of medium-assisted modes. This approach has been shown to work in general nanophotonic EM environments in any coupling regime~\cite{Medina2021,Sanchez-Barquilla2022,Lednev2024}.

In summary, the method presented in this section provides access to photon correlations in complex nanophotonic setups via three steps: (i) fit the real system's spectral density with $\mathbf{J}^\mathrm{mod}(\omega)$, (ii) compute the density matrix of the system with the corresponding spectral density using \autoref{eq:Hmodel}, (iii) compute the correlations between the detectors with \autoref{eq:equivalence2}. With this straightforward recipe, our theoretical framework
(i) treats the interaction of the emitters with realistic dispersive and absorbing nanophotonic environments; (ii) accounts for the spatial dependence of the field propagation from the emitters to the detectors, e.g., the out-coupling of the field from the nanophotonic setup to the detectors incorporates the whole complexity of the environment and comprises fundamental physical properties, such as retardation; (iii) treats the interaction between the emitters and EM modes in arbitrary coupling regimes, including strong and ultrastrong coupling; and (iv) effectively models the experimental setup by allowing to detect not only normalized correlation functions, but also the absolute magnitude of unnormalized correlation functions with their corresponding physical units.

While this concludes the description of the general theory behind our method, in \autoref{sec.implementation} we next present several practical considerations that help to improve convergence of the fitting procedure, reduce the computational cost, and ensure the accuracy of the results. We then proceed to benchmark and explore its capabilities in \autoref{sec.test} and \autoref{sec.multimode}.

\begin{figure*}[th!]
   \subfloat{\label{fig:J_func_sketch}}
   \subfloat{\label{fig:J_func_ee}}
   \subfloat{\label{fig:J_func_ed}}
   \includegraphics[width=\linewidth]{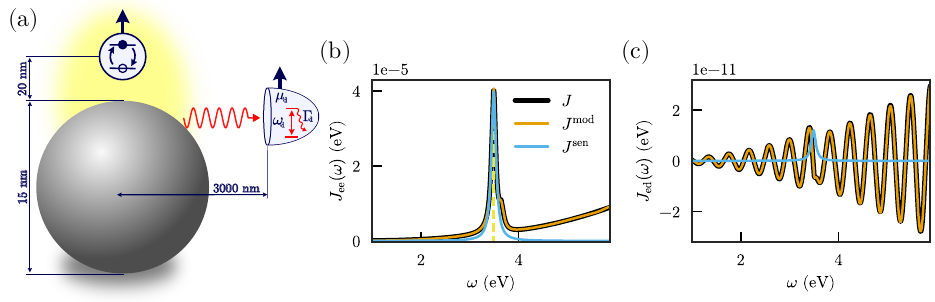}
   \caption{(a) Sketch of the geometrical configuration of the system. The dark blue arrows near the emitter and detector indicate the direction of the corresponding dipole moments. (b)-(c) Spectral densities relevant for light detection (black). Orange lines represent the fits provided by our method, while blue line shows spectral densities retrieved from the single-mode sensor method (see \appref{appendix.sensor_method}).
   The spectral position of the emitter transition is marked by the yellow line in (b).\label{fig:J_func}}
\end{figure*}

\section{Implementation}
\label{sec.implementation}

In this section, we discuss some practical aspects of the implementation of the theory described in \autoref{sec.theory}. In order to make some of the discussions more concrete, we consider a specific example system, which we will later also use to benchmark the method and compare to alternative formulations. The system, sketched in \autoref{fig:J_func_sketch}, consists of a two-level emitter close to a spherical metallic nanoparticle, for which $\boldsymbol{\mathcal{G}}$ can be obtained analytically via a multipolar expansion~\cite{Chew1999Waves,Scheel2008}. The nanoparticle's diameter is 15~nm, and the emitter at $\mathbf{r}_\mathrm{e}$ is located 20~nm away from its surface. Additionally, we place two detectors at the same position $\mathbf{r}_\mathrm{d}$, 3~$\mathrm{\mu m}$ away from the nanophotonic system, at an angle of $90^\circ$ with respect to the nanoparticle-emitter axis.

\subsection{General features of the spectral densities}
The first step to implement our method is the computation of the spectral density matrices corresponding to the system. They can be separated into three blocks: the emitter-emitter spectral density $\mathbf{J}_\mathrm{ee}(\omega)$ of size $N_\mathrm{e} \times N_\mathrm{e}$, the detector-detector spectral density $\mathbf{J}_\mathrm{dd}(\omega)$ of size $N_\mathrm{d} \times N_\mathrm{d}$, and the crossed emitter-detector spectral density $\mathbf{J}_\mathrm{ed}(\omega) = \mathbf{J}_\mathrm{de}^T(\omega)$ of size $N_\mathrm{e} \times N_\mathrm{d}$. They describe different physical processes due to the EM field and thus generically behave quite differently: The emitter-emitter spectral density $\mathbf{J}_\mathrm{ee}(\omega)$ fully determines the dynamics of the emitters due to their interaction with the EM field, while the crossed emitter-detector spectral density $\mathbf{J}_\mathrm{ed}(\omega)$ describes the propagation of the signal from the emitter to the detectors.
The detector-detector spectral density $\mathbf{J}_\mathrm{dd}(\omega)$, which describes the back-action of the EM field on the detectors, is quadratic in the small parameter $\mu^{\mathrm{d}}$. Its effect on the dynamics of the detectors is thus negligible, such that we can exclude $\mathbf{J}_\mathrm{dd}(\omega)$ from our analysis and do not need to fit it, which greatly simplifies the practical implementation of our method.

In \autoref{fig:J_func_ee} and \autoref{fig:J_func_ed}, we show the spectral densities for the system described above. Due to the chosen setup (one emitter, two identical detectors), the two relevant blocks are described by scalar functions $J_\mathrm{ee}$ and $J_\mathrm{ed}$. The black lines represent the spectral densities obtained from the full EM environment, while the orange lines show the fits provided by our method (with $N_\mathrm{m}=21$ modes). The blue lines show the spectral densities obtained from the single-mode sensor method (see \appref{appendix.sensor_method}).
As expected~\cite{Delga2014}, the emitter spectral density $J_\mathrm{ee}(\omega)$ in \autoref{fig:J_func_ee} shows a clear peak due to the dipolar surface plasmon resonance, with a small shoulder due to higher-order multipole resonances. 

The information about the signal's propagation from the emitter to the detectors is contained in the crossed emitter-detector spectral density $J_\mathrm{ed}(\omega)$ (\autoref{fig:J_func_ed}). Its most salient feature are oscillations inherent to the free-space spectral density. Their amplitude grows with $\omega^2$ and their period is $2\pi c/|\mathbf{r}_\mathrm{e} - \mathbf{r}_\mathrm{d}|$, which is essentially the inverse time the signal takes to reach the detectors from the system. Indeed, the oscillations encode retardation effects due to the finite value of the speed of light. Since our method can capture these oscillations, it follows that the signal will automatically arrive at the detectors only after the appropriate temporal delay (see \autoref{fig:ww} for an example). However, for large emitter-detector separations, and thus rapid oscillations, a faithful reproduction of $J_\mathrm{ed}$ with $J^{\mathrm{mod}}_\mathrm{ed}$ requires a significant number of modes. Behind this fact hides a deeper understanding of how retardation can be captured by the few-mode model. Essentially, the coupled modes of the model form a network or graph that delays the signal propagation from the emitter to the detectors. Thus, this network of modes must be large enough to accommodate the physical retardation. This simple intuition also serves to illustrate the incapacity of single-mode models to capture retardation effects.

\subsection{Importance of the real part of the Green's tensor}

As discussed in \autoref{sec.theory}, in theory it is sufficient to only fit the (generalized) spectral density of a given system to obtain a fully equivalent few-mode model. These spectral densities are proportional to the imaginary part of the relevant response functions (determined by the retarded Green's tensor in the current case). This seems to imply that their real part does not matter. Indeed, due to the causal nature of the response, the real and imaginary parts are linked by a Kramers-Kronig relation, and determining the imaginary part thus also fully determines the real one.

However, while this is in principle true, it should be kept in mind that the Kramers-Kronig relations are highly nonlocal in frequency, i.e., the real part at any given frequency is determined by the imaginary part at \emph{all} frequencies. In situations where the spectral density is appreciable only in a limited frequency range (i.e., close to the resonances of plasmonic nanostructures), it is usually possible to perform the fit in this range and obtain a good approximation to the full dynamics~\cite{Medina2021,Sanchez-Barquilla2022}, possibly after including any remaining contributions perturbatively~\cite{SanchezMartinez2024Mixed,Chuang2024}. However, in the current case of the emitter-detector spectral density, the oscillations due to the finite emitter-detector separation lead to a spectral density that is nonzero over a wide frequency range and not dominated by only a few resonances, but by the free-space contribution due to propagating photons. Furthermore, the relative timing information about photon propagation is encoded in the phase of the Green's tensor, which is directly related to its real part. At the same time, for a practical implementation, it is desirable to truncate the frequency range over which the fitting must be performed to reduce the numerical complexity of the problem. When performing such a truncation, it is thus crucial to ensure that the real part of the Green's tensor is also accurately captured in the truncated frequency range in order to preserve the correct timing relations. In order to achieve this, we implement an extension of the method explained so far that removes this problem and allows for a straightforward frequency truncation. This is achieved by fitting the \emph{full retarded response function} $\mathbf{R}(\omega)$, given by
\begin{subequations}
    \begin{equation}
        \label{eq:response_function}
        R_{ij}(\omega)=\frac{\omega^2}{\hbar\epsilon_0 c^2}\boldsymbol{\mu}_i\cdot\boldsymbol{\mathcal{G}}(\mathbf{r}_i,\mathbf{r}_j,\omega)\cdot\boldsymbol{\mu}_j,
    \end{equation}
    which fulfills $\Im[\mathbf{R}(\omega)] = \pi \mathbf{J}(\omega)$, in the frequency range of the emitters~\cite{Dung2002Resonant,Asenjo-Garcia2017,KamandarDezfouli2017,Chang2018Colloquium,Zhang2021Addressing,Esteban2022,Sheremet2023,Dura-Azorin2024}. Then, matching the few-mode model response function
    \begin{equation}
        \label{eq:model_response_function}
        \mathbf{R}^\mathrm{mod}(\omega)= \boldsymbol{\Lambda} \cdot \frac{1}{\widetilde{\mathbf{H}}-\omega} \cdot \boldsymbol{\Lambda}^T
    \end{equation}
\end{subequations}
to the physical one, $\mathbf{R}(\omega)$, greatly mitigates the problem of the finite fitting range for $\mathbf{J}(\omega)$ and permits a straightforward truncation of the fitted frequency region to the physically relevant frequencies.
In \appref{appendix.re_resp_func}, we show the real parts of the response functions for the model system introduced above, together with fits provided by our method, and discuss the results and details of the joint fitting procedure. Furthermore, in \appref{appendix.fit_opt} we show how the fitting procedure can be considerably simplified and its computational cost reduced by exploiting that the detectors are only weakly coupled to the EM field.

\subsection{Implementation of other common approaches}
\label{sec:other_methods}

Below, we will benchmark our method and compare its performance with three approaches commonly used to describe similar systems, which we summarize here:

(i) The single-mode sensor method, which is widely used in the quantum optics community to study cavity mode correlations~\cite{delValle2012,Gonzalez-Tudela2013Two-photon,SanchezMunoz2014Violation,SanchezMunoz2018Filtering,Schmidt2021,Martinez-Garcia2024,Juan-Delgado2024}. Specifically, this model corresponds to a Jaynes-Cummings Hamiltonian describing the interaction of the two-level emitter with a single ``cavity'' mode. Additionally, we include one or two two-level detectors to probe the mode correlations. Details concerning the implementation of this method are presented in \appref{appendix.sensor_method}. 

(ii) A direct calculation using the MQED Hamiltonian in \autoref{eq:H_d} in the Born-Markov approximation~\cite{Dung2002Resonant,KamandarDezfouli2017,Asenjo-Garcia2017,Chang2018Colloquium,Sheremet2023,Dura-Azorin2024}. The final form of the Hamiltonian is presented in \appref{appendix.mark_approx}. This approach neglects any retardation effects, and is restricted to the weak coupling regime and resonant interactions ($\omega_\mathrm{e}\approx\omega_\mathrm{d}$).\footnote{Note that the Markovian approximation highlights the importance of both the real and imaginary parts of the response function at the emitter and detector frequencies, see \appref{appendix.mark_approx} for details.}

(iii) An extension of the Wigner-Weisskopf theory for spontaneous emission to the setup considered here, as described in \appref{appendix.ww}, leads to an explicit expression for the electric field intensity, given by \autoref{eq:ww_an}. This approach can only be applied for the spontaneous emission problem and results from a perturbative treatment of the light-matter interaction. It is thus expected to accurately describe the setup in the weak coupling regime only.

\section{Single-mode system}
\label{sec.test}

In this section, we employ our approach to analyze the dynamics of the quasi-single-mode system with one emitter and one or two detectors described in \autoref{sec.implementation} and sketched in \autoref{fig:J_func_sketch}. This simple system allows us to benchmark our method against the commonly used approaches discussed in \autoref{sec:other_methods}, and to explore the capabilities of our approach in the weak and strong coupling regimes.

\subsection{Resonance fluorescence}
\label{subsec.res_fluor}

\begin{figure}[tb]
   \includegraphics[width=\linewidth]{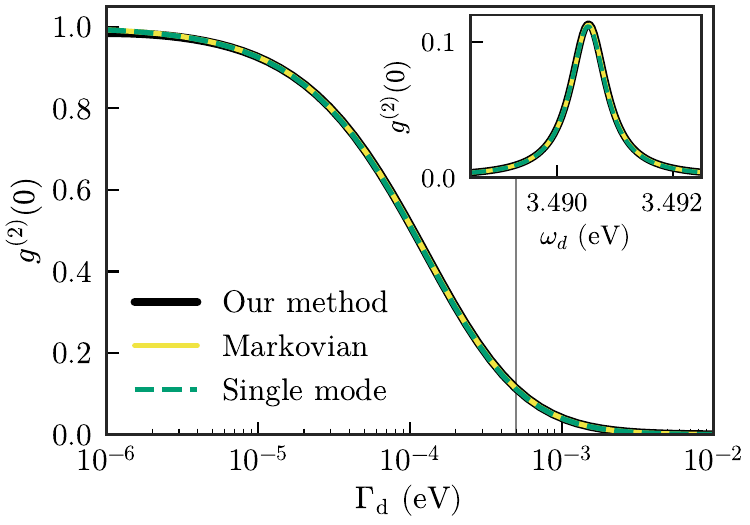}
   \caption{Resonance fluorescence: Dependence of the steady-state $g^{(2)}(0)$ on the detector linewidth $\Gamma_\mathrm{d}$ (at $\omega_\mathrm{d}=\wlas=\omega_\mathrm{e}$). Inset: dependence of steady-state $g^{(2)}(0)$ on detector frequency $\omega_\mathrm{d}$ (at $\Gamma_\mathrm{d}=5\times10^{-4}$~eV, indicated by the vertical gray line).}
   \label{fig:fluorescence}
\end{figure}

We begin by treating the resonance fluorescence of a two-level system in the weak-coupling regime, which has been extensively explored over the past decades~\cite{Mollow1969,Carmichael1976,Kimble1977,Hanschke2020,Phillips2020,ZubizarretaCasalengua2020Tuning}. This system, therefore, is a perfect candidate for benchmarking our method. In particular, we study the situation when the emitter, weakly interacting with the EM environment ($\mu_\mathrm{e}=50~\mathrm{D}$), is additionally weakly driven by a resonant laser. We note that the action of the laser can be represented fully through a driving term acting only on the emitters and detectors, as long as the field used is that obtained at the emitter position taking into account the scattering by the nanosphere~\cite{Sanchez-Barquilla2020,Feist2021}. For simplicity, we assume that the driving laser is oriented and focused such that the classical field does not scatter to the detectors.
Accordingly, the Hamiltonian of the system includes a coherent pumping term only for the emitter, given in rotating-wave approximation by $\hat{H}_{\las} = - \mathbf{E}_{\las}(\mathbf{r}_\mathrm{e})\cdot\boldsymbol{\mu}_\mathrm{e}\left(\hat{\sigma}^\dagger_\mathrm{e}e^{-i\wlas t} + \hat{\sigma}_\mathrm{e} e^{i\wlas t}\right)$ (with $\mathbf{E}_{\las}(\mathbf{r}_\mathrm{e})\cdot\boldsymbol{\mu}_\mathrm{e}=10^{-6}~\mathrm{eV}$). 
The explicit time dependence can be removed by a standard unitary transformation into the rotating frame, $\hat{U}(t)=\exp\left[-i\left(\hat{\sigma}^\dagger_\mathrm{e}\hat{\sigma}_\mathrm{e}+\sum_\alpha \hat{a}_\alpha^{\dagger} \hat{a}_\alpha + \sum_l \hat{d}_l^\dagger \hat{d}_l\right)\wlas t\right]$. The laser frequency is resonant with the emitter transition and the peak of the spectral density, $\wlas = \omega_\mathrm{e}=3.4905$~eV (marked as the vertical dashed line in \autoref{fig:J_func_ee}).

We include in these simulations two detectors to compute the normalized second-order correlation function $g^{(2)}(t_1,t_2)=G^{(2)}(t_1,t_2)/\left[G^{(1)}_1(t_1) G^{(1)}_2(t_2)\right]$, where $G^{(1)}_1$ and $G^{(1)}_2$ are the intensities probed by the first and the second detector, respectively. \autoref{fig:fluorescence} illustrates the dependence of the steady-state correlation function $g^{(2)}(t_2-t_1=0)$ on the linewidth of the detectors. It shows that $g^{(2)}(0)$ gradually changes from 1 to 0 as $\Gamma_\mathrm{d}$ increases, indicating a transition from coherent to antibunched light. This phenomenon is known to arise from the interference of the coherent scattering and incoherent signal components~\cite{ZubizarretaCasalengua2020Tuning,Hanschke2020,Phillips2020}. For this simple case and these observables, the Markovian and single-mode sensor methods are expected to be accurate. Our method gives results in agreement with these simpler approaches, which serves as a first important benchmark of its correctness. Moreover, the inset of \autoref{fig:fluorescence} shows the dependence of the steady-state $g^{(2)}(0)$ correlation function on the frequency of the detector. It shows a resonance around the laser frequency, which can also be accurately captured by the single-mode sensor method. Interestingly, the Markovian approach is also capable of reproducing the effect, despite the violation of the resonance condition ($\omega_\mathrm{d}\neq\omega_\mathrm{e}$). Nonetheless, since the range over which the detector frequency is varied is small, all the relevant couplings barely change across this spectral region, and, consequently, the statistics can still be well described by this approach.

\subsection{Spontaneous emission}

After showing that our method is capable of reproducing standard results from the field of cavity QED, we turn our attention to the spontaneous emission of a two-level system. Since only a single excitation is present in the system, at most a single photon is detected, and only one detector is required in the simulations. Again, we will compare the performance of our method with other commonly used approaches, as discussed in~\autoref{sec:other_methods}. We show that in this case, our method clearly outperforms the others and captures dynamics that are not otherwise accessible, even in the single-excitation regime.

\subsubsection{Weak coupling}

First, we focus on spontaneous emission in the weak coupling regime ($\mu_\mathrm{e}=50~\mathrm{D}$). We fix the transition frequency of the emitter at $\omega_\mathrm{e}=3.4905$~eV, in resonance with the peak of the spectral density (vertical dashed line in \autoref{fig:J_func_ee}). The detector, resonant with the emitter ($\omega_\mathrm{d}=\omega_\mathrm{e}$) and with a bandwidth of 3~eV, probes the polarization component perpendicular to the light propagation direction that lies in the emitter-nanosphere-detector plane sketched in \autoref{fig:J_func_sketch}. 

\begin{figure}[tb]
   \subfloat{\label{fig:ww_weak}}
   \subfloat{\label{fig:ww_strong}}
   \includegraphics[width=\linewidth]{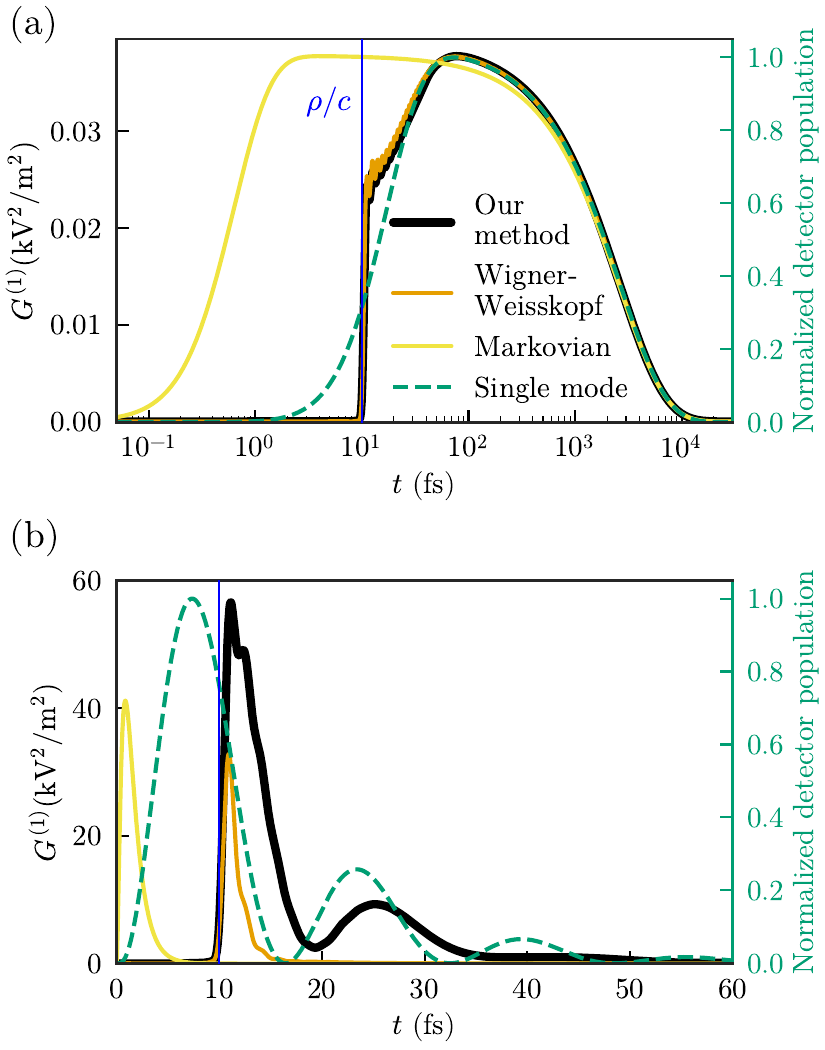}
   \caption{Spontaneous emission: the electric field intensity detected from (a) a weakly coupled and (b) a strongly coupled system, obtained with different methods. In the single-mode case, we show the detector population normalized to its maximum value.\label{fig:ww}}
\end{figure}

In \autoref{fig:ww_weak} we show the dynamics of the electric field intensity computed with our approach (black line). Since retardation effects are fully captured by our approach, the resulting signal appears only after the time $\tau=\rho/c$, where $\rho = |\mathbf{r}_\mathrm{d}-\mathbf{r}_\mathrm{e}|$. In contrast, the dynamics obtained via the single-mode sensor method (dashed green line) and via MQED in the Born-Markov approximation (solid yellow line) are not retarded. This is because retardation is essentially encoded in the oscillations of $J_\mathrm{ed}(\omega)$, as mentioned before. While our technique explicitly captures them, the associated spectral density of the single-mode sensor method, shown in blue in \autoref{fig:J_func_ee} and \autoref{fig:J_func_ed}, fails to do so (see \appref{appendix.sensor_method} for the details). It further cannot provide the absolute value of the intensity at any given point in space, such that we compare with the normalized $G^{(1)}$ function. Within the Markovian MQED method, the spectral density and real part of the response function are evaluated only at $\omega=\omega_\mathrm{e}=\omega_\mathrm{d}$ and the direct dipole-dipole interaction between the emitter and the detector is instantaneous. Nevertheless, both the single-mode sensor method and MQED in the Born-Markov approximation show general agreement in the temporal shape of the detected signal, especially at later times $t \gg \rho/c$, since the emitter lifetime is much longer than the propagation time from emitter to detector. Last, the Wigner-Weisskopf solution is represented by the orange line in \autoref{fig:ww_weak}. It clearly shows retardation effects, while deviating only slightly from our result at times slightly bigger than $\rho/c$. We attribute this to a small fitting inaccuracy inherent to our method and to the finite integration limits used in the numerical computation of \autoref{eq:ww_an}, which the method is very sensitive to, as we discuss in \appref{appendix.ww}. While the Wigner-Weisskopf solution works very well here, it is only applicable to the spontaneous emission problem where at most a single photon is present in the system, and furthermore limited to the weak-coupling regime as we will see in the next section.

\subsubsection{Strong coupling}

We now explore light detection from a strongly coupled system. To achieve this regime within the same model setup studied up to now, we increase the dipole moment of the emitter to an unrealistically large value ($\mu_\mathrm{e}=2500~\mathrm{D}$). Note, however, that room-temperature strong coupling with real emitters can be reached in more sophisticated nanostructures~\cite{Chikkaraddy2016,Gross2018,Heintz2021,Li2022Room,ThuHaDo2024,Hu2024,Liu2024Deterministic}. The intensity probed by the detector is shown in \autoref{fig:ww_strong}. The dynamics calculated with our method display two main features: retardation at times $\tau<\rho/c$, as in the weak-coupling regime, and Rabi oscillations characteristic of the strong coupling regime at times $\tau>\rho/c$. Rather remarkably, our method is the only one able to reproduce both features. The single-mode sensor description is able to capture Rabi oscillations with similar frequency, but it does not reproduce the delay and still only provides the \emph{normalized} intensity. The Wigner-Weisskopf result, which was very accurate in the weak coupling regime, keeps the retardation effects but fails to exhibit any Rabi oscillations. Last, the Markovian MQED approach completely fails on both accounts.

\section{Multimode system}
\label{sec.multimode}

\begin{figure}[tb]
   \subfloat{\label{fig:R_large_sketch}}
   \subfloat{\label{fig:R_large_spectral}}
   \subfloat{\label{fig:R_large_classical}}
   \includegraphics[width=0.91\linewidth]{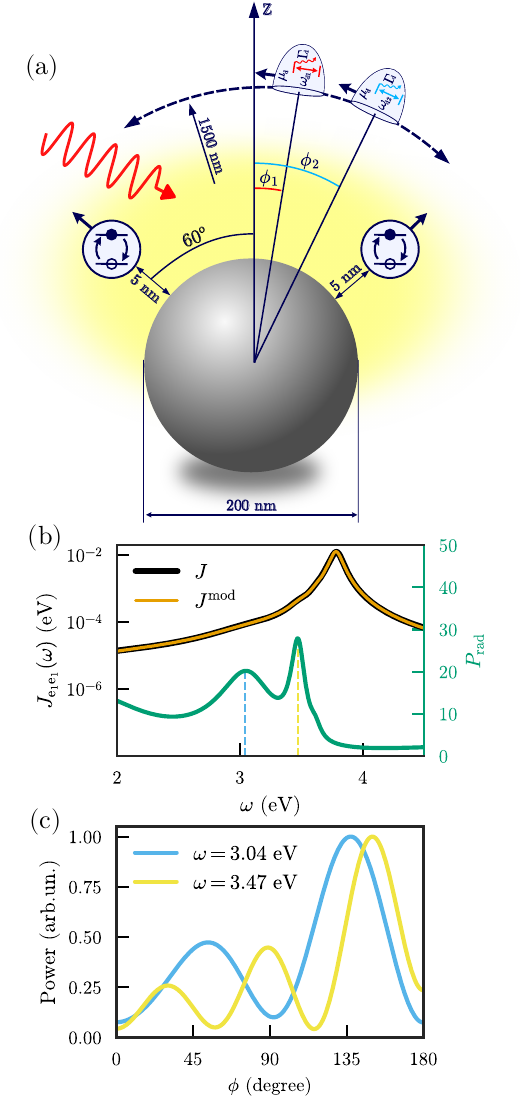}
   \caption{(a) Sketch of the system. The dark blue arrows near the emitter and detector indicate the direction of the corresponding dipole moments. (b) Spectral density of the single emitter (black) together with the fit provided by our method (orange); the green line shows the radiative Purcell factor $P_\mathrm{rad}$ for the single emitter as a function of the emitter frequency. (c) Far-field angular distribution of the power emitted by a classical point dipole 5~nm away from the sphere with radial dipole moment at different frequencies.}
\end{figure}

In the previous section, we studied a simple system dominated by a single resonance in order to be able to compare our method with the single-mode sensor method prevalent in the fields of quantum optics and cavity QED\@. In this section, we use our method to describe a more complex system, where the single-mode approximation is not applicable. To that end, we explore the emission characteristics of a system consisting of two quantum emitters interacting with the multimode electromagnetic environment provided by a \emph{large} silver sphere that supports several multipolar resonances. The schematic geometry of this setup is illustrated in \autoref{fig:R_large_sketch}. Both emitters are positioned equidistantly from the sphere (5~nm away from the surface), with a relative angle of $120^\circ$ between them. Their dipole moments are aligned along the radial directions of the sphere and have a magnitude of $50$~D. The detectors are located at a distance of 1.5~$\mu$m away from the nanophotonic system, have a bandwidth $\Gamma_\mathrm{d}=0.1~\mathrm{eV}$, and are oriented to detect the transverse field component at their respective locations. For definiteness, we are again interested in the field polarization orthogonal to the light propagation direction and contained in the plane of the sketch, \autoref{fig:R_large_sketch}.

The spectral density associated with each emitter is identical due to their symmetric placement relative to the sphere. It is dominated by a single pseudomode peak (\autoref{fig:R_large_spectral}), which arises from overlapping higher-order multipole resonances~\cite{Delga2014,Li2016Transformation}. These contributions become significant when the emitters are in close proximity to the sphere. Conversely, lower-order multipole resonances are largely hidden in the near field. Their presence, however, can be easily inferred from the radiative Purcell factor, which quantifies the radiative decay enhancement~\cite{Novotny2012}, that is, the part of the total power leaving the emitter that reaches the far field. We show this quantity in \autoref{fig:R_large_spectral}, where the peaks correspond to the quadrupolar (for $\omega=3.04$~eV) and octupolar (for $\omega=3.47$~eV) resonances, distinguished by their emission patterns in the far field (\autoref{fig:R_large_classical}).

\subsection{Resonant case}
\label{sec.res_case}

\begin{figure}[tb]
    \subfloat{\label{fig:R_large_quadrupole}}
    \subfloat{\label{fig:R_large_octupole}}
    \includegraphics[width=\linewidth]{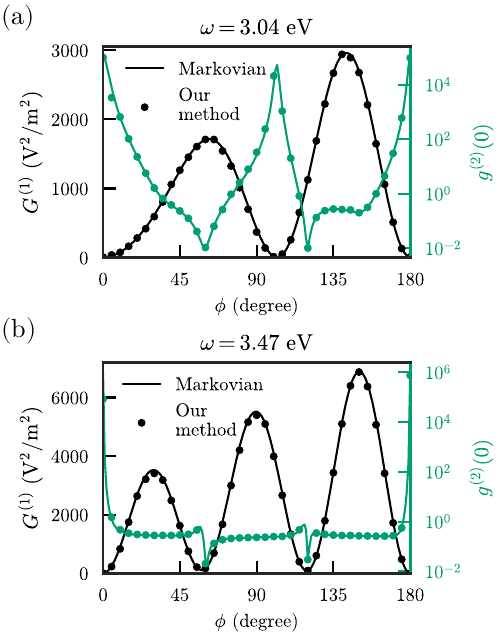}
    \caption{(a)-(b) The angular dependence of the steady state correlation functions $G^{(1)}$ and $g^{(2)}(0)$ when the emitters, detectors and laser are tuned to (a) $\omega=3.04$~eV and (b) $\omega=3.47$~eV.\label{fig:R_large}}
\end{figure}

We first explore the steady-state fluorescence of the system described above under continuous-wave laser excitation in the fully resonant scenario, where all components---emitters, detectors, and laser---are tuned to be resonant with the same sphere mode, either the quadrupole at $\omega=3.04$~eV or the octupole at $\omega=3.47$~eV. The resonant condition was deliberately imposed here in order to compare the performance of our method with the Markovian MQED approach. Both detectors are positioned at the same location to compute the conventional one-point $g^{(2)}(0)$ function. Due to the mirror symmetry of our setup, our analysis focuses on the angular range $\phi=\phi_1=\phi_2$ from $0^\circ$ to $180^\circ$. 

The results of the implementation of our method are the dots in \autoref{fig:R_large_quadrupole} and \autoref{fig:R_large_octupole}. The steady-state correlation function $G^{(1)}$ (black dots) exhibits 2 (3) peaks corresponding to the interaction with the quadrupolar (octupolar) mode, with a similar radiation pattern as the one shown in \autoref{fig:R_large_classical}. The minima observed at $\phi=0^\circ$ and $180^\circ$ reflect the symmetry of the emitters-nanoparticle subsystem: In this configuration, the detectors are symmetrically positioned between the two emitters and the dipoles of the two emitters along the detected polarization direction have opposite sign. As a result, the detected fields of the two emitters exactly cancel. 
As for the steady-state correlation function $g^{(2)}(0)$ (green dots), it shows a strong angular dependence at both frequencies, with regions of superbunching [$g^{(2)}(0)>10^3$] and antibunching [$g^{(2)}(0)<10^{-1}$]. Notably, for both frequencies there are pronounced dips at identical angles $\phi=60^\circ$ and $120^\circ$. These features can be attributed to the geometrical suppression of emitted photons from one of the emitters. In particular, the detector positioned at $\phi=60^\circ$ is aligned with the dipole orientation of the right emitter (see \autoref{fig:R_large_sketch}), where the dipole is known not to emit photons. Thus, it is expected that light emission in this direction will exhibit antibunching, a characteristic signature of single-photon emission. We note that, given the resonant and weak-coupling nature of the problem under consideration, the correlations can also be obtained to high accuracy with the Markovian MQED approach presented in the previous section, as shown by the continuous lines in \autoref{fig:R_large_quadrupole} and \autoref{fig:R_large_octupole}.

\subsection{Off-resonant case}
\label{sec.offres_case}

\begin{figure}[tb]
   \subfloat{\label{fig:map_g1}}
   \subfloat{\label{fig:map_G2}}
   \subfloat{\label{fig:map_g2}}
   \subfloat{\label{fig:map_12}}
   \includegraphics[width=\linewidth]{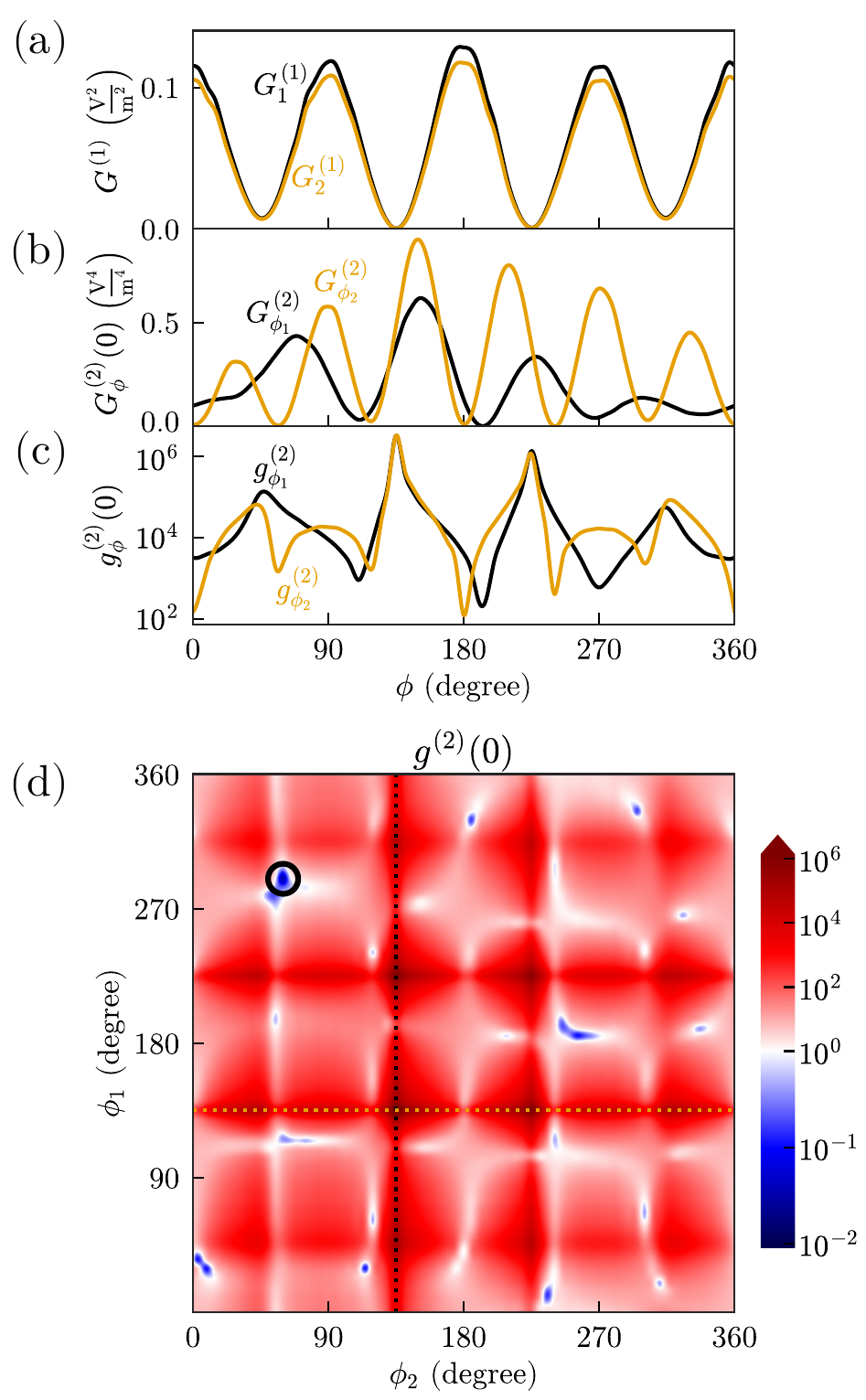}
   \caption{(a) Steady-state field intensities probed by the first ($G^{(1)}_1$) and second ($G^{(1)}_2$) detector depending on the angular position of the respective detector. (b) and (c) Steady-state $g^{(2)}(0)$ and $G^{(2)}(0)$, respectively, as functions of $\phi_1$, while $\phi_2=135^\circ$ (black) and vice versa (orange).(d) Steady-state $g^{(2)}(0)$ as a function of the angular position of the detectors. The black circle indicates the angular positions $(\phi_1,\phi_2)=(290^\circ,60^\circ)$ corresponding to \autoref{fig:freq_map}. Dotted lines correspond to the cuts shown in panel (c). }
   \label{fig:map}
\end{figure}

For the last illustration of the full capabilities of our method, we explore the multimode system in a regime beyond the Born-Markov approximation, where none of the previously used simplified models can be applied. For this, we investigate a scenario where emitters and detectors are spectrally detuned from each other. Specifically, we set the frequency of the left emitter and the first detector to be resonant with the nanosphere's quadrupolar mode ($\omega=3.04$~eV), while the right emitter and the second detector are tuned to the octupolar mode frequency ($\omega=3.47$~eV). In order to obtain a similar probability of excitation for both emitters, the laser frequency is chosen to be $\wlas=3.25$~eV, equally detuned from both emitters. Moreover, we allow the detectors to be in different positions, in contrast to the case considered in the previous subsection. The results for this situation are presented in~\autoref{fig:map}.

First, we show in \autoref{fig:map_g1} the field intensities probed by both detectors ($G^{(1)}_1$ and $G^{(1)}_2$ respectively). As one can see, $G^{(1)}_1$ and $G^{(1)}_2$ are similar and differ only slightly in magnitude. Indeed, this implies that the angular pattern of the field intensity is roughly independent of the detector frequency in this particular setup and for these specific parameters, which can be understood as due to the symmetry of the problem in frequency space. 
Moreover, the $G^{(1)}$ correlation function exhibits a notable deviation from the case considered in the previous subsection, which reflects the influence of varying the emitter and laser frequencies. In contrast to the fully resonant case, the $G^{(1)}$ correlation function does not demonstrate minima at $0^\circ$ and $180^\circ$, indicating the breaking of the mirror symmetry inherent to the previous system.

The steady-state second order correlation function $g^{(2)}(0)$, shown in \autoref{fig:map_12}, demonstrates significant angular dependence, manifesting both bunching and antibunching as a function of the angles of the detectors $\phi_1$ and $\phi_2$. Bunching (antibunching) in this context means that when detecting a photon at angular position $\phi_1$ with frequency $\omega^\mathrm{d}_1$, the probability of detecting another photon in the direction $\phi_2$ with frequency $\omega^\mathrm{d}_1$ is enhanced (suppressed) relative to Poissonian statistics. Most bunching is found at the angular positions of the detectors corresponding to global minima of $G^{(1)}$, i.e. $\phi_1,\phi_2=135^\circ$ or $225^\circ$. The steady-state $g^{(2)}(0)$ corresponding to $\phi_1=135^\circ$ and $\phi_2=135^\circ$ (horizontal and vertical dotted lines at \autoref{fig:map_12}) are presented in \autoref{fig:map_g2}. Along these lines, $g^{(2)}(0)$ does not drop below a value of $100$. Importantly, these bunching features are not ``normalization artifacts'' where almost no photons are emitted, because also the unnormalized $G^{(2)}$ function, determining the overall count rate of photon pairs, reaches nearly maximal values at these angles, as seen at \autoref{fig:map_G2}.

Furthermore, the second-order correlation function (both normalized and unnormalized) highlights the differences between the detectors: $G^{(2)}(0)$ and $g^{(2)}(0)$ exhibit four peaks across an angular sweep along the angle of the first detector $\phi_1$ (resonant with the quadrupolar mode), whereas a sweep along the angle $\phi_2$ of detector $2$ (resonant with the octupolar mode) displays six peaks, clearly distinguished in \autoref{fig:map_G2} and \autoref{fig:map_g2}.
This distinction is a clear indication of the quadrupolar and octupolar modes of the nanosphere, which exhibit similar emission patterns in the far field (see \autoref{fig:R_large_classical}). An interesting observation is that while in the fully resonant case (\autoref{sec.res_case}) the fingerprints of different nanosphere modes were already clearly visible in $G^{(1)}$, in the off-resonant system they can be observed only by considering $G^{(2)}(0)$($g^{(2)}(0)$). 

In addition to exhibiting strong bunching features, the $g^{(2)}$ function also demonstrates regions with antibunching, the most extended of which is situated in the vicinity of the coordinates $(\phi_1,\phi_2)=(300^\circ,60^\circ)$. This angular configuration is special, because these detector positions resemble the angular positions of the emitters in the studied setup (see \autoref{fig:R_large_sketch}). Therefore, the observed antibunching feature is analogous to that observed in the resonant case: the first detector, resonant with the left emitter and aligned with its dipole orientation, does not probe emission from it. The same holds for the second detector and the right emitter.

\begin{figure}[tb]
   \subfloat{\label{fig:freq_map_g1}}
   \subfloat{\label{fig:freq_map_12}}
   \subfloat{\label{fig:freq_map_g2}}
   \includegraphics[width=\linewidth]{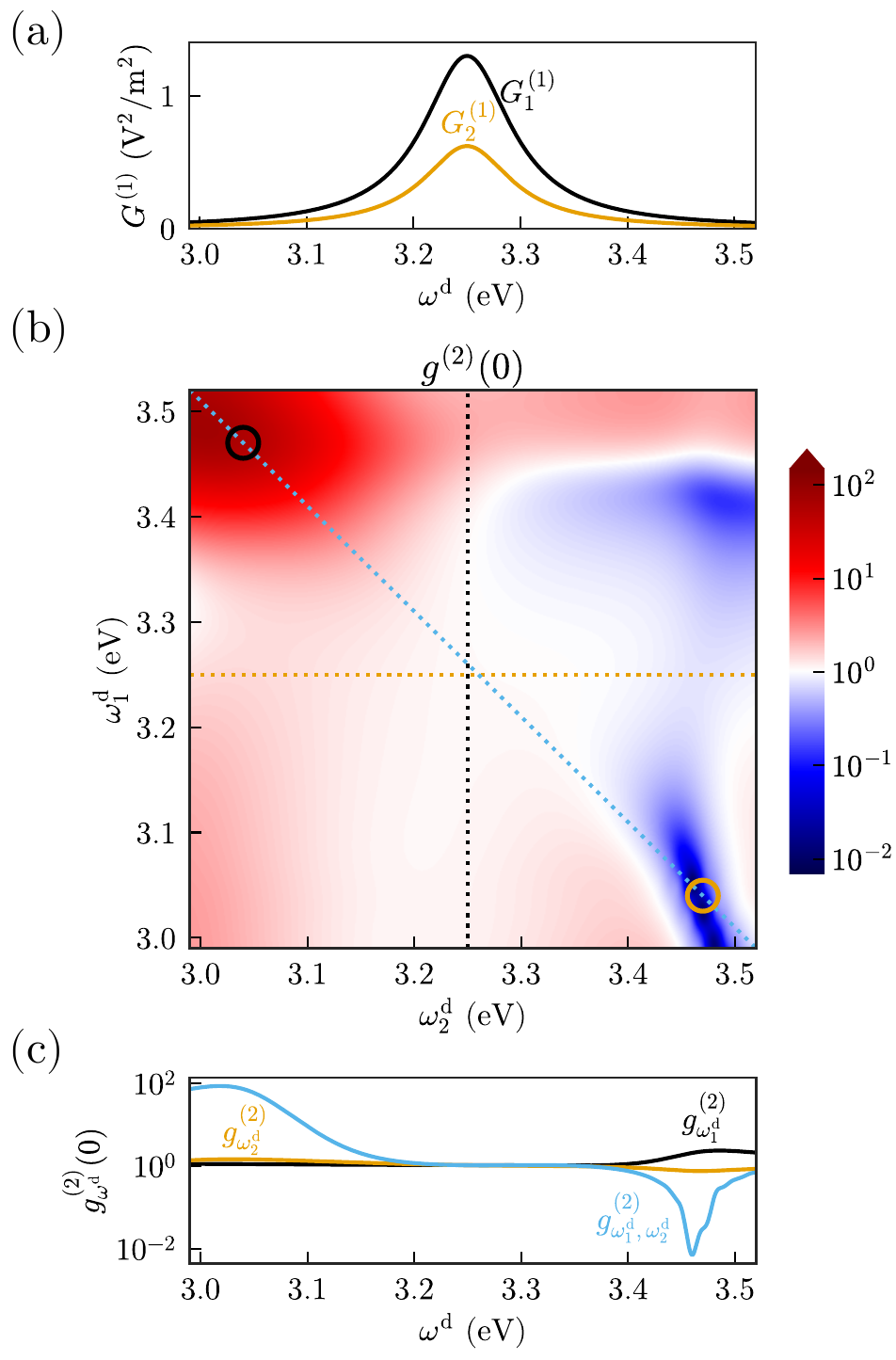}
   \caption{(a) The steady-state field intensities probed by the first ($G^{(1)}_1$) and second ($G^{(1)}_2$) detectors depending on the detector frequency. (b) Steady-state $g^{(2)}(0)$ as a function of detectors frequencies. Dotted lines correspond to the cuts shown on panel (c). Black and orange circles correspond to complementary spectral positions of paired detectors $(\omega^{\mathrm{d}}_1,\omega^{\mathrm{d}}_2)=(3.47,3.04)~\mathrm{eV}$ and $(3.04,3.47)~\mathrm{eV}$, respectively.  (c) Steady-state $g^{(2)}(0)$ as a function of $\omega^{\mathrm{d}}_1$, while $\omega^{\mathrm{d}}_2=\wlas=3.25~\mathrm{eV}$ (black) and vice versa (orange). The blue line corresponds to the $g^{(2)}(0)$ along the blue line at panel (b).}
   \label{fig:freq_map}
\end{figure}

To gain further insight into this phenomenon, we fixed the angular positions of the detectors at ($\phi_1$, $\phi_2$) = ($290^\circ$, $60^\circ$) [the global minimum of $g^{(2)}(0)$], represented by the black circle in \autoref{fig:map_12}, and swept over the detectors' frequencies. The results of these simulations are presented in \autoref{fig:freq_map}. The intensities probed by both detectors along the frequency sweep exhibit a clear Lorentzian line shape, centered at the laser frequency with a linewidth equal to $\Gamma_\mathrm{d}$ (see \autoref{fig:freq_map_g1}). Remembering that there is no direct laser driving of the detectors, this observable is thus clearly sensitive to the laser photons that were elastically scattered by the emitters.

The dependence of $g^{(2)}(0)$ on the detector frequencies (\autoref{fig:freq_map_12}) demonstrates quite extended regions of bunching and antibunching. When the frequency of one detector is set to the laser frequency, the contrast in $g^{(2)}(0)$ over a frequency sweep of the other detector is minimal (black and orange lines in \autoref{fig:freq_map_g2}). Furthermore, $g^{(2)}(0)$ remains close to unity in this case, indicating that the emission is nearly coherent, a feature inherent to the laser. When both detectors are detuned from the laser frequency, highly bunched or antibunched emission is observed (blue line in \autoref{fig:freq_map_g2}). For example, at $(\omega^{\mathrm{d}}_1,\omega^{\mathrm{d}}_2)=(3.47,3.04)$~eV (represented by a yellow circle), corresponding to \autoref{fig:map}, the value of $g^{(2)}(0)$ drops close to a level of $10^{-2}$. Conversely, in the complementary situation, when the detector frequencies are swapped, $(\omega^{\mathrm{d}}_1,\omega^{\mathrm{d}}_2)=(3.04,3.47)$~eV, $g^{(2)}(0)$ reaches approximately $\approx10^{2}$ (black circle). This demonstrates that the photon emission from even a simple system such as two emitters close to a spherical nanoparticle can demonstrate complex space- and frequency-dependent patterns of bunching and antibunching.
We finish by emphasizing the novelty of the obtained results, as, to the best of our knowledge, there are currently no alternative techniques that can effectively address systems of this level of complexity and detail. Accordingly, this example highlights the power and generality of our method, which can be used to explore and understand photon correlations in complex nanophotonic setups.

\section{Conclusions}

In this work, we have introduced a theoretical framework and computational implementation that makes traditional quantum-optical observables, such as photon correlations, accessible in complex nanophotonic devices.
To achieve this, we have integrated the idea of using weakly-coupled, lossy two-level systems as detectors into the few-mode description of macroscopic quantum electrodynamics. This provides an affordable tool to study the statistical properties of output light in nanophotonics. The resulting framework provides direct access to photon correlations of any order resolved in space, frequency, polarization, and time in general lossy and dispersive nanophotonic environments and in any coupling regime of light-matter interactions.

We have then implemented the approach to analyze electric field correlations in two exemplary nanoplasmonic systems. First, we studied the scenario of a single emitter interacting with the modes of a small nanosphere. Here, our model provided comprehensive access to field correlation functions, accounting for retardation effects, absolute physical units and strong-coupling features. In addition, we validated our results by finding agreement with previously reported approaches in certain limiting cases.
Secondly, we demonstrated the generality of the method by studying driven two-emitter setups interacting with a multimode electromagnetic environment. In both resonant and off-resonant driving configurations, our results unveiled a rich landscape of spatially resolved light statistics.

The demonstrated capabilities of our method open new avenues for the exploration and control of field correlations in complex nanophotonic systems. This will grant a deeper understanding and enable the design of novel non-classical light sources that exploit quantum light-matter interactions at the nanoscale. These directions promise to advance both fundamental research and practical applications in quantum optics and nanophotonics.

\label{sec.conclusions}

\begin{acknowledgments}

This work was funded by the Spanish Ministry of Science, Innovation and Universities-Agencia Estatal de Investigación through the FPI Grants PRE2021-098978 and PRE2019-090589, as well as Grants PID2021-125894NB-I00, EUR2023-143478 and CEX2023-001316-M (through the María de Maeztu program for Units of Excellence in R\&D). We also acknowledge financial support from the Proyecto Sinérgico CAM 2020 Project No.~Y2020/TCS-6545 (NanoQuCo-CM) of the Community of Madrid, and from the European Union’s Horizon Europe Research and Innovation Programme through agreements 101070700 (MIRAQLS) and 101098813 (SCOLED).

\end{acknowledgments}

\appendix
\section{Electric field detection in general EM environments}
\label{appendix.Efield_op}

The MQED electric field operator in the Heisenberg picture is given by:
\begin{align}
    &\hat{\mathbf{E}}(\mathbf{r},t) = \hat{\mathbf{E}}^+(\mathbf{r},t) + \hat{\mathbf{E}}^-(\mathbf{r},t) \nonumber\\
    &= \sum_{\alpha=e,m} \int \mathrm{d}^3r' \int_0^\infty \mathrm{d}\omega\ \boldsymbol{\mathcal{G}}_\alpha(\mathbf{r},\mathbf{r}',\omega)\cdot\hat{\mathbf{f}}_\alpha(\mathbf{r}', \omega,t)  + \mathrm{H.c.},
    \label{eq:E_mqed}
\end{align}
where $\hat{\mathbf{E}}^+(\mathbf{r})$ and $\hat{\mathbf{E}}^-(\mathbf{r})$ are positive and negative frequency components of the electric field operator, and
$\boldsymbol{\mathcal{G}}_e(\mathbf{r},\mathbf{r}',\omega) = i\frac{\omega^2}{c^2}\sqrt{\frac{\hbar}{\pi\epsilon_0}\mathrm{Im}[\epsilon(\mathbf{r'},\omega)]}\boldsymbol{\mathcal{G}}(\mathbf{r},\mathbf{r}',\omega),$  
$\boldsymbol{\mathcal{G}}_m(\mathbf{r},\mathbf{r}',\omega) = i\frac{\omega}{c}\sqrt{\frac{\hbar}{\pi\epsilon_0}\frac{\mathrm{Im}[\mu(\mathbf{r'},\omega)]}{\left|\mu(\mathbf{r'},\omega)\right|^2}}\left[{\nabla}'\times\boldsymbol{\mathcal{G}}(\mathbf{r}',\mathbf{r},\omega)\right]^T$. The medium-assisted electric field in~\autoref{eq:E_mqed} accounts for general linearly responding and absorbing optical environments given by the complex permittivity $\epsilon(\mathbf{r},\omega)$ and permeability $\mu(\mathbf{r},\omega)$ via the dyadic Green's tensor $ \boldsymbol{\mathcal{G}}$ of the vector Helmholtz equation~\cite{Scheel2008,Buhmann2012I,Buhmann2012II}. $\boldsymbol{\mathcal{G}}$ can be split into a free-space part $\boldsymbol{\mathcal{G}}^\mathrm{fs}$ and a scattering part $\boldsymbol{\mathcal{G}}^\mathrm{sc}$ that depends on the nanophotonic structure: $\boldsymbol{\mathcal{G}} = \boldsymbol{\mathcal{G}}^\mathrm{fs} + \boldsymbol{\mathcal{G}}^\mathrm{sc}$.

To realistically model photodetection events and their correlations, the finite detection bandwidth must be incorporated. This can be done by implementing the theory developed by Eberly and W\'{o}dkiewicz~\cite{Eberly1977, Knoll1984, Arnoldus1984}. Following their work, the electric field operator transmitted through the filter of the detector can be written as:
\begin{equation}
\hat{E}_\mathrm{d}^+(t) = \frac{\Gamma_\mathrm{d}}{2} \int_0^t \mathrm{d}\tau\ \mathbf{n}_\mathrm{d}\cdot\hat{\mathbf{E}}^+(\mathbf{r}_\mathrm{d},\tau) e^{-i(\omega_\mathrm{d} - \frac{i}{2} \Gamma_\mathrm{d})(t-\tau)},
\label{eq:E_d_simple}
\end{equation}
where $\omega_{\mathrm{d}}$ and $\Gamma_{\mathrm{d}}$ are the central frequency and FWHM linewidth of the detector transmission function, and $\mathbf{n}_\mathrm{d}$ is a unit vector aligned with the field polarization probed by the detector at position $\mathbf{r}_\mathrm{d}$. From this result, it is straightforward to derive an expression for the $N$-th order correlation functions probed by $N$ detectors as:
\begin{equation} 
    G^{(N)}\left(\{\omega_i,\Gamma_i; t_i\}_{i=1}^N\right) = \langle :\prod_{i=1}^N \hat{E}^-_{\mathrm{d}_i}(t_i)\hat{E}^+_{\mathrm{d}_i}(t_i): \rangle.
    \label{eq:corr}
\end{equation}
The quantities $\hat{E}^+_{\mathrm{d}_i}$ and $\hat{E}^-_{\mathrm{d}_i}$ are operators corresponding to the positive and negative frequency parts of the electric field probed by the detector located at $\mathbf{r}_i$. Therefore, following this straightforward approach for obtaining the $N$-th order correlations, one needs to evaluate $2N$ time integrals, which becomes computationally unfeasible already for $N=2$. A computationally more feasible alternative is presented in the main text. 

\begin{figure}[tb]
   \includegraphics[width=\linewidth]{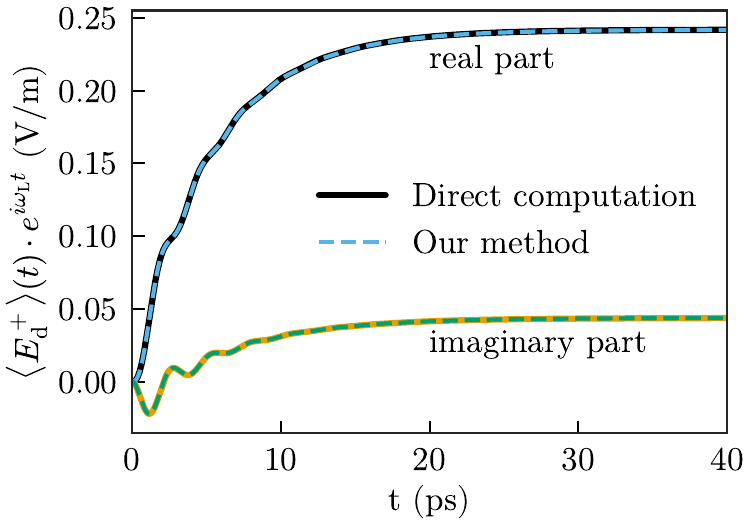}
   \caption{Frequency-resolved electric field $\langle E^+_{\mathrm{d}}\rangle(t)\cdot e^{i\wlas t}$ with $\omega_{\mathrm{d}} = 3.489~\mathrm{eV}$ and $\Gamma_{\mathrm{d}}=5\times10^{-4}~\mathrm{eV}$ calculated via direct computation and our method.}
   \label{fig:comparison}
\end{figure}

In \autoref{fig:comparison}, we show the frequency-resolved electric field $\langle E^+_{\mathrm{d}}\rangle(t) e^{i\wlas t}$ calculated via the method proposed in the main text and a direct evaluation of \autoref{eq:E_d_simple}. The latter was obtained using the technique described in~\cite{Medina2021} to get the unfiltered electric field, $\hat{\mathbf{E}}^+(\mathbf{r}_\mathrm{d}, t)$, and then \autoref{eq:E_d_simple} to account for the finite detection bandwidth. To show the equivalence between the two approaches, we used the system described in \autoref{subsec.res_fluor} with $\omega_{\mathrm{d}} = 3.489~\mathrm{eV}$ and $\Gamma_{\mathrm{d}}= 5\times10^{-4}~\mathrm{eV}$. The results in \autoref{fig:comparison} demonstrate perfect agreement. Still, it is important to note that the direct evaluation of the temporal integrals that implicitly appear in \autoref{eq:corr} is only numerically tractable for low order electric field observables.

\section{Inclusion of classical fields}
\label{appendix.class}

In this section we show that, with the introduced method, it is possible to probe both quantum and classical sources. First, we note that the field generated by classical sources can be easily integrated within the formalism of macroscopic QED~\cite{Feist2021}. In the presence of external classical fields, the electric field operator becomes $\hat{\mathbf{E}}'(\mathbf{r})=\hat{\mathbf{E}}(\mathbf{r})+\mathbf{E}_\mathrm{cl}(\mathbf{r},t)$. If we then model the detection bandwidth in the same manner as in \autoref{eq:E_d_simple} we get:
\begin{align}
    \hat{E}'_\mathrm{d}(\omega_\mathrm{d}&,\mathbf{r}_\mathrm{d},t) = \hat{E}_\mathrm{d}(\omega_\mathrm{d},\mathbf{r}_\mathrm{d},t) + \nonumber\\ &+ \frac{\Gamma_\mathrm{d}}{2}\int_0^t \mathrm{d}\tau\ \mathbf{n}_\mathrm{d}\cdot\mathbf{E}_\mathrm{cl}(\mathbf{r}_\mathrm{d},\tau)\ e^{-i(\omega_\mathrm{d}-\frac{i}{2}\Gamma_\mathrm{d}) (t-\tau)}
    \label{eq:e_comp}
\end{align}
Let us now show how the expression for the annihilation operator of the auxiliary lossy TLS will change if we take classical fields into consideration. In this case, the Hamiltonian \autoref{eq:H_d} will change to $\hat{H}'_\mathrm{d}= \hat{H}_\mathrm{d}-\sum_{i=1}^N \mu_i\mathbf{n}_i\cdot\mathbf{E}_\mathrm{cl}(\mathbf{r}_i,t) (\hat{d}_i+\hat{d}_i^\dagger)$. Following the same procedure with the new Hamiltonian, the Heisenberg equation for the operator $\hat{d}_i^\prime$ becomes:
\begin{equation}
    \dot{\hat{d}}'_i(t) = \dot{\hat{d}}_i(t) + i\frac{\mu_i}{\hbar}\mathbf{n}_i\cdot\mathbf{E}_\mathrm{cl}(\mathbf{r}_i,t).
\end{equation}
Then, the integration of this expression directly gives:
\begin{equation}
    \hat{d}'_i(t) = \hat{d}_i(t) + i\frac{\mu_i}{\hbar}\int_0^t \mathrm{d}\tau\ \mathbf{n}_i\cdot\mathbf{E}_\mathrm{cl}(\mathbf{r}_i,\tau)\  e^{-i\widetilde{\omega}_i (t-\tau)},
\end{equation}
where $\widetilde{\omega}_i =\omega_i - \frac{i}{2}\Gamma_i$.
As this formula corresponds to \autoref{eq:e_comp} up to the prefactor, we can write the following equality:
\begin{equation}
    \hat{E'}_{\mathrm{d}_i}^+(t) = -i\frac{\hbar\Gamma_i}{2\mu_i}\hat{d}'_i(t).
\end{equation}
This makes all our conclusions about the correspondence between the detected electric field and the annihilation operator of weakly coupled detectors still valid for classical fields.

\section{Fitting the real part of the response function}
\label{appendix.re_resp_func}

\begin{figure}[tb]
   \subfloat{\label{fig:real_resp_func_ed}}
   \subfloat{\label{fig:real_resp_func_ee}}
   \includegraphics[width=\linewidth]{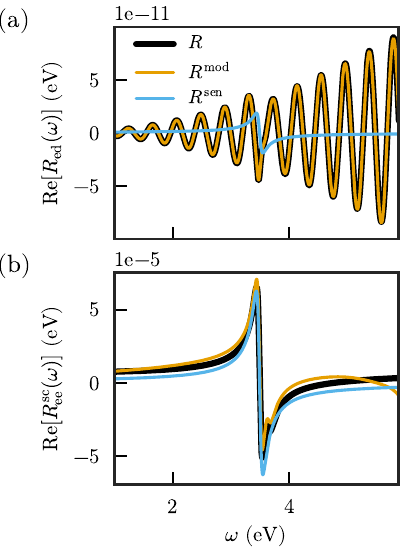}
   \caption{Real parts of the response functions relevant for light detection (black). Orange lines represent the fits provided by our method, while blue lines show corresponding functions retrieved from the single-mode sensor method (see \appref{appendix.sensor_method}).}
   \label{fig:real_resp_func}
\end{figure}
As discussed in the main text, to perform an accurate frequency truncation of the spectral density we actually fit the response function of the model \autoref{eq:model_response_function} to the physical \emph{fully retarded} response function~\autoref{eq:response_function}. In \autoref{fig:real_resp_func_ed} we show the real part of the crossed emitter-detector response function, $R_\mathrm{ed}$, for the setup under consideration in \autoref{sec.test} together with the fits provided by our method. It behaves similarly to the corresponding spectral density $J_\mathrm{ed}$ (\autoref{fig:J_func_ed}) and by taking it into account during the fitting procedure, we ensure accuracy of the spectral truncation.

In \autoref{fig:real_resp_func_ee} we only render the real part of the scattering contribution to the emitter response function, $\mathrm{Re}[R^{\mathrm{sc}}_\mathrm{ee}]$. We first note that the free-space contribution of $\mathrm{Re}[R_\mathrm{ee}]$ diverges because the real part of $\boldsymbol{\mathcal{G}}(\mathbf{r}_i,\mathbf{r}_j,\omega)$ diverges in the coincidence limit ($\mathbf{r}_i=\mathbf{r}_j$) [see \autoref{eq:response_function}]. $\mathrm{Re}[R^\mathrm{fs}_\mathrm{ee}]$ is known to be responsible for the free-space Lamb shift of the emitter's energy levels~\cite{Buhmann2012I}. The result is a very small frequency shift that we consider to be included in the bare frequencies of the emitter. Therefore, \autoref{fig:real_resp_func_ee} shows only its finite scattering part $\mathrm{Re}[R^\mathrm{sc}_\mathrm{ee}]$, which provides an environment-induced Lamb shift. However, because $\pi J_\mathrm{ee}(\omega)$ and $\mathrm{Re}[R^\mathrm{sc}_\mathrm{ee}]$ are not connected through the Kramers-Kronig relations, $\mathrm{Re}[R^\mathrm{mod}_\mathrm{ee}]$ cannot perfectly reproduce the target function $\mathrm{Re}[R^\mathrm{sc}_\mathrm{ee}]$. This can lead to undesirable deviations when computing the dynamics of the system. To mitigate this, we use a recently developed approach~\cite{SanchezMartinez2024Mixed}, which allows to minimize the error due to the inaccurate fit of $\mathrm{Re}[R^\mathrm{sc}_\mathrm{ee}]$ by adding an appropriate perturbative correction.

\section{Optimization of the fitting procedure}
\label{appendix.fit_opt}
The procedure giving access to the correlations outlined in \autoref{sec.theory} implies fitting an $(N_\mathrm{e}+N_\mathrm{d})\times(N_\mathrm{e}+N_\mathrm{d})$ matrix-valued function for every unique geometrical configuration of emitters and detectors. This complicates the implementation of the proposed approach. However, as we show in this appendix, there are ways to significantly simplify this technical obstacle. 

For this, let us decompose the set of field modes into two subsets, which interact weakly with each other with couplings~$\epsilon_{ij}$. Moreover, we impose that the first subset only interacts with the emitters, and the second one only with the detectors, as depicted in \autoref{fig:scheme}. It is thus natural to name the first subset ``emitter modes'' and the second one ``detector modes''. This separation allows us to rewrite the matrices $\widetilde{\mathbf{H}}$ and $\boldsymbol{\Lambda}$ in block form:
\begin{figure}[tb]
   \includegraphics[width=\linewidth]{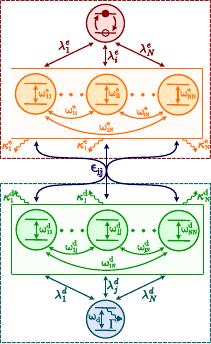}
   \caption{Schematic representation of the quantized modes decomposition.}
   \label{fig:scheme}
\end{figure}
\begin{subequations}
\begin{align}
    \widetilde{\mathbf{H}} &= 
    \begin{pmatrix}
        \widetilde{\mathbf{H}}_\mathrm{e} & \boldsymbol{\epsilon} \\
        \boldsymbol{\epsilon}^T & \widetilde{\mathbf{H}}_\mathrm{d}
    \end{pmatrix},\\
    \boldsymbol{\Lambda} &=
    \begin{pmatrix}
        \boldsymbol{\lambda}_\mathrm{ee} & 0 \\
        0 & \boldsymbol{\lambda}_\mathrm{dd}
    \end{pmatrix},
\end{align}
\end{subequations}
where $\widetilde{H}_{k,ij}=\omega_{k,ij}-\frac{i}{2} \kappa_{k,i}\delta_{ij}$ ($k=\mathrm{e,d}$) are the matrices defining emitter and detector modes, respectively. Since the emitter modes interact only with the emitters, and the detector modes are coupled only to the detectors, $\boldsymbol{\Lambda}$ takes the block-diagonal form with $\boldsymbol{\lambda}_\mathrm{ee}$ and $\boldsymbol{\lambda}_\mathrm{dd}$ being the couplings between emitter and emitter modes and between detector and detector modes, respectively. Here, $\boldsymbol{\lambda}_\mathrm{ee}$ and $\boldsymbol{\lambda}_\mathrm{dd}$ are matrices with shapes $\left(N_\mathrm{e}\times N_\mathrm{em}\right)$ and $\left(N_\mathrm{d}\times N_\mathrm{dm}\right)$, respectively, where $N_\mathrm{em}$($N_\mathrm{dm}$) is the number of ``emitter''(``detector'') modes.

Writing the model response function also in block form,
\begin{equation}
    \mathbf{R}^\mathrm{mod}(\omega) = \begin{pmatrix} \mathbf{R}^\mathrm{mod}_\mathrm{ee}(\omega) & \mathbf{R}^\mathrm{mod}_\mathrm{ed}(\omega) \\
        \mathbf{R}^\mathrm{mod}_\mathrm{ed}(\omega)^T & \mathbf{R}^\mathrm{mod}_\mathrm{dd}(\omega) \end{pmatrix},    
\end{equation}
and performing a perturbative expansion to lowest order in the offdiagonal coupling matrix $\boldsymbol{\epsilon}$, we obtain
\begin{subequations}
\begin{align}
    \mathbf{R}^\mathrm{mod}_\mathrm{ee}(\omega) &\approx \boldsymbol{\lambda}_\mathrm{ee} \frac{1}{\widetilde{\mathbf{H}}_\mathrm{e}-\omega} \boldsymbol{\lambda}_\mathrm{ee}^T,\\
    \mathbf{R}^\mathrm{mod}_\mathrm{dd}(\omega) &\approx \boldsymbol{\lambda}_\mathrm{dd} \frac{1}{\widetilde{\mathbf{H}}_\mathrm{d}-\omega} \boldsymbol{\lambda}_\mathrm{dd}^T,\\
    \mathbf{R}^\mathrm{mod}_\mathrm{ed}(\omega) &\approx -\boldsymbol{\lambda}_\mathrm{ee} \frac{1}{\widetilde{\mathbf{H}}_\mathrm{e}-\omega} \boldsymbol{\epsilon} \frac{1}{\widetilde{\mathbf{H}}_\mathrm{d}-\omega} \boldsymbol{\lambda}_\mathrm{dd}^T.
    \label{eq:R_sd}
\end{align}
\end{subequations}
As argued in the main text, the response function $\mathbf{R}^\mathrm{mod}_\mathrm{dd}(\omega)$ of the detector subsystem can be neglected due to the small dipole moment of the detectors, and can thus be ignored in the fitting procedure.

Another advantage of this separation of response functions is that it enables fitting them separately. This, in fact, means that to analyze the spatial dependence of correlations, it is not needed to fit the whole $(N_\mathrm{e}+N_\mathrm{d})\times(N_\mathrm{e}+N_\mathrm{d})$ matrix-valued response function anew for each detector configuration. Instead, one can fit $\mathbf{R}^\mathrm{mod}_\mathrm{ee}(\omega)$, with size $(N_\mathrm{e}\times N_\mathrm{e})$, only once, and then fit the remaining $\left(N_\mathrm{e}\times N_\mathrm{d}\right)$ matrix that corresponds to $\mathbf{R}^\mathrm{mod}_\mathrm{ed}(\omega)$ for every new configuration of the detectors. Moreover, if one wants to sweep over different spatial positions with a small distance between detection points, the smooth spatial behavior of the response function means that the fit parameters obtained for the previous detector position are a good initial guess for the fit of the next nearby detection point, improving the convergence of the procedure.

Last, we note that, due to the foundational role of the single-mode sensor method~\cite{delValle2012}, further optimization techniques presented in that context~\cite{LopezCarreno2018,Holdaway2018} can also be straightforwardly applied here.

\section{Single-mode sensor method}
\label{appendix.sensor_method}
In \autoref{sec.test}, we used the sensor method~\cite{delValle2012} to describe the detection of light from a system composed of an emitter coupled to a quasi-single-mode EM environment. Due to the simplicity of the EM environment, the system under study can be effectively described by the standard Jaynes-Cummings model, where a TLS interacts with a single generic cavity mode that represents the dipolar resonance of the nanosphere. To retrieve mode parameters of the cavity, we approximate the emitter spectral density (\autoref{fig:J_func_ee}) by a Lorentzian $F(\omega)=\frac{\lambda_\mathrm{e}^2}{2\pi}\frac{\kappa}{(\omega-\omega_\mathrm{c})^2 + \kappa_\mathrm{c}^2/4}$, where $\omega_\mathrm{c}$ and $\kappa_\mathrm{c}$ are the mode's frequency and losses, and $\lambda_\mathrm{e}$ is the coupling to the emitter. The dynamics provided by this method are given by:
\begin{subequations}
\label{eq:sensor_method}
\begin{align}
    \hat{H}^{\mathrm{sen}} &= \omega_\mathrm{e} \hat{\sigma}^\dagger_\mathrm{e}\hat{\sigma}_\mathrm{e} + \omega_\mathrm{d} \hat{d}^\dagger\hat{d} + \omega_\mathrm{c}\hat{a}^\dagger\hat{a} \nonumber \\
    &\phantom{=\ } + \lambda_\mathrm{e} (\hat{a}^\dagger\hat{\sigma}_\mathrm{e}+\hat{a}\hat{\sigma}^\dagger_\mathrm{e}) + \lambda_\mathrm{d}(\hat{a}^\dagger\hat{d}+\hat{a}\hat{d}^\dagger),\\
    \dot{\hat{\rho}} &= -\frac{i}{\hbar}[\hat{H}^\mathrm{sen},\rho] + \kappa_\mathrm{d} L_{\hat{a}}[\hat{\rho}] + \Gamma_\mathrm{d} L_{\hat{d}}[\hat{\rho}],
\end{align}
\end{subequations}
where $\lambda_\mathrm{d}$ is the coupling of the cavity mode to the detector, which can be set arbitrarily small to prevent back-action~\cite{delValle2012}. Due to the arbitrariness of $\lambda_\mathrm{d}$, it is not possible to analyze the spatial dependence of correlations using this method and only normalized correlations obtained with this method are meaningful.

\section{Markovian MQED}
\label{appendix.mark_approx}
In the weak-coupling limit, the MQED Hamiltonian \autoref{eq:H_d} can be considerably simplified by using the Born-Markov approximation for the light-matter interaction. As a result of the approximation, all EM modes are traced out, which leads to the following master equation:
\begin{subequations}
    \label{eq:mark_method}
    \begin{align}
        \hat{H}^\mathrm{M} &= \hbar\Omega_\mathrm{e} \hat{\sigma}^\dagger_\mathrm{e}\hat{\sigma}_\mathrm{e} + \hbar\omega_\mathrm{d} \hat{d}^\dagger\hat{d}+ \hbar t_\mathrm{ed} (\hat{d}^\dagger\hat{\sigma}_\mathrm{e}+\hat{d}\hat{\sigma}^\dagger_\mathrm{e})\\
        \dot{\hat{\rho}} &= -\frac{i}{\hbar}[\hat{H}^\mathrm{M},\hat{\rho}]+ \sum_{i,j=\mathrm{e,d}} \gamma_{ij} L_{ij}[\hat\rho].
    \end{align}
\end{subequations}
In $\hat{H}^\mathrm{M}$, the emitter frequency includes the scattering Lamb shift $\Omega_\mathrm{e} = \omega_\mathrm{e}-\mathrm{Re}\left[R^\mathrm{sc}_\mathrm{ee}(\omega_\mathrm{e})\right]$, and the direct dipole-dipole interaction strength is $t_\mathrm{ed}=-\mathrm{Re}\left[R_\mathrm{ed}(\omega_\mathrm{e})\right]$. On the dissipative side, the coefficients $\gamma_{ij}$ are given by $\gamma_\mathrm{ee}=2\mathrm{Im}\left[R_\mathrm{ee}(\omega_\mathrm{e})\right]$ and $\gamma_\mathrm{ed}=\gamma_\mathrm{de}=2\mathrm{Im}\left[R_\mathrm{ed}(\omega_\mathrm{e})\right]$, while $\gamma_\mathrm{dd}$ is set to the detector bandwidth $\Gamma_\mathrm{d}$ as it is assumed to be due to ``internal'' detector properties and not induced by the interaction with the local EM environment. The dissipator is defined as $L_{ij}[\hat\rho] = \hat{O}_i\hat{\rho}\hat{O}_j^\dagger - \frac{1}{2} \{ \hat{O}_j^\dagger\hat{O}_i, \hat{\rho}\}$ with $\hat{O}_\mathrm{e} = \hat{\sigma}_\mathrm{e}$ and  $\hat{O}_\mathrm{d} = \hat{d}$.

\section{Wigner-Weisskopf approximation}
\label{appendix.ww}

In this appendix, we consider the particular problem of detecting the spontaneously emitted field due to an initially excited two-level emitter, weakly coupled to an arbitrary electromagnetic environment. For convenience in the derivation, we will use the formulation of MQED in terms of ``emitter-centered'' modes~\cite{Buhmann2012II,Feist2021}, in which the full EM environment coupling to $N$ emitters is reexpressed in a basis of $N$ continua. Accordingly, the Hamiltonian can be written as:
\begin{multline}\label{eq:H_1_d}
    \hat{H} = \hbar\omega_\mathrm{e} \hat{\sigma}_\mathrm{e}^\dagger \hat{\sigma}_\mathrm{e} + \hbar\widetilde{\omega}_\mathrm{d}\hat{d}^\dagger\hat{d}
    + \sum_{i=\mathrm{e,d}} \int_0^\infty \mathrm{d} \omega\, \hbar\omega \hat{C}_i^\dagger (\omega) \hat{C}_i(\omega) \\
    + \sum_{i,j=\mathrm{e,d}} \int_0^\infty \mathrm{d}\omega\, \hbar h_{ij}(\omega) \left[\hat{C}_i(\omega)\hat{O}_j^\dagger + \hat{C}_i^\dagger(\omega)\hat{O}_j\right].
\end{multline}
As in the main text, $\widetilde{\omega}_\mathrm{d} = \omega_\mathrm{d} - \frac{i}{2}\Gamma_\mathrm{d}$ is the complex frequency of the detector, while the $\hat{O}_i$ are defined as in \appref{appendix.mark_approx} and $\hat{C}_i(\omega)$ [$\hat{C}_i^\dagger(\omega)$] are annihilation [creation] operators for the two orthogonal photonic continua, $[\hat{C}_i(\omega), \hat{C}_j^\dagger(\omega')] = \delta_{ij}\delta(\omega-\omega')$. 
The couplings $h_{ij}(\omega)$ must fulfill $\mathbf{h}(\omega)^T \mathbf{h}(\omega) = \mathbf{J}(\omega)$ and are chosen here as $h_\mathrm{ee}(\omega) = \sqrt{J_\mathrm{ee}(\omega)}$, $h_\mathrm{ed}(\omega) = \frac{J_\mathrm{ed}(\omega)}{h_\mathrm{ee}(\omega)}$, $h_\mathrm{de}(\omega) = 0$, and $h_\mathrm{dd}(\omega) = \sqrt{J_\mathrm{dd}(\omega)-h_\mathrm{ed}^2(\omega)}$, corresponding to a Cholesky decomposition of $\mathbf{J}(\omega)$ and ensuring that the emitter only couples to a single continuum $\hat{C}_\mathrm{e}(\omega)$. Since we restrict the coupling to be weak, the rotating wave approximation has been safely applied. Furthermore, the second continuum $\hat{C}_\mathrm{d}(\omega)$ is only coupled to the detector, which has a negligible dipole moment, so it can be neglected completely and dropped from the Hamiltonian.

For the considered problem, the initial state is $\ket{\psi(0)}=\hat{\sigma}_\mathrm{e}^+ \ket{0}$, where $\ket{0}$ is the vacuum state. Because the Hamiltonian conserves the total number of excitations, the wave function $\ket{\psi(t)}$ can be decomposed in the basis of singly excited states:
\begin{align}
\ket{\psi(t)} &= \sum_{i=\mathrm{e,d}} \alpha_i(t) \hat{O}_{i}^\dagger\ket{0} + \int_0^\infty \mathrm{d}\omega\, \beta_\omega(t)\hat{C}_\mathrm{e}^\dagger (\omega) \ket{0}.
\label{eq:psi_d}
\end{align}

Applying the Schrödinger equation with \autoref{eq:H_1_d} and defining new variables $\widetilde{\alpha}_i(t) = \alpha_i(t) e^{i\bar{\omega}_i t}$, where $\bar{\omega}_\mathrm{e} = \omega_\mathrm{e}$ and $\bar{\omega}_\mathrm{d} = \widetilde{\omega}_\mathrm{d}$, and $\widetilde{\beta}_\omega(t)=\beta_\omega(t) e^{i\omega t}$, we obtain a closed set of equations:
\begin{subequations}
    \begin{align}
    \dot{\widetilde{\alpha}}_i(t) &= -i\int_0^\infty \mathrm{d}\omega\, h_{\mathrm{e}i}(\omega) \widetilde{\beta}_\omega(t) e^{-i(\omega - \bar{\omega}_i) t}, \\
    \dot{\widetilde{\beta}}_\omega(t) &= -i \sum_{i=\mathrm{e,d}} h_{\mathrm{e}i}(\omega) \widetilde{\alpha}_i(t) e^{i(\omega - \bar{\omega}_i) t}.
    \end{align}
\end{subequations}
In the second equation, the coupling to the detector can again be neglected, giving the formal solution $\widetilde{\beta}_\omega(t)=-i h_\mathrm{ee}(\omega) \int_0^t\mathrm{d}\tau\, e^{i(\omega - \omega_\mathrm{e}) \tau}\widetilde{\alpha}_\mathrm{e}(\tau)$. Inserting this into the first equation for $i=\mathrm{e}$ gives
\begin{equation}
\dot{\widetilde{\alpha}}_\mathrm{e}(t) = -\int_0^t \mathrm{d}\tau \int_0^\infty\mathrm{d}\omega\, J_\mathrm{ee}(\omega) e^{-i(\omega - \omega_\mathrm{e}) (t-\tau)}\widetilde{\alpha}_\mathrm{e}(\tau).
\end{equation}
Assuming that the spectral density is sufficiently smooth compared to the complex exponential, we can replace it by its value at $\omega = \omega_\mathrm{e}$. Due to the rapidly oscillating exponential, we can also extend the lower limit of the frequency integral to $-\infty$, giving
\begin{align}
\dot{\widetilde{\alpha}}_\mathrm{e}(t) &\approx -J_\mathrm{ee}(\omega_\mathrm{e})\int_0^t \mathrm{d}\tau \int_{-\infty}^\infty \mathrm{d}\omega\, e^{-i(\omega - \omega_\mathrm{e}) (t-\tau)}\widetilde{\alpha}_\mathrm{e}(\tau) \nonumber\\ 
&= -\pi J_\mathrm{ee}(\omega_\mathrm{e})\widetilde{\alpha}_\mathrm{e}(t),
\end{align}
where we used that $\int_{-\infty}^{\infty} \mathrm{d}k\, e^{-ikx} = 2\pi \delta(x)$. Solving the last differential equation yields $\alpha_\mathrm{e}(t) = e^{-i\Tilde{\omega}_\mathrm{e}t}$, where $\Tilde{\omega}_\mathrm{e} = \omega_\mathrm{e} - i\pi J_\mathrm{ee}(\omega_\mathrm{e})$, giving exponential decay of the excited state population of the emitter, $\langle\hat{\sigma}^\dagger_\mathrm{e}\hat{\sigma}_\mathrm{e}\rangle(t) = \left|\alpha_\mathrm{e}(t)\right|^2 = e^{-2\pi J_\mathrm{ee}(\omega_\mathrm{e})t}$. The photon and detector amplitudes can then be obtained by simple integration as
\begin{subequations}
    \begin{align}
    \beta_\omega(t) &= \frac{\sqrt{J_\mathrm{ee}(\omega)}}{\omega-\Tilde{\omega}_\mathrm{e}}\left(e^{-i\omega t} - e^{-i\Tilde{\omega}_{e}t}\right),\label{eq:beta}\\
    \alpha_\mathrm{d}(t) &= \frac{r_\mathrm{ed}(\Tilde\omega_\mathrm{d}) e^{-i\Tilde\omega_\mathrm{d} t} - r_\mathrm{ed}(\Tilde\omega_e) e^{-i\Tilde\omega_e t}}{\Tilde\omega_e - \Tilde\omega_\mathrm{d}} \nonumber\\
    &+ \int_0^\infty \mathrm{d}\omega\, \frac{J_\mathrm{ed}(\omega) e^{-i\omega t}}{(\omega - \widetilde{\omega}_\mathrm{e})(\omega - \widetilde{\omega}_\mathrm{d})}\label{eq:alpha_d}
    \end{align}
\end{subequations}
where $r_\mathrm{ed}(\omega) = \int_0^\infty \mathrm{d}\omega'\, \frac{J_\mathrm{ed}(\omega')}{\omega' - \omega}$.
Using \autoref{eq:equivalence2}, the light intensity at the detector is then
\begin{equation}\label{eq:ww_an}
    \langle \hat{E}_\mathrm{d}^\dagger \hat{E}_\mathrm{d} \rangle = \left(\frac{\hbar\Gamma_\mathrm{d}}{2\mu^\mathrm{d}}\right)^2 \left|\alpha_\mathrm{d}(t)\right|^2.
\end{equation}

\begin{figure}[t]
   \includegraphics[width=\linewidth]{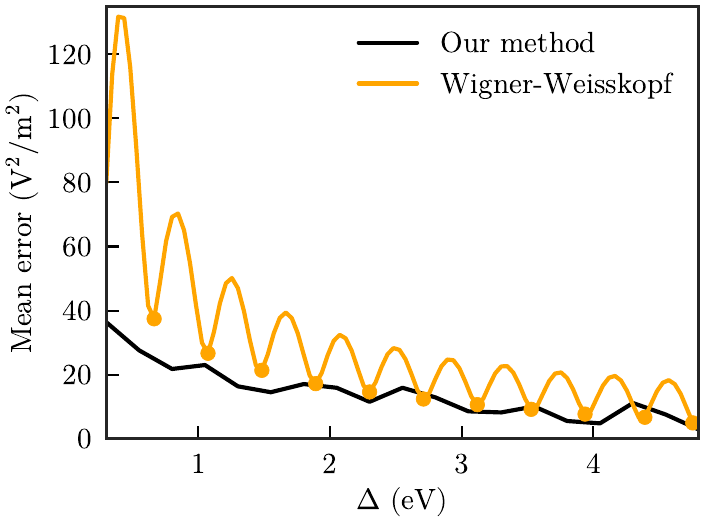}
   \caption{Dependence of the mean error of $G^{(1)}$ on $\Delta$ for different methods. For our method developed in \autoref{sec.theory}, $\Delta$ indicates the extent of the fitted frequency range, whereas for the Wigner-Weisskopf treatment it defines the frequency range over which the integral in \autoref{eq:ww_an} is evaluated. The orange dots mark the values of $\Delta$ for which the truncation was accurate.}
   \label{fig:error}
\end{figure}

The detector amplitude involves integration over the whole positive frequency axis, which implies that some kind of truncation of the spectral region has to be performed for numerical computation. As described in the main text, this truncation is not straightforward. In \autoref{fig:error}, we show the convergence of this method as a function of the range $\Delta$ taken for numerical integration. The error is defined as $\frac{1}{N_p}\sum_i^{N_p} \left|G^{(1)}_{\Delta}(t_i) - G^{(1)}_\mathrm{ref}(t_i)\right|$, where $t_i$ are the calculation points and $N_p$ is the number of points. The distribution of the $t_i$ is equidistant on a logarithmic scale. Here, $G^{(1)}_\mathrm{ref}$ is the result obtained with our method developed in \autoref{sec.theory} with the widest fitted frequency range that we considered, from 1 to 6~eV. As expected, our method clearly converges as the range $\Delta$ is increased, with the small oscillations related to fitting inaccuracies. The Wigner-Weisskopf approach also converges, but does so non-uniformly with strongly oscillatory behavior. The reason for these oscillations can be understood by realizing that the functions $r_\mathrm{ed}(\omega)$ defined above correspond (up to a factor of $\pi$) to the (dominant) positive-frequency part of the Kramers-Kronig integral that gives the real part of the response function, $\mathrm{Re}\left[R_\mathrm{ed}(\omega)\right]$. This again highlights the importance of the real part of the response function $\mathrm{Re}[R_\mathrm{ed}(\omega)]$, an observation that we can use to accurately truncate the spectral region to evaluate \autoref{eq:ww_an}. Due to the oscillating nature of $J_\mathrm{ed}(\omega)$, the integrals in \autoref{eq:alpha_d} will also oscillate as a function of the frequency cutoff, and the error will tend to be minimal for those values of the cutoff for which the value of the integrals is close to the real part of the response function. The orange dots in \autoref{fig:error} correspond to values of the frequency cutoff $\Delta$ of the integral for which $\Re[R_\mathrm{ed}(\omega_\mathrm{e})]$ is close to its asymptotic value, and these positions obviously correlate with the minima of the convergence plot. 

\bibliography{references}

\begin{thebibliography}{92}%
\makeatletter
\providecommand \@ifxundefined [1]{%
 \@ifx{#1\undefined}
}%
\providecommand \@ifnum [1]{%
 \ifnum #1\expandafter \@firstoftwo
 \else \expandafter \@secondoftwo
 \fi
}%
\providecommand \@ifx [1]{%
 \ifx #1\expandafter \@firstoftwo
 \else \expandafter \@secondoftwo
 \fi
}%
\providecommand \natexlab [1]{#1}%
\providecommand \enquote  [1]{``#1''}%
\providecommand \bibnamefont  [1]{#1}%
\providecommand \bibfnamefont [1]{#1}%
\providecommand \citenamefont [1]{#1}%
\providecommand \href@noop [0]{\@secondoftwo}%
\providecommand \href [0]{\begingroup \@sanitize@url \@href}%
\providecommand \@href[1]{\@@startlink{#1}\@@href}%
\providecommand \@@href[1]{\endgroup#1\@@endlink}%
\providecommand \@sanitize@url [0]{\catcode `\\12\catcode `\$12\catcode `\&12\catcode `\#12\catcode `\^12\catcode `\_12\catcode `\%12\relax}%
\providecommand \@@startlink[1]{}%
\providecommand \@@endlink[0]{}%
\providecommand \url  [0]{\begingroup\@sanitize@url \@url }%
\providecommand \@url [1]{\endgroup\@href {#1}{\urlprefix }}%
\providecommand \urlprefix  [0]{URL }%
\providecommand \Eprint [0]{\href }%
\providecommand \doibase [0]{https://doi.org/}%
\providecommand \selectlanguage [0]{\@gobble}%
\providecommand \bibinfo  [0]{\@secondoftwo}%
\providecommand \bibfield  [0]{\@secondoftwo}%
\providecommand \translation [1]{[#1]}%
\providecommand \BibitemOpen [0]{}%
\providecommand \bibitemStop [0]{}%
\providecommand \bibitemNoStop [0]{.\EOS\space}%
\providecommand \EOS [0]{\spacefactor3000\relax}%
\providecommand \BibitemShut  [1]{\csname bibitem#1\endcsname}%
\let\auto@bib@innerbib\@empty
%</preamble>
\bibitem [{\citenamefont {O'Brien}\ \emph {et~al.}(2009)\citenamefont {O'Brien}, \citenamefont {Furusawa},\ and\ \citenamefont {Vu{\v c}kovi{\'c}}}]{OBrien2009}%
  \BibitemOpen
  \bibfield  {author} {\bibinfo {author} {\bibfnamefont {J.~L.}\ \bibnamefont {O'Brien}}, \bibinfo {author} {\bibfnamefont {A.}~\bibnamefont {Furusawa}},\ and\ \bibinfo {author} {\bibfnamefont {J.}~\bibnamefont {Vu{\v c}kovi{\'c}}},\ }\bibfield  {title} {\bibinfo {title} {{Photonic Quantum Technologies}},\ }\href {https://doi.org/10.1038/nphoton.2009.229} {\bibfield  {journal} {\bibinfo  {journal} {Nat. Photonics}\ }\textbf {\bibinfo {volume} {3}},\ \bibinfo {pages} {687} (\bibinfo {year} {2009})}\BibitemShut {NoStop}%
\bibitem [{\citenamefont {Yao}\ \emph {et~al.}(2012)\citenamefont {Yao}, \citenamefont {Wang}, \citenamefont {Xu}, \citenamefont {Lu}, \citenamefont {Pan}, \citenamefont {Bao}, \citenamefont {Peng}, \citenamefont {Lu}, \citenamefont {Chen},\ and\ \citenamefont {Pan}}]{Yao2012}%
  \BibitemOpen
  \bibfield  {author} {\bibinfo {author} {\bibfnamefont {X.-C.}\ \bibnamefont {Yao}}, \bibinfo {author} {\bibfnamefont {T.-X.}\ \bibnamefont {Wang}}, \bibinfo {author} {\bibfnamefont {P.}~\bibnamefont {Xu}}, \bibinfo {author} {\bibfnamefont {H.}~\bibnamefont {Lu}}, \bibinfo {author} {\bibfnamefont {G.-S.}\ \bibnamefont {Pan}}, \bibinfo {author} {\bibfnamefont {X.-H.}\ \bibnamefont {Bao}}, \bibinfo {author} {\bibfnamefont {C.-Z.}\ \bibnamefont {Peng}}, \bibinfo {author} {\bibfnamefont {C.-Y.}\ \bibnamefont {Lu}}, \bibinfo {author} {\bibfnamefont {Y.-A.}\ \bibnamefont {Chen}},\ and\ \bibinfo {author} {\bibfnamefont {J.-W.}\ \bibnamefont {Pan}},\ }\bibfield  {title} {\bibinfo {title} {{Observation of Eight-Photon Entanglement}},\ }\href {https://doi.org/10.1038/nphoton.2011.354} {\bibfield  {journal} {\bibinfo  {journal} {Nat. Photonics}\ }\textbf {\bibinfo {volume} {6}},\ \bibinfo {pages} {225} (\bibinfo {year} {2012})}\BibitemShut {NoStop}%
\bibitem [{\citenamefont {Hiesmayr}\ \emph {et~al.}(2016)\citenamefont {Hiesmayr}, \citenamefont {De~Dood},\ and\ \citenamefont {L{\"o}ffler}}]{Hiesmayr2016}%
  \BibitemOpen
  \bibfield  {author} {\bibinfo {author} {\bibfnamefont {B.~C.}\ \bibnamefont {Hiesmayr}}, \bibinfo {author} {\bibfnamefont {M.~J.~A.}\ \bibnamefont {De~Dood}},\ and\ \bibinfo {author} {\bibfnamefont {W.}~\bibnamefont {L{\"o}ffler}},\ }\bibfield  {title} {\bibinfo {title} {{Observation of {{Four-Photon Orbital Angular Momentum Entanglement}}}},\ }\href {https://doi.org/10.1103/PhysRevLett.116.073601} {\bibfield  {journal} {\bibinfo  {journal} {Phys. Rev. Lett.}\ }\textbf {\bibinfo {volume} {116}},\ \bibinfo {pages} {073601} (\bibinfo {year} {2016})}\BibitemShut {NoStop}%
\bibitem [{\citenamefont {Huber}\ \emph {et~al.}(2017)\citenamefont {Huber}, \citenamefont {Reindl}, \citenamefont {Huo}, \citenamefont {Huang}, \citenamefont {Wildmann}, \citenamefont {Schmidt}, \citenamefont {Rastelli},\ and\ \citenamefont {Trotta}}]{Huber2017}%
  \BibitemOpen
  \bibfield  {author} {\bibinfo {author} {\bibfnamefont {D.}~\bibnamefont {Huber}}, \bibinfo {author} {\bibfnamefont {M.}~\bibnamefont {Reindl}}, \bibinfo {author} {\bibfnamefont {Y.}~\bibnamefont {Huo}}, \bibinfo {author} {\bibfnamefont {H.}~\bibnamefont {Huang}}, \bibinfo {author} {\bibfnamefont {J.~S.}\ \bibnamefont {Wildmann}}, \bibinfo {author} {\bibfnamefont {O.~G.}\ \bibnamefont {Schmidt}}, \bibinfo {author} {\bibfnamefont {A.}~\bibnamefont {Rastelli}},\ and\ \bibinfo {author} {\bibfnamefont {R.}~\bibnamefont {Trotta}},\ }\bibfield  {title} {\bibinfo {title} {{Highly Indistinguishable and Strongly Entangled Photons from Symmetric {{GaAs}} Quantum Dots}},\ }\href {https://doi.org/10.1038/ncomms15506} {\bibfield  {journal} {\bibinfo  {journal} {Nat. Commun.}\ }\textbf {\bibinfo {volume} {8}},\ \bibinfo {pages} {15506} (\bibinfo {year} {2017})}\BibitemShut {NoStop}%
\bibitem [{\citenamefont {Prasad}\ \emph {et~al.}(2020)\citenamefont {Prasad}, \citenamefont {Hinney}, \citenamefont {Mahmoodian}, \citenamefont {Hammerer}, \citenamefont {Rind}, \citenamefont {Schneeweiss}, \citenamefont {S{\o}rensen}, \citenamefont {Volz},\ and\ \citenamefont {Rauschenbeutel}}]{Prasad2020}%
  \BibitemOpen
  \bibfield  {author} {\bibinfo {author} {\bibfnamefont {A.~S.}\ \bibnamefont {Prasad}}, \bibinfo {author} {\bibfnamefont {J.}~\bibnamefont {Hinney}}, \bibinfo {author} {\bibfnamefont {S.}~\bibnamefont {Mahmoodian}}, \bibinfo {author} {\bibfnamefont {K.}~\bibnamefont {Hammerer}}, \bibinfo {author} {\bibfnamefont {S.}~\bibnamefont {Rind}}, \bibinfo {author} {\bibfnamefont {P.}~\bibnamefont {Schneeweiss}}, \bibinfo {author} {\bibfnamefont {A.~S.}\ \bibnamefont {S{\o}rensen}}, \bibinfo {author} {\bibfnamefont {J.}~\bibnamefont {Volz}},\ and\ \bibinfo {author} {\bibfnamefont {A.}~\bibnamefont {Rauschenbeutel}},\ }\bibfield  {title} {\bibinfo {title} {{Correlating Photons Using the Collective Nonlinear Response of Atoms Weakly Coupled to an Optical Mode}},\ }\href {https://doi.org/10.1038/s41566-020-0692-z} {\bibfield  {journal} {\bibinfo  {journal} {Nat. Photonics}\ }\textbf {\bibinfo {volume} {14}},\ \bibinfo {pages} {719} (\bibinfo {year} {2020})}\BibitemShut {NoStop}%
\bibitem [{\citenamefont {Schimpf}\ \emph {et~al.}(2021)\citenamefont {Schimpf}, \citenamefont {Reindl}, \citenamefont {Basso~Basset}, \citenamefont {J{\"o}ns}, \citenamefont {Trotta},\ and\ \citenamefont {Rastelli}}]{Schimpf2021}%
  \BibitemOpen
  \bibfield  {author} {\bibinfo {author} {\bibfnamefont {C.}~\bibnamefont {Schimpf}}, \bibinfo {author} {\bibfnamefont {M.}~\bibnamefont {Reindl}}, \bibinfo {author} {\bibfnamefont {F.}~\bibnamefont {Basso~Basset}}, \bibinfo {author} {\bibfnamefont {K.~D.}\ \bibnamefont {J{\"o}ns}}, \bibinfo {author} {\bibfnamefont {R.}~\bibnamefont {Trotta}},\ and\ \bibinfo {author} {\bibfnamefont {A.}~\bibnamefont {Rastelli}},\ }\bibfield  {title} {\bibinfo {title} {{Quantum Dots as Potential Sources of Strongly Entangled Photons: {{Perspectives}} and Challenges for Applications in Quantum Networks}},\ }\href {https://doi.org/10.1063/5.0038729} {\bibfield  {journal} {\bibinfo  {journal} {Appl. Phys. Lett.}\ }\textbf {\bibinfo {volume} {118}},\ \bibinfo {pages} {100502} (\bibinfo {year} {2021})}\BibitemShut {NoStop}%
\bibitem [{\citenamefont {Wein}\ \emph {et~al.}(2022)\citenamefont {Wein}, \citenamefont {Loredo}, \citenamefont {Maffei}, \citenamefont {Hilaire}, \citenamefont {Harouri}, \citenamefont {Somaschi}, \citenamefont {Lema{\^i}tre}, \citenamefont {Sagnes}, \citenamefont {Lanco}, \citenamefont {Krebs}, \citenamefont {Auff{\`e}ves}, \citenamefont {Simon}, \citenamefont {Senellart},\ and\ \citenamefont {{Ant{\'o}n-Solanas}}}]{Wein2022}%
  \BibitemOpen
  \bibfield  {author} {\bibinfo {author} {\bibfnamefont {S.~C.}\ \bibnamefont {Wein}}, \bibinfo {author} {\bibfnamefont {J.~C.}\ \bibnamefont {Loredo}}, \bibinfo {author} {\bibfnamefont {M.}~\bibnamefont {Maffei}}, \bibinfo {author} {\bibfnamefont {P.}~\bibnamefont {Hilaire}}, \bibinfo {author} {\bibfnamefont {A.}~\bibnamefont {Harouri}}, \bibinfo {author} {\bibfnamefont {N.}~\bibnamefont {Somaschi}}, \bibinfo {author} {\bibfnamefont {A.}~\bibnamefont {Lema{\^i}tre}}, \bibinfo {author} {\bibfnamefont {I.}~\bibnamefont {Sagnes}}, \bibinfo {author} {\bibfnamefont {L.}~\bibnamefont {Lanco}}, \bibinfo {author} {\bibfnamefont {O.}~\bibnamefont {Krebs}}, \bibinfo {author} {\bibfnamefont {A.}~\bibnamefont {Auff{\`e}ves}}, \bibinfo {author} {\bibfnamefont {C.}~\bibnamefont {Simon}}, \bibinfo {author} {\bibfnamefont {P.}~\bibnamefont {Senellart}},\ and\ \bibinfo {author} {\bibfnamefont {C.}~\bibnamefont {{Ant{\'o}n-Solanas}}},\ }\bibfield  {title} {\bibinfo {title} {{Photon-Number Entanglement Generated by Sequential Excitation of a Two-Level Atom}},\ }\href {https://doi.org/10.1038/s41566-022-00979-z} {\bibfield  {journal} {\bibinfo  {journal} {Nat. Photonics}\ }\textbf {\bibinfo {volume} {16}},\ \bibinfo {pages} {374} (\bibinfo {year} {2022})}\BibitemShut {NoStop}%
\bibitem [{\citenamefont {Zhang}\ \emph {et~al.}(2022)\citenamefont {Zhang}, \citenamefont {Ma}, \citenamefont {Parry}, \citenamefont {Cai}, \citenamefont {{Camacho-Morales}}, \citenamefont {Xu}, \citenamefont {Neshev},\ and\ \citenamefont {Sukhorukov}}]{Zhang2022}%
  \BibitemOpen
  \bibfield  {author} {\bibinfo {author} {\bibfnamefont {J.}~\bibnamefont {Zhang}}, \bibinfo {author} {\bibfnamefont {J.}~\bibnamefont {Ma}}, \bibinfo {author} {\bibfnamefont {M.}~\bibnamefont {Parry}}, \bibinfo {author} {\bibfnamefont {M.}~\bibnamefont {Cai}}, \bibinfo {author} {\bibfnamefont {R.}~\bibnamefont {{Camacho-Morales}}}, \bibinfo {author} {\bibfnamefont {L.}~\bibnamefont {Xu}}, \bibinfo {author} {\bibfnamefont {D.~N.}\ \bibnamefont {Neshev}},\ and\ \bibinfo {author} {\bibfnamefont {A.~A.}\ \bibnamefont {Sukhorukov}},\ }\bibfield  {title} {\bibinfo {title} {{Spatially Entangled Photon Pairs from Lithium Niobate Nonlocal Metasurfaces}},\ }\href {https://doi.org/10.1126/sciadv.abq4240} {\bibfield  {journal} {\bibinfo  {journal} {Science Adv.}\ }\textbf {\bibinfo {volume} {8}},\ \bibinfo {pages} {eabq4240} (\bibinfo {year} {2022})}\BibitemShut {NoStop}%
\bibitem [{\citenamefont {Vento}\ \emph {et~al.}(2023)\citenamefont {Vento}, \citenamefont {Tarrago~Velez}, \citenamefont {Pogrebna},\ and\ \citenamefont {Galland}}]{Vento2023}%
  \BibitemOpen
  \bibfield  {author} {\bibinfo {author} {\bibfnamefont {V.}~\bibnamefont {Vento}}, \bibinfo {author} {\bibfnamefont {S.}~\bibnamefont {Tarrago~Velez}}, \bibinfo {author} {\bibfnamefont {A.}~\bibnamefont {Pogrebna}},\ and\ \bibinfo {author} {\bibfnamefont {C.}~\bibnamefont {Galland}},\ }\bibfield  {title} {\bibinfo {title} {{Measurement-Induced Collective Vibrational Quantum Coherence under Spontaneous {{Raman}} Scattering in a Liquid}},\ }\href {https://doi.org/10.1038/s41467-023-38483-9} {\bibfield  {journal} {\bibinfo  {journal} {Nat. Commun.}\ }\textbf {\bibinfo {volume} {14}},\ \bibinfo {pages} {2818} (\bibinfo {year} {2023})}\BibitemShut {NoStop}%
\bibitem [{\citenamefont {Darqui{\'e}}\ \emph {et~al.}(2005)\citenamefont {Darqui{\'e}}, \citenamefont {Jones}, \citenamefont {Dingjan}, \citenamefont {Beugnon}, \citenamefont {Bergamini}, \citenamefont {Sortais}, \citenamefont {Messin}, \citenamefont {Browaeys},\ and\ \citenamefont {Grangier}}]{Darquie2005}%
  \BibitemOpen
  \bibfield  {author} {\bibinfo {author} {\bibfnamefont {B.}~\bibnamefont {Darqui{\'e}}}, \bibinfo {author} {\bibfnamefont {M.~P.~A.}\ \bibnamefont {Jones}}, \bibinfo {author} {\bibfnamefont {J.}~\bibnamefont {Dingjan}}, \bibinfo {author} {\bibfnamefont {J.}~\bibnamefont {Beugnon}}, \bibinfo {author} {\bibfnamefont {S.}~\bibnamefont {Bergamini}}, \bibinfo {author} {\bibfnamefont {Y.}~\bibnamefont {Sortais}}, \bibinfo {author} {\bibfnamefont {G.}~\bibnamefont {Messin}}, \bibinfo {author} {\bibfnamefont {A.}~\bibnamefont {Browaeys}},\ and\ \bibinfo {author} {\bibfnamefont {P.}~\bibnamefont {Grangier}},\ }\bibfield  {title} {\bibinfo {title} {{Controlled {{Single-Photon Emission}} from a {{Single Trapped Two-Level Atom}}}},\ }\href {https://doi.org/10.1126/science.1113394} {\bibfield  {journal} {\bibinfo  {journal} {Science}\ }\textbf {\bibinfo {volume} {309}},\ \bibinfo {pages} {454} (\bibinfo {year} {2005})}\BibitemShut {NoStop}%
\bibitem [{\citenamefont {Lounis}\ and\ \citenamefont {Orrit}(2005)}]{Lounis2005}%
  \BibitemOpen
  \bibfield  {author} {\bibinfo {author} {\bibfnamefont {B.}~\bibnamefont {Lounis}}\ and\ \bibinfo {author} {\bibfnamefont {M.}~\bibnamefont {Orrit}},\ }\bibfield  {title} {\bibinfo {title} {{Single-Photon Sources}},\ }\href {https://doi.org/10.1088/0034-4885/68/5/R04} {\bibfield  {journal} {\bibinfo  {journal} {Rep. Prog. Phys.}\ }\textbf {\bibinfo {volume} {68}},\ \bibinfo {pages} {1129} (\bibinfo {year} {2005})}\BibitemShut {NoStop}%
\bibitem [{\citenamefont {Reindl}\ \emph {et~al.}(2017)\citenamefont {Reindl}, \citenamefont {J{\"o}ns}, \citenamefont {Huber}, \citenamefont {Schimpf}, \citenamefont {Huo}, \citenamefont {Zwiller}, \citenamefont {Rastelli},\ and\ \citenamefont {Trotta}}]{Reindl2017}%
  \BibitemOpen
  \bibfield  {author} {\bibinfo {author} {\bibfnamefont {M.}~\bibnamefont {Reindl}}, \bibinfo {author} {\bibfnamefont {K.~D.}\ \bibnamefont {J{\"o}ns}}, \bibinfo {author} {\bibfnamefont {D.}~\bibnamefont {Huber}}, \bibinfo {author} {\bibfnamefont {C.}~\bibnamefont {Schimpf}}, \bibinfo {author} {\bibfnamefont {Y.}~\bibnamefont {Huo}}, \bibinfo {author} {\bibfnamefont {V.}~\bibnamefont {Zwiller}}, \bibinfo {author} {\bibfnamefont {A.}~\bibnamefont {Rastelli}},\ and\ \bibinfo {author} {\bibfnamefont {R.}~\bibnamefont {Trotta}},\ }\bibfield  {title} {\bibinfo {title} {{Phonon-{{Assisted Two-Photon Interference}} from {{Remote Quantum Emitters}}}},\ }\href {https://doi.org/10.1021/acs.nanolett.7b00777} {\bibfield  {journal} {\bibinfo  {journal} {Nano Lett.}\ }\textbf {\bibinfo {volume} {17}},\ \bibinfo {pages} {4090} (\bibinfo {year} {2017})}\BibitemShut {NoStop}%
\bibitem [{\citenamefont {Gao}\ \emph {et~al.}(2023)\citenamefont {Gao}, \citenamefont {{von Helversen}}, \citenamefont {{Ant{\'o}n-Solanas}}, \citenamefont {Schneider},\ and\ \citenamefont {Heindel}}]{Gao2023}%
  \BibitemOpen
  \bibfield  {author} {\bibinfo {author} {\bibfnamefont {T.}~\bibnamefont {Gao}}, \bibinfo {author} {\bibfnamefont {M.}~\bibnamefont {{von Helversen}}}, \bibinfo {author} {\bibfnamefont {C.}~\bibnamefont {{Ant{\'o}n-Solanas}}}, \bibinfo {author} {\bibfnamefont {C.}~\bibnamefont {Schneider}},\ and\ \bibinfo {author} {\bibfnamefont {T.}~\bibnamefont {Heindel}},\ }\bibfield  {title} {\bibinfo {title} {{Atomically-Thin Single-Photon Sources for Quantum Communication}},\ }\href {https://doi.org/10.1038/s41699-023-00366-4} {\bibfield  {journal} {\bibinfo  {journal} {npj 2D Mater. Appl.}\ }\textbf {\bibinfo {volume} {7}},\ \bibinfo {pages} {4} (\bibinfo {year} {2023})}\BibitemShut {NoStop}%
\bibitem [{\citenamefont {Olejniczak}\ \emph {et~al.}(2024)\citenamefont {Olejniczak}, \citenamefont {Lawera}, \citenamefont {{Zapata-Herrera}}, \citenamefont {Chuvilin}, \citenamefont {Samokhvalov}, \citenamefont {Nabiev}, \citenamefont {Grzelczak}, \citenamefont {Rakovich},\ and\ \citenamefont {Krivenkov}}]{Olejniczak2024}%
  \BibitemOpen
  \bibfield  {author} {\bibinfo {author} {\bibfnamefont {A.}~\bibnamefont {Olejniczak}}, \bibinfo {author} {\bibfnamefont {Z.}~\bibnamefont {Lawera}}, \bibinfo {author} {\bibfnamefont {M.}~\bibnamefont {{Zapata-Herrera}}}, \bibinfo {author} {\bibfnamefont {A.}~\bibnamefont {Chuvilin}}, \bibinfo {author} {\bibfnamefont {P.}~\bibnamefont {Samokhvalov}}, \bibinfo {author} {\bibfnamefont {I.}~\bibnamefont {Nabiev}}, \bibinfo {author} {\bibfnamefont {M.}~\bibnamefont {Grzelczak}}, \bibinfo {author} {\bibfnamefont {Y.}~\bibnamefont {Rakovich}},\ and\ \bibinfo {author} {\bibfnamefont {V.}~\bibnamefont {Krivenkov}},\ }\bibfield  {title} {\bibinfo {title} {{On-Demand Reversible Switching of the Emission Mode of Individual Semiconductor Quantum Emitters Using Plasmonic Metasurfaces}},\ }\href {https://doi.org/10.1063/5.0170535} {\bibfield  {journal} {\bibinfo  {journal} {APL Photonics}\ }\textbf {\bibinfo {volume} {9}},\ \bibinfo {pages} {016107} (\bibinfo {year} {2024})}\BibitemShut {NoStop}%
\bibitem [{\citenamefont {Khalid}\ and\ \citenamefont {Laussy}(2024)}]{Khalid2024}%
  \BibitemOpen
  \bibfield  {author} {\bibinfo {author} {\bibfnamefont {S.}~\bibnamefont {Khalid}}\ and\ \bibinfo {author} {\bibfnamefont {F.~P.}\ \bibnamefont {Laussy}},\ }\bibfield  {title} {\bibinfo {title} {{Perfect Single-Photon Sources}},\ }\href {https://doi.org/10.1038/s41598-023-47585-9} {\bibfield  {journal} {\bibinfo  {journal} {Sci. Rep.}\ }\textbf {\bibinfo {volume} {14}},\ \bibinfo {pages} {2684} (\bibinfo {year} {2024})}\BibitemShut {NoStop}%
\bibitem [{\citenamefont {Esmann}\ \emph {et~al.}(2024)\citenamefont {Esmann}, \citenamefont {Wein},\ and\ \citenamefont {{Ant{\'o}n-Solanas}}}]{Esmann2024}%
  \BibitemOpen
  \bibfield  {author} {\bibinfo {author} {\bibfnamefont {M.}~\bibnamefont {Esmann}}, \bibinfo {author} {\bibfnamefont {S.~C.}\ \bibnamefont {Wein}},\ and\ \bibinfo {author} {\bibfnamefont {C.}~\bibnamefont {{Ant{\'o}n-Solanas}}},\ }\bibfield  {title} {\bibinfo {title} {{Solid-{{State Single-Photon Sources}}: {{Recent Advances}} for {{Novel Quantum Materials}}}},\ }\href {https://doi.org/10.1002/adfm.202315936} {\bibfield  {journal} {\bibinfo  {journal} {Adv. Funct. Mater.}\ }\textbf {\bibinfo {volume} {34}},\ \bibinfo {pages} {2315936} (\bibinfo {year} {2024})}\BibitemShut {NoStop}%
\bibitem [{\citenamefont {{Ben-Asher}}\ \emph {et~al.}(2023)\citenamefont {{Ben-Asher}}, \citenamefont {{Fern{\'a}ndez-Dom{\'i}nguez}},\ and\ \citenamefont {Feist}}]{Ben-Asher2023}%
  \BibitemOpen
  \bibfield  {author} {\bibinfo {author} {\bibfnamefont {A.}~\bibnamefont {{Ben-Asher}}}, \bibinfo {author} {\bibfnamefont {A.~I.}\ \bibnamefont {{Fern{\'a}ndez-Dom{\'i}nguez}}},\ and\ \bibinfo {author} {\bibfnamefont {J.}~\bibnamefont {Feist}},\ }\bibfield  {title} {\bibinfo {title} {{Non-{{Hermitian Anharmonicity Induces Single-Photon Emission}}}},\ }\href {https://doi.org/10.1103/PhysRevLett.130.243601} {\bibfield  {journal} {\bibinfo  {journal} {Phys. Rev. Lett.}\ }\textbf {\bibinfo {volume} {130}},\ \bibinfo {pages} {243601} (\bibinfo {year} {2023})}\BibitemShut {NoStop}%
\bibitem [{\citenamefont {Shan}\ \emph {et~al.}(2023)\citenamefont {Shan}, \citenamefont {Drawer}, \citenamefont {Sun}, \citenamefont {{Anton-Solanas}}, \citenamefont {Esmann}, \citenamefont {Yumigeta}, \citenamefont {Watanabe}, \citenamefont {Taniguchi}, \citenamefont {Tongay}, \citenamefont {H{\"o}fling}, \citenamefont {Savenko},\ and\ \citenamefont {Schneider}}]{Shan2023}%
  \BibitemOpen
  \bibfield  {author} {\bibinfo {author} {\bibfnamefont {H.}~\bibnamefont {Shan}}, \bibinfo {author} {\bibfnamefont {J.-C.}\ \bibnamefont {Drawer}}, \bibinfo {author} {\bibfnamefont {M.}~\bibnamefont {Sun}}, \bibinfo {author} {\bibfnamefont {C.}~\bibnamefont {{Anton-Solanas}}}, \bibinfo {author} {\bibfnamefont {M.}~\bibnamefont {Esmann}}, \bibinfo {author} {\bibfnamefont {K.}~\bibnamefont {Yumigeta}}, \bibinfo {author} {\bibfnamefont {K.}~\bibnamefont {Watanabe}}, \bibinfo {author} {\bibfnamefont {T.}~\bibnamefont {Taniguchi}}, \bibinfo {author} {\bibfnamefont {S.}~\bibnamefont {Tongay}}, \bibinfo {author} {\bibfnamefont {S.}~\bibnamefont {H{\"o}fling}}, \bibinfo {author} {\bibfnamefont {I.}~\bibnamefont {Savenko}},\ and\ \bibinfo {author} {\bibfnamefont {C.}~\bibnamefont {Schneider}},\ }\bibfield  {title} {\bibinfo {title} {{Second-{{Order Temporal Coherence}} of {{Polariton Lasers Based}} on an {{Atomically Thin Crystal}} in a {{Microcavity}}}},\ }\href {https://doi.org/10.1103/PhysRevLett.131.206901} {\bibfield  {journal} {\bibinfo  {journal} {Phys. Rev. Lett.}\ }\textbf {\bibinfo {volume} {131}},\ \bibinfo {pages} {206901} (\bibinfo {year} {2023})}\BibitemShut {NoStop}%
\bibitem [{\citenamefont {Groiseau}\ \emph {et~al.}(2024)\citenamefont {Groiseau}, \citenamefont {{Fern{\'a}ndez-Dom{\'i}nguez}}, \citenamefont {{Mart{\'i}n-Cano}},\ and\ \citenamefont {Mu{\~n}oz}}]{Groiseau2024}%
  \BibitemOpen
  \bibfield  {author} {\bibinfo {author} {\bibfnamefont {C.}~\bibnamefont {Groiseau}}, \bibinfo {author} {\bibfnamefont {A.~I.}\ \bibnamefont {{Fern{\'a}ndez-Dom{\'i}nguez}}}, \bibinfo {author} {\bibfnamefont {D.}~\bibnamefont {{Mart{\'i}n-Cano}}},\ and\ \bibinfo {author} {\bibfnamefont {C.~S.}\ \bibnamefont {Mu{\~n}oz}},\ }\bibfield  {title} {\bibinfo {title} {{Single-{{Photon Source Over}} the {{Terahertz Regime}}}},\ }\href {https://doi.org/10.1103/PRXQuantum.5.010312} {\bibfield  {journal} {\bibinfo  {journal} {PRX Quantum}\ }\textbf {\bibinfo {volume} {5}},\ \bibinfo {pages} {010312} (\bibinfo {year} {2024})}\BibitemShut {NoStop}%
\bibitem [{\citenamefont {Putintsev}\ \emph {et~al.}(2024)\citenamefont {Putintsev}, \citenamefont {Zasedatelev}, \citenamefont {Shishkov}, \citenamefont {Misko}, \citenamefont {Sannikov}, \citenamefont {Andrianov}, \citenamefont {Lozovik}, \citenamefont {Scherf},\ and\ \citenamefont {Lagoudakis}}]{Putintsev2024}%
  \BibitemOpen
  \bibfield  {author} {\bibinfo {author} {\bibfnamefont {A.~D.}\ \bibnamefont {Putintsev}}, \bibinfo {author} {\bibfnamefont {A.~V.}\ \bibnamefont {Zasedatelev}}, \bibinfo {author} {\bibfnamefont {V.~Y.}\ \bibnamefont {Shishkov}}, \bibinfo {author} {\bibfnamefont {M.}~\bibnamefont {Misko}}, \bibinfo {author} {\bibfnamefont {D.~A.}\ \bibnamefont {Sannikov}}, \bibinfo {author} {\bibfnamefont {E.~S.}\ \bibnamefont {Andrianov}}, \bibinfo {author} {\bibfnamefont {Y.~E.}\ \bibnamefont {Lozovik}}, \bibinfo {author} {\bibfnamefont {U.}~\bibnamefont {Scherf}},\ and\ \bibinfo {author} {\bibfnamefont {P.~G.}\ \bibnamefont {Lagoudakis}},\ }\bibfield  {title} {\bibinfo {title} {{Photon Statistics of Organic Polariton Condensates}},\ }\href {https://doi.org/10.1103/PhysRevB.110.045125} {\bibfield  {journal} {\bibinfo  {journal} {Phys. Rev. B}\ }\textbf {\bibinfo {volume} {110}},\ \bibinfo {pages} {045125} (\bibinfo {year} {2024})}\BibitemShut {NoStop}%
\bibitem [{\citenamefont {Faraon}\ \emph {et~al.}(2008)\citenamefont {Faraon}, \citenamefont {Fushman}, \citenamefont {Englund}, \citenamefont {Stoltz}, \citenamefont {Petroff},\ and\ \citenamefont {Vu{\v c}kovi{\'c}}}]{Faraon2008}%
  \BibitemOpen
  \bibfield  {author} {\bibinfo {author} {\bibfnamefont {A.}~\bibnamefont {Faraon}}, \bibinfo {author} {\bibfnamefont {I.}~\bibnamefont {Fushman}}, \bibinfo {author} {\bibfnamefont {D.}~\bibnamefont {Englund}}, \bibinfo {author} {\bibfnamefont {N.}~\bibnamefont {Stoltz}}, \bibinfo {author} {\bibfnamefont {P.}~\bibnamefont {Petroff}},\ and\ \bibinfo {author} {\bibfnamefont {J.}~\bibnamefont {Vu{\v c}kovi{\'c}}},\ }\bibfield  {title} {\bibinfo {title} {{Coherent Generation of Non-Classical Light on a Chip via Photon-Induced Tunnelling and Blockade}},\ }\href {https://doi.org/10.1038/nphys1078} {\bibfield  {journal} {\bibinfo  {journal} {Nat. Phys.}\ }\textbf {\bibinfo {volume} {4}},\ \bibinfo {pages} {859} (\bibinfo {year} {2008})}\BibitemShut {NoStop}%
\bibitem [{\citenamefont {Chang}\ \emph {et~al.}(2014)\citenamefont {Chang}, \citenamefont {Vuleti{\'c}},\ and\ \citenamefont {Lukin}}]{Chang2014Quantum}%
  \BibitemOpen
  \bibfield  {author} {\bibinfo {author} {\bibfnamefont {D.~E.}\ \bibnamefont {Chang}}, \bibinfo {author} {\bibfnamefont {V.}~\bibnamefont {Vuleti{\'c}}},\ and\ \bibinfo {author} {\bibfnamefont {M.~D.}\ \bibnamefont {Lukin}},\ }\bibfield  {title} {\bibinfo {title} {{Quantum Nonlinear Optics --- Photon by Photon}},\ }\href {https://doi.org/10.1038/nphoton.2014.192} {\bibfield  {journal} {\bibinfo  {journal} {Nat. Photonics}\ }\textbf {\bibinfo {volume} {8}},\ \bibinfo {pages} {685} (\bibinfo {year} {2014})}\BibitemShut {NoStop}%
\bibitem [{\citenamefont {Loredo}\ \emph {et~al.}(2019)\citenamefont {Loredo}, \citenamefont {Ant{\'o}n}, \citenamefont {Reznychenko}, \citenamefont {Hilaire}, \citenamefont {Harouri}, \citenamefont {Millet}, \citenamefont {Ollivier}, \citenamefont {Somaschi}, \citenamefont {De~Santis}, \citenamefont {Lema{\^i}tre}, \citenamefont {Sagnes}, \citenamefont {Lanco}, \citenamefont {Auff{\`e}ves}, \citenamefont {Krebs},\ and\ \citenamefont {Senellart}}]{Loredo2019}%
  \BibitemOpen
  \bibfield  {author} {\bibinfo {author} {\bibfnamefont {J.~C.}\ \bibnamefont {Loredo}}, \bibinfo {author} {\bibfnamefont {C.}~\bibnamefont {Ant{\'o}n}}, \bibinfo {author} {\bibfnamefont {B.}~\bibnamefont {Reznychenko}}, \bibinfo {author} {\bibfnamefont {P.}~\bibnamefont {Hilaire}}, \bibinfo {author} {\bibfnamefont {A.}~\bibnamefont {Harouri}}, \bibinfo {author} {\bibfnamefont {C.}~\bibnamefont {Millet}}, \bibinfo {author} {\bibfnamefont {H.}~\bibnamefont {Ollivier}}, \bibinfo {author} {\bibfnamefont {N.}~\bibnamefont {Somaschi}}, \bibinfo {author} {\bibfnamefont {L.}~\bibnamefont {De~Santis}}, \bibinfo {author} {\bibfnamefont {A.}~\bibnamefont {Lema{\^i}tre}}, \bibinfo {author} {\bibfnamefont {I.}~\bibnamefont {Sagnes}}, \bibinfo {author} {\bibfnamefont {L.}~\bibnamefont {Lanco}}, \bibinfo {author} {\bibfnamefont {A.}~\bibnamefont {Auff{\`e}ves}}, \bibinfo {author} {\bibfnamefont {O.}~\bibnamefont {Krebs}},\ and\ \bibinfo {author} {\bibfnamefont {P.}~\bibnamefont {Senellart}},\ }\bibfield  {title} {\bibinfo {title} {{Generation of Non-Classical Light in a Photon-Number Superposition}},\ }\href {https://doi.org/10.1038/s41566-019-0506-3} {\bibfield  {journal} {\bibinfo  {journal} {Nat. Photonics}\ }\textbf {\bibinfo {volume} {13}},\ \bibinfo {pages} {803} (\bibinfo {year} {2019})}\BibitemShut {NoStop}%
\bibitem [{\citenamefont {Tomm}\ \emph {et~al.}(2021)\citenamefont {Tomm}, \citenamefont {Javadi}, \citenamefont {Antoniadis}, \citenamefont {Najer}, \citenamefont {L{\"o}bl}, \citenamefont {Korsch}, \citenamefont {Schott}, \citenamefont {Valentin}, \citenamefont {Wieck}, \citenamefont {Ludwig},\ and\ \citenamefont {Warburton}}]{Tomm2021}%
  \BibitemOpen
  \bibfield  {author} {\bibinfo {author} {\bibfnamefont {N.}~\bibnamefont {Tomm}}, \bibinfo {author} {\bibfnamefont {A.}~\bibnamefont {Javadi}}, \bibinfo {author} {\bibfnamefont {N.~O.}\ \bibnamefont {Antoniadis}}, \bibinfo {author} {\bibfnamefont {D.}~\bibnamefont {Najer}}, \bibinfo {author} {\bibfnamefont {M.~C.}\ \bibnamefont {L{\"o}bl}}, \bibinfo {author} {\bibfnamefont {A.~R.}\ \bibnamefont {Korsch}}, \bibinfo {author} {\bibfnamefont {R.}~\bibnamefont {Schott}}, \bibinfo {author} {\bibfnamefont {S.~R.}\ \bibnamefont {Valentin}}, \bibinfo {author} {\bibfnamefont {A.~D.}\ \bibnamefont {Wieck}}, \bibinfo {author} {\bibfnamefont {A.}~\bibnamefont {Ludwig}},\ and\ \bibinfo {author} {\bibfnamefont {R.~J.}\ \bibnamefont {Warburton}},\ }\bibfield  {title} {\bibinfo {title} {{A Bright and Fast Source of Coherent Single Photons}},\ }\href {https://doi.org/10.1038/s41565-020-00831-x} {\bibfield  {journal} {\bibinfo  {journal} {Nat. Nanotechnol.}\ }\textbf {\bibinfo {volume} {16}},\ \bibinfo {pages} {399} (\bibinfo {year} {2021})}\BibitemShut {NoStop}%
\bibitem [{\citenamefont {Lukin}\ \emph {et~al.}(2023)\citenamefont {Lukin}, \citenamefont {Guidry}, \citenamefont {Yang}, \citenamefont {Ghezellou}, \citenamefont {Deb~Mishra}, \citenamefont {Abe}, \citenamefont {Ohshima}, \citenamefont {{Ul-Hassan}},\ and\ \citenamefont {Vu{\v c}kovi{\'c}}}]{Lukin2023}%
  \BibitemOpen
  \bibfield  {author} {\bibinfo {author} {\bibfnamefont {D.~M.}\ \bibnamefont {Lukin}}, \bibinfo {author} {\bibfnamefont {M.~A.}\ \bibnamefont {Guidry}}, \bibinfo {author} {\bibfnamefont {J.}~\bibnamefont {Yang}}, \bibinfo {author} {\bibfnamefont {M.}~\bibnamefont {Ghezellou}}, \bibinfo {author} {\bibfnamefont {S.}~\bibnamefont {Deb~Mishra}}, \bibinfo {author} {\bibfnamefont {H.}~\bibnamefont {Abe}}, \bibinfo {author} {\bibfnamefont {T.}~\bibnamefont {Ohshima}}, \bibinfo {author} {\bibfnamefont {J.}~\bibnamefont {{Ul-Hassan}}},\ and\ \bibinfo {author} {\bibfnamefont {J.}~\bibnamefont {Vu{\v c}kovi{\'c}}},\ }\bibfield  {title} {\bibinfo {title} {{Two-{{Emitter Multimode Cavity Quantum Electrodynamics}} in {{Thin-Film Silicon Carbide Photonics}}}},\ }\href {https://doi.org/10.1103/PhysRevX.13.011005} {\bibfield  {journal} {\bibinfo  {journal} {Phys. Rev. X}\ }\textbf {\bibinfo {volume} {13}},\ \bibinfo {pages} {011005} (\bibinfo {year} {2023})}\BibitemShut {NoStop}%
\bibitem [{\citenamefont {Ridolfo}\ \emph {et~al.}(2010)\citenamefont {Ridolfo}, \citenamefont {Di~Stefano}, \citenamefont {Fina}, \citenamefont {Saija},\ and\ \citenamefont {Savasta}}]{Ridolfo2010}%
  \BibitemOpen
  \bibfield  {author} {\bibinfo {author} {\bibfnamefont {A.}~\bibnamefont {Ridolfo}}, \bibinfo {author} {\bibfnamefont {O.}~\bibnamefont {Di~Stefano}}, \bibinfo {author} {\bibfnamefont {N.}~\bibnamefont {Fina}}, \bibinfo {author} {\bibfnamefont {R.}~\bibnamefont {Saija}},\ and\ \bibinfo {author} {\bibfnamefont {S.}~\bibnamefont {Savasta}},\ }\bibfield  {title} {\bibinfo {title} {{Quantum {{Plasmonics}} with {{Quantum Dot-Metal Nanoparticle Molecules}}: {{Influence}} of the {{Fano Effect}} on {{Photon Statistics}}}},\ }\href {https://doi.org/10.1103/PhysRevLett.105.263601} {\bibfield  {journal} {\bibinfo  {journal} {Phys. Rev. Lett.}\ }\textbf {\bibinfo {volume} {105}},\ \bibinfo {pages} {263601} (\bibinfo {year} {2010})}\BibitemShut {NoStop}%
\bibitem [{\citenamefont {Auff{\`e}ves}\ \emph {et~al.}(2011)\citenamefont {Auff{\`e}ves}, \citenamefont {Gerace}, \citenamefont {Portolan}, \citenamefont {Drezet},\ and\ \citenamefont {Santos}}]{Auffeves2011}%
  \BibitemOpen
  \bibfield  {author} {\bibinfo {author} {\bibfnamefont {A.}~\bibnamefont {Auff{\`e}ves}}, \bibinfo {author} {\bibfnamefont {D.}~\bibnamefont {Gerace}}, \bibinfo {author} {\bibfnamefont {S.}~\bibnamefont {Portolan}}, \bibinfo {author} {\bibfnamefont {A.}~\bibnamefont {Drezet}},\ and\ \bibinfo {author} {\bibfnamefont {M.~F.}\ \bibnamefont {Santos}},\ }\bibfield  {title} {\bibinfo {title} {{Few Emitters in a Cavity: From Cooperative Emission to Individualization}},\ }\href {https://doi.org/10.1088/1367-2630/13/9/093020} {\bibfield  {journal} {\bibinfo  {journal} {New J. Phys.}\ }\textbf {\bibinfo {volume} {13}},\ \bibinfo {pages} {093020} (\bibinfo {year} {2011})}\BibitemShut {NoStop}%
\bibitem [{\citenamefont {{del Valle}}\ \emph {et~al.}(2012)\citenamefont {{del Valle}}, \citenamefont {{Gonzalez-Tudela}}, \citenamefont {Laussy}, \citenamefont {Tejedor},\ and\ \citenamefont {Hartmann}}]{delValle2012}%
  \BibitemOpen
  \bibfield  {author} {\bibinfo {author} {\bibfnamefont {E.}~\bibnamefont {{del Valle}}}, \bibinfo {author} {\bibfnamefont {A.}~\bibnamefont {{Gonzalez-Tudela}}}, \bibinfo {author} {\bibfnamefont {F.~P.}\ \bibnamefont {Laussy}}, \bibinfo {author} {\bibfnamefont {C.}~\bibnamefont {Tejedor}},\ and\ \bibinfo {author} {\bibfnamefont {M.~J.}\ \bibnamefont {Hartmann}},\ }\bibfield  {title} {\bibinfo {title} {{Theory of {{Frequency-Filtered}} and {{Time-Resolved}} ${{N}}$-{{Photon Correlations}}}},\ }\href {https://doi.org/10.1103/PhysRevLett.109.183601} {\bibfield  {journal} {\bibinfo  {journal} {Phys. Rev. Lett.}\ }\textbf {\bibinfo {volume} {109}},\ \bibinfo {pages} {183601} (\bibinfo {year} {2012})}\BibitemShut {NoStop}%
\bibitem [{\citenamefont {{S{\'a}ez-Bl{\'a}zquez}}\ \emph {et~al.}(2017)\citenamefont {{S{\'a}ez-Bl{\'a}zquez}}, \citenamefont {Feist}, \citenamefont {{Fern{\'a}ndez-Dom{\'i}nguez}},\ and\ \citenamefont {{Garc{\'i}a-Vidal}}}]{Saez-Blazquez2017}%
  \BibitemOpen
  \bibfield  {author} {\bibinfo {author} {\bibfnamefont {R.}~\bibnamefont {{S{\'a}ez-Bl{\'a}zquez}}}, \bibinfo {author} {\bibfnamefont {J.}~\bibnamefont {Feist}}, \bibinfo {author} {\bibfnamefont {A.~I.}\ \bibnamefont {{Fern{\'a}ndez-Dom{\'i}nguez}}},\ and\ \bibinfo {author} {\bibfnamefont {F.~J.}\ \bibnamefont {{Garc{\'i}a-Vidal}}},\ }\bibfield  {title} {\bibinfo {title} {{Enhancing Photon Correlations through Plasmonic Strong Coupling}},\ }\href {https://doi.org/10.1364/OPTICA.4.001363} {\bibfield  {journal} {\bibinfo  {journal} {Optica}\ }\textbf {\bibinfo {volume} {4}},\ \bibinfo {pages} {1363} (\bibinfo {year} {2017})}\BibitemShut {NoStop}%
\bibitem [{\citenamefont {{S{\'a}ez-Bl{\'a}zquez}}\ \emph {et~al.}(2018)\citenamefont {{S{\'a}ez-Bl{\'a}zquez}}, \citenamefont {Feist}, \citenamefont {{Garc{\'i}a-Vidal}},\ and\ \citenamefont {{Fern{\'a}ndez-Dom{\'i}nguez}}}]{Saez-Blazquez2018Photon}%
  \BibitemOpen
  \bibfield  {author} {\bibinfo {author} {\bibfnamefont {R.}~\bibnamefont {{S{\'a}ez-Bl{\'a}zquez}}}, \bibinfo {author} {\bibfnamefont {J.}~\bibnamefont {Feist}}, \bibinfo {author} {\bibfnamefont {F.~J.}\ \bibnamefont {{Garc{\'i}a-Vidal}}},\ and\ \bibinfo {author} {\bibfnamefont {A.~I.}\ \bibnamefont {{Fern{\'a}ndez-Dom{\'i}nguez}}},\ }\bibfield  {title} {\bibinfo {title} {{Photon Statistics in Collective Strong Coupling: {{Nanocavities}} and Microcavities}},\ }\href {https://doi.org/10.1103/PhysRevA.98.013839} {\bibfield  {journal} {\bibinfo  {journal} {Phys. Rev. A}\ }\textbf {\bibinfo {volume} {98}},\ \bibinfo {pages} {013839} (\bibinfo {year} {2018})}\BibitemShut {NoStop}%
\bibitem [{\citenamefont {Peyskens}\ and\ \citenamefont {Englund}(2018)}]{Peyskens2018}%
  \BibitemOpen
  \bibfield  {author} {\bibinfo {author} {\bibfnamefont {F.}~\bibnamefont {Peyskens}}\ and\ \bibinfo {author} {\bibfnamefont {D.}~\bibnamefont {Englund}},\ }\bibfield  {title} {\bibinfo {title} {{Quantum Photonics Model for Nonclassical Light Generation Using Integrated Nanoplasmonic Cavity-Emitter Systems}},\ }\href {https://doi.org/10.1103/PhysRevA.97.063844} {\bibfield  {journal} {\bibinfo  {journal} {Phys. Rev. A}\ }\textbf {\bibinfo {volume} {97}},\ \bibinfo {pages} {063844} (\bibinfo {year} {2018})}\BibitemShut {NoStop}%
\bibitem [{\citenamefont {Rousseaux}\ \emph {et~al.}(2020)\citenamefont {Rousseaux}, \citenamefont {Baranov}, \citenamefont {Antosiewicz}, \citenamefont {Shegai},\ and\ \citenamefont {Johansson}}]{Rousseaux2020}%
  \BibitemOpen
  \bibfield  {author} {\bibinfo {author} {\bibfnamefont {B.}~\bibnamefont {Rousseaux}}, \bibinfo {author} {\bibfnamefont {D.~G.}\ \bibnamefont {Baranov}}, \bibinfo {author} {\bibfnamefont {T.~J.}\ \bibnamefont {Antosiewicz}}, \bibinfo {author} {\bibfnamefont {T.}~\bibnamefont {Shegai}},\ and\ \bibinfo {author} {\bibfnamefont {G.}~\bibnamefont {Johansson}},\ }\bibfield  {title} {\bibinfo {title} {{Strong Coupling as an Interplay of Quantum Emitter Hybridization with Plasmonic Dark and Bright Modes}},\ }\href {https://doi.org/10.1103/PhysRevResearch.2.033056} {\bibfield  {journal} {\bibinfo  {journal} {Phys. Rev. Res.}\ }\textbf {\bibinfo {volume} {2}},\ \bibinfo {pages} {033056} (\bibinfo {year} {2020})}\BibitemShut {NoStop}%
\bibitem [{\citenamefont {Chikkaraddy}\ \emph {et~al.}(2016)\citenamefont {Chikkaraddy}, \citenamefont {{de Nijs}}, \citenamefont {Benz}, \citenamefont {Barrow}, \citenamefont {Scherman}, \citenamefont {Rosta}, \citenamefont {Demetriadou}, \citenamefont {Fox}, \citenamefont {Hess},\ and\ \citenamefont {Baumberg}}]{Chikkaraddy2016}%
  \BibitemOpen
  \bibfield  {author} {\bibinfo {author} {\bibfnamefont {R.}~\bibnamefont {Chikkaraddy}}, \bibinfo {author} {\bibfnamefont {B.}~\bibnamefont {{de Nijs}}}, \bibinfo {author} {\bibfnamefont {F.}~\bibnamefont {Benz}}, \bibinfo {author} {\bibfnamefont {S.~J.}\ \bibnamefont {Barrow}}, \bibinfo {author} {\bibfnamefont {O.~A.}\ \bibnamefont {Scherman}}, \bibinfo {author} {\bibfnamefont {E.}~\bibnamefont {Rosta}}, \bibinfo {author} {\bibfnamefont {A.}~\bibnamefont {Demetriadou}}, \bibinfo {author} {\bibfnamefont {P.}~\bibnamefont {Fox}}, \bibinfo {author} {\bibfnamefont {O.}~\bibnamefont {Hess}},\ and\ \bibinfo {author} {\bibfnamefont {J.~J.}\ \bibnamefont {Baumberg}},\ }\bibfield  {title} {\bibinfo {title} {{Single-Molecule Strong Coupling at Room Temperature in Plasmonic Nanocavities}},\ }\href {https://doi.org/10.1038/nature17974} {\bibfield  {journal} {\bibinfo  {journal} {Nature}\ }\textbf {\bibinfo {volume} {535}},\ \bibinfo {pages} {127} (\bibinfo {year} {2016})}\BibitemShut {NoStop}%
\bibitem [{\citenamefont {Benz}\ \emph {et~al.}(2016)\citenamefont {Benz}, \citenamefont {Schmidt}, \citenamefont {Dreismann}, \citenamefont {Chikkaraddy}, \citenamefont {Zhang}, \citenamefont {Demetriadou}, \citenamefont {Carnegie}, \citenamefont {Ohadi}, \citenamefont {{de Nijs}}, \citenamefont {Esteban}, \citenamefont {Aizpurua},\ and\ \citenamefont {Baumberg}}]{Benz2016}%
  \BibitemOpen
  \bibfield  {author} {\bibinfo {author} {\bibfnamefont {F.}~\bibnamefont {Benz}}, \bibinfo {author} {\bibfnamefont {M.~K.}\ \bibnamefont {Schmidt}}, \bibinfo {author} {\bibfnamefont {A.}~\bibnamefont {Dreismann}}, \bibinfo {author} {\bibfnamefont {R.}~\bibnamefont {Chikkaraddy}}, \bibinfo {author} {\bibfnamefont {Y.}~\bibnamefont {Zhang}}, \bibinfo {author} {\bibfnamefont {A.}~\bibnamefont {Demetriadou}}, \bibinfo {author} {\bibfnamefont {C.}~\bibnamefont {Carnegie}}, \bibinfo {author} {\bibfnamefont {H.}~\bibnamefont {Ohadi}}, \bibinfo {author} {\bibfnamefont {B.}~\bibnamefont {{de Nijs}}}, \bibinfo {author} {\bibfnamefont {R.}~\bibnamefont {Esteban}}, \bibinfo {author} {\bibfnamefont {J.}~\bibnamefont {Aizpurua}},\ and\ \bibinfo {author} {\bibfnamefont {J.~J.}\ \bibnamefont {Baumberg}},\ }\bibfield  {title} {\bibinfo {title} {{Single-Molecule Optomechanics in ``Picocavities''}},\ }\href {https://doi.org/10.1126/science.aah5243} {\bibfield  {journal} {\bibinfo  {journal} {Science}\ }\textbf {\bibinfo {volume} {354}},\ \bibinfo {pages} {726} (\bibinfo {year} {2016})}\BibitemShut {NoStop}%
\bibitem [{\citenamefont {Franke}\ \emph {et~al.}(2019)\citenamefont {Franke}, \citenamefont {Hughes}, \citenamefont {Kamandar~Dezfouli}, \citenamefont {Kristensen}, \citenamefont {Busch}, \citenamefont {Knorr},\ and\ \citenamefont {Richter}}]{Franke2019}%
  \BibitemOpen
  \bibfield  {author} {\bibinfo {author} {\bibfnamefont {S.}~\bibnamefont {Franke}}, \bibinfo {author} {\bibfnamefont {S.}~\bibnamefont {Hughes}}, \bibinfo {author} {\bibfnamefont {M.}~\bibnamefont {Kamandar~Dezfouli}}, \bibinfo {author} {\bibfnamefont {P.~T.}\ \bibnamefont {Kristensen}}, \bibinfo {author} {\bibfnamefont {K.}~\bibnamefont {Busch}}, \bibinfo {author} {\bibfnamefont {A.}~\bibnamefont {Knorr}},\ and\ \bibinfo {author} {\bibfnamefont {M.}~\bibnamefont {Richter}},\ }\bibfield  {title} {\bibinfo {title} {{Quantization of {{Quasinormal Modes}} for {{Open Cavities}} and {{Plasmonic Cavity Quantum Electrodynamics}}}},\ }\href {https://doi.org/10.1103/PhysRevLett.122.213901} {\bibfield  {journal} {\bibinfo  {journal} {Phys. Rev. Lett.}\ }\textbf {\bibinfo {volume} {122}},\ \bibinfo {pages} {213901} (\bibinfo {year} {2019})}\BibitemShut {NoStop}%
\bibitem [{\citenamefont {Heintz}\ \emph {et~al.}(2021)\citenamefont {Heintz}, \citenamefont {Marke{\v s}evi{\'c}}, \citenamefont {Gayet}, \citenamefont {Bonod},\ and\ \citenamefont {Bidault}}]{Heintz2021}%
  \BibitemOpen
  \bibfield  {author} {\bibinfo {author} {\bibfnamefont {J.}~\bibnamefont {Heintz}}, \bibinfo {author} {\bibfnamefont {N.}~\bibnamefont {Marke{\v s}evi{\'c}}}, \bibinfo {author} {\bibfnamefont {E.~Y.}\ \bibnamefont {Gayet}}, \bibinfo {author} {\bibfnamefont {N.}~\bibnamefont {Bonod}},\ and\ \bibinfo {author} {\bibfnamefont {S.}~\bibnamefont {Bidault}},\ }\bibfield  {title} {\bibinfo {title} {{Few-{{Molecule Strong Coupling}} with {{Dimers}} of {{Plasmonic Nanoparticles Assembled}} on {{DNA}}}},\ }\href {https://doi.org/10.1021/acsnano.1c04552} {\bibfield  {journal} {\bibinfo  {journal} {ACS Nano}\ }\textbf {\bibinfo {volume} {15}},\ \bibinfo {pages} {14732} (\bibinfo {year} {2021})}\BibitemShut {NoStop}%
\bibitem [{\citenamefont {{Gonz{\'a}lez-Tudela}}\ \emph {et~al.}(2024)\citenamefont {{Gonz{\'a}lez-Tudela}}, \citenamefont {Reiserer}, \citenamefont {{Garc{\'i}a-Ripoll}},\ and\ \citenamefont {{Garc{\'i}a-Vidal}}}]{Gonzalez-Tudela2024}%
  \BibitemOpen
  \bibfield  {author} {\bibinfo {author} {\bibfnamefont {A.}~\bibnamefont {{Gonz{\'a}lez-Tudela}}}, \bibinfo {author} {\bibfnamefont {A.}~\bibnamefont {Reiserer}}, \bibinfo {author} {\bibfnamefont {J.~J.}\ \bibnamefont {{Garc{\'i}a-Ripoll}}},\ and\ \bibinfo {author} {\bibfnamefont {F.~J.}\ \bibnamefont {{Garc{\'i}a-Vidal}}},\ }\bibfield  {title} {\bibinfo {title} {{Light--Matter Interactions in Quantum Nanophotonic Devices}},\ }\href {https://doi.org/10.1038/s42254-023-00681-1} {\bibfield  {journal} {\bibinfo  {journal} {Nat. Rev. Phys.}\ }\textbf {\bibinfo {volume} {6}},\ \bibinfo {pages} {166} (\bibinfo {year} {2024})}\BibitemShut {NoStop}%
\bibitem [{\citenamefont {Huttner}\ and\ \citenamefont {Barnett}(1992)}]{Huttner1992}%
  \BibitemOpen
  \bibfield  {author} {\bibinfo {author} {\bibfnamefont {B.}~\bibnamefont {Huttner}}\ and\ \bibinfo {author} {\bibfnamefont {S.~M.}\ \bibnamefont {Barnett}},\ }\bibfield  {title} {\bibinfo {title} {{Quantization of the Electromagnetic Field in Dielectrics}},\ }\href {https://doi.org/10.1103/PhysRevA.46.4306} {\bibfield  {journal} {\bibinfo  {journal} {Phys. Rev. A}\ }\textbf {\bibinfo {volume} {46}},\ \bibinfo {pages} {4306} (\bibinfo {year} {1992})}\BibitemShut {NoStop}%
\bibitem [{\citenamefont {Scheel}\ and\ \citenamefont {Buhmann}(2008)}]{Scheel2008}%
  \BibitemOpen
  \bibfield  {author} {\bibinfo {author} {\bibfnamefont {S.}~\bibnamefont {Scheel}}\ and\ \bibinfo {author} {\bibfnamefont {S.~Y.}\ \bibnamefont {Buhmann}},\ }\bibfield  {title} {\bibinfo {title} {{Macroscopic Quantum Electrodynamics - {{Concepts}} and Applications}},\ }\href {https://doi.org/10.2478/v10155-010-0092-x} {\bibfield  {journal} {\bibinfo  {journal} {Acta Phys. Slovaca}\ }\textbf {\bibinfo {volume} {58}},\ \bibinfo {pages} {675} (\bibinfo {year} {2008})}\BibitemShut {NoStop}%
\bibitem [{\citenamefont {Buhmann}(2012{\natexlab{a}})}]{Buhmann2012I}%
  \BibitemOpen
  \bibfield  {author} {\bibinfo {author} {\bibfnamefont {S.~Y.}\ \bibnamefont {Buhmann}},\ }\href {https://doi.org/10.1007/978-3-642-32484-0} {\emph {\bibinfo {title} {{Dispersion {{Forces I}}}}}},\ \bibinfo {series} {Springer {{Tracts}} in {{Modern Physics}}}, Vol.\ \bibinfo {volume} {247}\ (\bibinfo  {publisher} {Springer Berlin Heidelberg},\ \bibinfo {address} {Berlin, Heidelberg},\ \bibinfo {year} {2012})\BibitemShut {NoStop}%
\bibitem [{\citenamefont {Buhmann}(2012{\natexlab{b}})}]{Buhmann2012II}%
  \BibitemOpen
  \bibfield  {author} {\bibinfo {author} {\bibfnamefont {S.~Y.}\ \bibnamefont {Buhmann}},\ }\href {https://doi.org/10.1007/978-3-642-32466-6} {\emph {\bibinfo {title} {{Dispersion {{Forces II}}}}}},\ \bibinfo {series} {Springer {{Tracts}} in {{Modern Physics}}}, Vol.\ \bibinfo {volume} {248}\ (\bibinfo  {publisher} {Springer Berlin Heidelberg},\ \bibinfo {address} {Berlin, Heidelberg},\ \bibinfo {year} {2012})\BibitemShut {NoStop}%
\bibitem [{\citenamefont {Hughes}\ \emph {et~al.}(2018)\citenamefont {Hughes}, \citenamefont {Richter},\ and\ \citenamefont {Knorr}}]{Hughes2018}%
  \BibitemOpen
  \bibfield  {author} {\bibinfo {author} {\bibfnamefont {S.}~\bibnamefont {Hughes}}, \bibinfo {author} {\bibfnamefont {M.}~\bibnamefont {Richter}},\ and\ \bibinfo {author} {\bibfnamefont {A.}~\bibnamefont {Knorr}},\ }\bibfield  {title} {\bibinfo {title} {{Quantized Pseudomodes for Plasmonic Cavity {{QED}}}},\ }\href {https://doi.org/10.1364/OL.43.001834} {\bibfield  {journal} {\bibinfo  {journal} {Opt. Lett.}\ }\textbf {\bibinfo {volume} {43}},\ \bibinfo {pages} {1834} (\bibinfo {year} {2018})}\BibitemShut {NoStop}%
\bibitem [{\citenamefont {Medina}\ \emph {et~al.}(2021)\citenamefont {Medina}, \citenamefont {{Garc{\'i}a-Vidal}}, \citenamefont {{Fern{\'a}ndez-Dom{\'i}nguez}},\ and\ \citenamefont {Feist}}]{Medina2021}%
  \BibitemOpen
  \bibfield  {author} {\bibinfo {author} {\bibfnamefont {I.}~\bibnamefont {Medina}}, \bibinfo {author} {\bibfnamefont {F.~J.}\ \bibnamefont {{Garc{\'i}a-Vidal}}}, \bibinfo {author} {\bibfnamefont {A.~I.}\ \bibnamefont {{Fern{\'a}ndez-Dom{\'i}nguez}}},\ and\ \bibinfo {author} {\bibfnamefont {J.}~\bibnamefont {Feist}},\ }\bibfield  {title} {\bibinfo {title} {{Few-{{Mode Field Quantization}} of {{Arbitrary Electromagnetic Spectral Densities}}}},\ }\href {https://doi.org/10.1103/PhysRevLett.126.093601} {\bibfield  {journal} {\bibinfo  {journal} {Phys. Rev. Lett.}\ }\textbf {\bibinfo {volume} {126}},\ \bibinfo {pages} {093601} (\bibinfo {year} {2021})}\BibitemShut {NoStop}%
\bibitem [{\citenamefont {{S{\'a}nchez-Barquilla}}\ and\ \citenamefont {Feist}(2021)}]{Sanchez-Barquilla2021}%
  \BibitemOpen
  \bibfield  {author} {\bibinfo {author} {\bibfnamefont {M.}~\bibnamefont {{S{\'a}nchez-Barquilla}}}\ and\ \bibinfo {author} {\bibfnamefont {J.}~\bibnamefont {Feist}},\ }\bibfield  {title} {\bibinfo {title} {{Accurate {{Truncations}} of {{Chain Mapping Models}} for {{Open Quantum Systems}}}},\ }\href {https://doi.org/10.3390/nano11082104} {\bibfield  {journal} {\bibinfo  {journal} {Nanomaterials}\ }\textbf {\bibinfo {volume} {11}},\ \bibinfo {pages} {2104} (\bibinfo {year} {2021})}\BibitemShut {NoStop}%
\bibitem [{\citenamefont {{S{\'a}nchez-Barquilla}}\ \emph {et~al.}(2022)\citenamefont {{S{\'a}nchez-Barquilla}}, \citenamefont {{Garc{\'i}a-Vidal}}, \citenamefont {{Fern{\'a}ndez-Dom{\'i}nguez}},\ and\ \citenamefont {Feist}}]{Sanchez-Barquilla2022}%
  \BibitemOpen
  \bibfield  {author} {\bibinfo {author} {\bibfnamefont {M.}~\bibnamefont {{S{\'a}nchez-Barquilla}}}, \bibinfo {author} {\bibfnamefont {F.~J.}\ \bibnamefont {{Garc{\'i}a-Vidal}}}, \bibinfo {author} {\bibfnamefont {A.~I.}\ \bibnamefont {{Fern{\'a}ndez-Dom{\'i}nguez}}},\ and\ \bibinfo {author} {\bibfnamefont {J.}~\bibnamefont {Feist}},\ }\bibfield  {title} {\bibinfo {title} {{Few-Mode Field Quantization for Multiple Emitters}},\ }\href {https://doi.org/10.1515/nanoph-2021-0795} {\bibfield  {journal} {\bibinfo  {journal} {Nanophotonics}\ }\textbf {\bibinfo {volume} {11}},\ \bibinfo {pages} {4363} (\bibinfo {year} {2022})}\BibitemShut {NoStop}%
\bibitem [{\citenamefont {Lednev}\ \emph {et~al.}(2024)\citenamefont {Lednev}, \citenamefont {{Garc{\'i}a-Vidal}},\ and\ \citenamefont {Feist}}]{Lednev2024}%
  \BibitemOpen
  \bibfield  {author} {\bibinfo {author} {\bibfnamefont {M.}~\bibnamefont {Lednev}}, \bibinfo {author} {\bibfnamefont {F.~J.}\ \bibnamefont {{Garc{\'i}a-Vidal}}},\ and\ \bibinfo {author} {\bibfnamefont {J.}~\bibnamefont {Feist}},\ }\bibfield  {title} {\bibinfo {title} {{Lindblad {{Master Equation Capable}} of {{Describing Hybrid Quantum Systems}} in the {{Ultrastrong Coupling Regime}}}},\ }\href {https://doi.org/10.1103/PhysRevLett.132.106902} {\bibfield  {journal} {\bibinfo  {journal} {Phys. Rev. Lett.}\ }\textbf {\bibinfo {volume} {132}},\ \bibinfo {pages} {106902} (\bibinfo {year} {2024})}\BibitemShut {NoStop}%
\bibitem [{\citenamefont {Kristensen}\ \emph {et~al.}(2020)\citenamefont {Kristensen}, \citenamefont {Herrmann}, \citenamefont {Intravaia},\ and\ \citenamefont {Busch}}]{Kristensen2020}%
  \BibitemOpen
  \bibfield  {author} {\bibinfo {author} {\bibfnamefont {P.~T.}\ \bibnamefont {Kristensen}}, \bibinfo {author} {\bibfnamefont {K.}~\bibnamefont {Herrmann}}, \bibinfo {author} {\bibfnamefont {F.}~\bibnamefont {Intravaia}},\ and\ \bibinfo {author} {\bibfnamefont {K.}~\bibnamefont {Busch}},\ }\bibfield  {title} {\bibinfo {title} {{Modeling Electromagnetic Resonators Using Quasinormal Modes}},\ }\href {https://doi.org/10.1364/AOP.377940} {\bibfield  {journal} {\bibinfo  {journal} {Adv. Opt. Photonics}\ }\textbf {\bibinfo {volume} {12}},\ \bibinfo {pages} {612} (\bibinfo {year} {2020})}\BibitemShut {NoStop}%
\bibitem [{\citenamefont {Loudon}(2000)}]{Loudon2000}%
  \BibitemOpen
  \bibfield  {author} {\bibinfo {author} {\bibfnamefont {R.}~\bibnamefont {Loudon}},\ }\href@noop {} {\emph {\bibinfo {title} {{The {{Quantum Theory}} of {{Light}}}}}},\ \bibinfo {edition} {third edition}\ ed.\ (\bibinfo  {publisher} {Oxford University Press},\ \bibinfo {address} {Oxford, New York},\ \bibinfo {year} {2000})\BibitemShut {NoStop}%
\bibitem [{\citenamefont {Grynberg}\ \emph {et~al.}(2010)\citenamefont {Grynberg}, \citenamefont {Aspect}, \citenamefont {Fabre},\ and\ \citenamefont {{Cohen-Tannoudji}}}]{Grynberg2010}%
  \BibitemOpen
  \bibfield  {author} {\bibinfo {author} {\bibfnamefont {G.}~\bibnamefont {Grynberg}}, \bibinfo {author} {\bibfnamefont {A.}~\bibnamefont {Aspect}}, \bibinfo {author} {\bibfnamefont {C.}~\bibnamefont {Fabre}},\ and\ \bibinfo {author} {\bibfnamefont {C.}~\bibnamefont {{Cohen-Tannoudji}}},\ }\href {https://doi.org/10.1017/CBO9780511778261} {\emph {\bibinfo {title} {{Introduction to {{Quantum Optics}}: {{From}} the {{Semi-classical Approach}} to {{Quantized Light}}}}}}\ (\bibinfo  {publisher} {Cambridge University Press},\ \bibinfo {address} {Cambridge},\ \bibinfo {year} {2010})\BibitemShut {NoStop}%
\bibitem [{\citenamefont {Holdaway}\ \emph {et~al.}(2018)\citenamefont {Holdaway}, \citenamefont {Notararigo},\ and\ \citenamefont {{Olaya-Castro}}}]{Holdaway2018}%
  \BibitemOpen
  \bibfield  {author} {\bibinfo {author} {\bibfnamefont {D.~I.~H.}\ \bibnamefont {Holdaway}}, \bibinfo {author} {\bibfnamefont {V.}~\bibnamefont {Notararigo}},\ and\ \bibinfo {author} {\bibfnamefont {A.}~\bibnamefont {{Olaya-Castro}}},\ }\bibfield  {title} {\bibinfo {title} {{Perturbation Approach for Computing Frequency- and Time-Resolved Photon Correlation Functions}},\ }\href {https://doi.org/10.1103/PhysRevA.98.063828} {\bibfield  {journal} {\bibinfo  {journal} {Phys. Rev. A}\ }\textbf {\bibinfo {volume} {98}},\ \bibinfo {pages} {063828} (\bibinfo {year} {2018})}\BibitemShut {NoStop}%
\bibitem [{\citenamefont {Nienhuis}(1993)}]{Nienhuis1993}%
  \BibitemOpen
  \bibfield  {author} {\bibinfo {author} {\bibfnamefont {G.}~\bibnamefont {Nienhuis}},\ }\bibfield  {title} {\bibinfo {title} {{Spectral Correlations in Resonance Fluorescence}},\ }\href {https://doi.org/10.1103/PhysRevA.47.510} {\bibfield  {journal} {\bibinfo  {journal} {Phys. Rev. A}\ }\textbf {\bibinfo {volume} {47}},\ \bibinfo {pages} {510} (\bibinfo {year} {1993})}\BibitemShut {NoStop}%
\bibitem [{\citenamefont {Joosten}\ and\ \citenamefont {Nienhuis}(2000)}]{Joosten2000}%
  \BibitemOpen
  \bibfield  {author} {\bibinfo {author} {\bibfnamefont {K.}~\bibnamefont {Joosten}}\ and\ \bibinfo {author} {\bibfnamefont {G.}~\bibnamefont {Nienhuis}},\ }\bibfield  {title} {\bibinfo {title} {{Influence of Spectral Filtering on the Quantum Nature of Light}},\ }\href {https://doi.org/10.1088/1464-4266/2/2/317} {\bibfield  {journal} {\bibinfo  {journal} {J. Opt. B}\ }\textbf {\bibinfo {volume} {2}},\ \bibinfo {pages} {158} (\bibinfo {year} {2000})}\BibitemShut {NoStop}%
\bibitem [{\citenamefont {Eberly}\ and\ \citenamefont {W{\'o}dkiewicz}(1977)}]{Eberly1977}%
  \BibitemOpen
  \bibfield  {author} {\bibinfo {author} {\bibfnamefont {J.~H.}\ \bibnamefont {Eberly}}\ and\ \bibinfo {author} {\bibfnamefont {K.}~\bibnamefont {W{\'o}dkiewicz}},\ }\bibfield  {title} {\bibinfo {title} {{The Time-Dependent Physical Spectrum of Light}},\ }\href {https://doi.org/10.1364/JOSA.67.001252} {\bibfield  {journal} {\bibinfo  {journal} {J. Opt. Soc. Am.}\ }\textbf {\bibinfo {volume} {67}},\ \bibinfo {pages} {1252} (\bibinfo {year} {1977})}\BibitemShut {NoStop}%
\bibitem [{\citenamefont {Knoll}\ \emph {et~al.}(1984)\citenamefont {Knoll}, \citenamefont {Weber},\ and\ \citenamefont {Schafer}}]{Knoll1984}%
  \BibitemOpen
  \bibfield  {author} {\bibinfo {author} {\bibfnamefont {L.}~\bibnamefont {Knoll}}, \bibinfo {author} {\bibfnamefont {G.}~\bibnamefont {Weber}},\ and\ \bibinfo {author} {\bibfnamefont {T.}~\bibnamefont {Schafer}},\ }\bibfield  {title} {\bibinfo {title} {{Theory of Time-Resolved Correlation Spectroscopy and Its Application to Resonance Fluorescence Radiation}},\ }\href {https://doi.org/10.1088/0022-3700/17/24/020} {\bibfield  {journal} {\bibinfo  {journal} {J. Phys. B}\ }\textbf {\bibinfo {volume} {17}},\ \bibinfo {pages} {4861} (\bibinfo {year} {1984})}\BibitemShut {NoStop}%
\bibitem [{\citenamefont {Arnoldus}\ and\ \citenamefont {Nienhuis}(1984)}]{Arnoldus1984}%
  \BibitemOpen
  \bibfield  {author} {\bibinfo {author} {\bibfnamefont {H.~F.}\ \bibnamefont {Arnoldus}}\ and\ \bibinfo {author} {\bibfnamefont {G.}~\bibnamefont {Nienhuis}},\ }\bibfield  {title} {\bibinfo {title} {{Photon Correlations between the Lines in the Spectrum of Resonance Fluorescence}},\ }\href {https://doi.org/10.1088/0022-3700/17/6/011} {\bibfield  {journal} {\bibinfo  {journal} {J. Phys. B}\ }\textbf {\bibinfo {volume} {17}},\ \bibinfo {pages} {963} (\bibinfo {year} {1984})}\BibitemShut {NoStop}%
\bibitem [{\citenamefont {Breuer}\ and\ \citenamefont {Petruccione}(2007)}]{Breuer2007}%
  \BibitemOpen
  \bibfield  {author} {\bibinfo {author} {\bibfnamefont {H.-P.}\ \bibnamefont {Breuer}}\ and\ \bibinfo {author} {\bibfnamefont {F.}~\bibnamefont {Petruccione}},\ }\href {https://doi.org/10.1093/acprof:oso/9780199213900.001.0001} {\emph {\bibinfo {title} {{The {{Theory}} of {{Open Quantum Systems}}}}}}\ (\bibinfo  {publisher} {Oxford University Press},\ \bibinfo {address} {Oxford},\ \bibinfo {year} {2007})\BibitemShut {NoStop}%
\bibitem [{\citenamefont {Tamascelli}\ \emph {et~al.}(2018)\citenamefont {Tamascelli}, \citenamefont {Smirne}, \citenamefont {Huelga},\ and\ \citenamefont {Plenio}}]{Tamascelli2018}%
  \BibitemOpen
  \bibfield  {author} {\bibinfo {author} {\bibfnamefont {D.}~\bibnamefont {Tamascelli}}, \bibinfo {author} {\bibfnamefont {A.}~\bibnamefont {Smirne}}, \bibinfo {author} {\bibfnamefont {S.~F.}\ \bibnamefont {Huelga}},\ and\ \bibinfo {author} {\bibfnamefont {M.~B.}\ \bibnamefont {Plenio}},\ }\bibfield  {title} {\bibinfo {title} {{Nonperturbative {{Treatment}} of Non-{{Markovian Dynamics}} of {{Open Quantum Systems}}}},\ }\href {https://doi.org/10.1103/PhysRevLett.120.030402} {\bibfield  {journal} {\bibinfo  {journal} {Phys. Rev. Lett.}\ }\textbf {\bibinfo {volume} {120}},\ \bibinfo {pages} {030402} (\bibinfo {year} {2018})}\BibitemShut {NoStop}%
\bibitem [{\citenamefont {Menczel}\ \emph {et~al.}(2024)\citenamefont {Menczel}, \citenamefont {Funo}, \citenamefont {Cirio}, \citenamefont {Lambert},\ and\ \citenamefont {Nori}}]{Menczel2024}%
  \BibitemOpen
  \bibfield  {author} {\bibinfo {author} {\bibfnamefont {P.}~\bibnamefont {Menczel}}, \bibinfo {author} {\bibfnamefont {K.}~\bibnamefont {Funo}}, \bibinfo {author} {\bibfnamefont {M.}~\bibnamefont {Cirio}}, \bibinfo {author} {\bibfnamefont {N.}~\bibnamefont {Lambert}},\ and\ \bibinfo {author} {\bibfnamefont {F.}~\bibnamefont {Nori}},\ }\bibfield  {title} {\bibinfo {title} {{Non-{{Hermitian}} Pseudomodes for Strongly Coupled Open Quantum Systems: {{Unravelings}}, Correlations, and Thermodynamics}},\ }\href {https://doi.org/10.1103/PhysRevResearch.6.033237} {\bibfield  {journal} {\bibinfo  {journal} {Phys. Rev. Res.}\ }\textbf {\bibinfo {volume} {6}},\ \bibinfo {pages} {033237} (\bibinfo {year} {2024})}\BibitemShut {NoStop}%
\bibitem [{\citenamefont {Chew}(1999)}]{Chew1999Waves}%
  \BibitemOpen
  \bibfield  {author} {\bibinfo {author} {\bibfnamefont {W.~C.}\ \bibnamefont {Chew}},\ }\href@noop {} {\emph {\bibinfo {title} {{Waves and {{Fields}} in {{Inhomogenous Media}}}}}}\ (\bibinfo  {publisher} {Wiley-IEEE Press},\ \bibinfo {address} {New York},\ \bibinfo {year} {1999})\BibitemShut {NoStop}%
\bibitem [{\citenamefont {Delga}\ \emph {et~al.}(2014)\citenamefont {Delga}, \citenamefont {Feist}, \citenamefont {{Bravo-Abad}},\ and\ \citenamefont {{Garcia-Vidal}}}]{Delga2014}%
  \BibitemOpen
  \bibfield  {author} {\bibinfo {author} {\bibfnamefont {A.}~\bibnamefont {Delga}}, \bibinfo {author} {\bibfnamefont {J.}~\bibnamefont {Feist}}, \bibinfo {author} {\bibfnamefont {J.}~\bibnamefont {{Bravo-Abad}}},\ and\ \bibinfo {author} {\bibfnamefont {F.~J.}\ \bibnamefont {{Garcia-Vidal}}},\ }\bibfield  {title} {\bibinfo {title} {{Quantum {{Emitters Near}} a {{Metal Nanoparticle}}: {{Strong Coupling}} and {{Quenching}}}},\ }\href {https://doi.org/10.1103/PhysRevLett.112.253601} {\bibfield  {journal} {\bibinfo  {journal} {Phys. Rev. Lett.}\ }\textbf {\bibinfo {volume} {112}},\ \bibinfo {pages} {253601} (\bibinfo {year} {2014})}\BibitemShut {NoStop}%
\bibitem [{\citenamefont {S{\'a}nchez~Mart{\'i}nez}\ \emph {et~al.}(2024)\citenamefont {S{\'a}nchez~Mart{\'i}nez}, \citenamefont {Feist},\ and\ \citenamefont {{Garc{\'i}a-Vidal}}}]{SanchezMartinez2024Mixed}%
  \BibitemOpen
  \bibfield  {author} {\bibinfo {author} {\bibfnamefont {C.~J.}\ \bibnamefont {S{\'a}nchez~Mart{\'i}nez}}, \bibinfo {author} {\bibfnamefont {J.}~\bibnamefont {Feist}},\ and\ \bibinfo {author} {\bibfnamefont {F.~J.}\ \bibnamefont {{Garc{\'i}a-Vidal}}},\ }\bibfield  {title} {\bibinfo {title} {{A Mixed Perturbative-Nonperturbative Treatment for Strong Light-Matter Interactions}},\ }\href {https://doi.org/10.1515/nanoph-2023-0863} {\bibfield  {journal} {\bibinfo  {journal} {Nanophotonics}\ }\textbf {\bibinfo {volume} {13}},\ \bibinfo {pages} {2669} (\bibinfo {year} {2024})}\BibitemShut {NoStop}%
\bibitem [{\citenamefont {Chuang}\ and\ \citenamefont {Hsu}(2024)}]{Chuang2024}%
  \BibitemOpen
  \bibfield  {author} {\bibinfo {author} {\bibfnamefont {Y.-T.}\ \bibnamefont {Chuang}}\ and\ \bibinfo {author} {\bibfnamefont {L.-Y.}\ \bibnamefont {Hsu}},\ }\bibfield  {title} {\bibinfo {title} {{Microscopic Theory of Exciton--Polariton Model Involving Multiple Molecules: {{Macroscopic}} Quantum Electrodynamics Formulation and Essence of Direct Intermolecular Interactions}},\ }\href {https://doi.org/10.1063/5.0192704} {\bibfield  {journal} {\bibinfo  {journal} {J. Chem. Phys.}\ }\textbf {\bibinfo {volume} {160}},\ \bibinfo {pages} {114105} (\bibinfo {year} {2024})}\BibitemShut {NoStop}%
\bibitem [{\citenamefont {Dung}\ \emph {et~al.}(2002)\citenamefont {Dung}, \citenamefont {Kn{\"o}ll},\ and\ \citenamefont {Welsch}}]{Dung2002Resonant}%
  \BibitemOpen
  \bibfield  {author} {\bibinfo {author} {\bibfnamefont {H.~T.}\ \bibnamefont {Dung}}, \bibinfo {author} {\bibfnamefont {L.}~\bibnamefont {Kn{\"o}ll}},\ and\ \bibinfo {author} {\bibfnamefont {D.-G.}\ \bibnamefont {Welsch}},\ }\bibfield  {title} {\bibinfo {title} {{Resonant Dipole-Dipole Interaction in the Presence of Dispersing and Absorbing Surroundings}},\ }\href {https://doi.org/10.1103/PhysRevA.66.063810} {\bibfield  {journal} {\bibinfo  {journal} {Phys. Rev. A}\ }\textbf {\bibinfo {volume} {66}},\ \bibinfo {pages} {063810} (\bibinfo {year} {2002})}\BibitemShut {NoStop}%
\bibitem [{\citenamefont {{Asenjo-Garcia}}\ \emph {et~al.}(2017)\citenamefont {{Asenjo-Garcia}}, \citenamefont {Hood}, \citenamefont {Chang},\ and\ \citenamefont {Kimble}}]{Asenjo-Garcia2017}%
  \BibitemOpen
  \bibfield  {author} {\bibinfo {author} {\bibfnamefont {A.}~\bibnamefont {{Asenjo-Garcia}}}, \bibinfo {author} {\bibfnamefont {J.~D.}\ \bibnamefont {Hood}}, \bibinfo {author} {\bibfnamefont {D.~E.}\ \bibnamefont {Chang}},\ and\ \bibinfo {author} {\bibfnamefont {H.~J.}\ \bibnamefont {Kimble}},\ }\bibfield  {title} {\bibinfo {title} {{Atom-Light Interactions in Quasi-One-Dimensional Nanostructures: {{A Green}}'s-Function Perspective}},\ }\href {https://doi.org/10.1103/PhysRevA.95.033818} {\bibfield  {journal} {\bibinfo  {journal} {Phys. Rev. A}\ }\textbf {\bibinfo {volume} {95}},\ \bibinfo {pages} {033818} (\bibinfo {year} {2017})}\BibitemShut {NoStop}%
\bibitem [{\citenamefont {Kamandar~Dezfouli}\ and\ \citenamefont {Hughes}(2017)}]{KamandarDezfouli2017}%
  \BibitemOpen
  \bibfield  {author} {\bibinfo {author} {\bibfnamefont {M.}~\bibnamefont {Kamandar~Dezfouli}}\ and\ \bibinfo {author} {\bibfnamefont {S.}~\bibnamefont {Hughes}},\ }\bibfield  {title} {\bibinfo {title} {{Quantum {{Optics Model}} of {{Surface-Enhanced Raman Spectroscopy}} for {{Arbitrarily Shaped Plasmonic Resonators}}}},\ }\href {https://doi.org/10.1021/acsphotonics.7b00157} {\bibfield  {journal} {\bibinfo  {journal} {ACS Photonics}\ }\textbf {\bibinfo {volume} {4}},\ \bibinfo {pages} {1245} (\bibinfo {year} {2017})}\BibitemShut {NoStop}%
\bibitem [{\citenamefont {Chang}\ \emph {et~al.}(2018)\citenamefont {Chang}, \citenamefont {Douglas}, \citenamefont {{Gonz{\'a}lez-Tudela}}, \citenamefont {Hung},\ and\ \citenamefont {Kimble}}]{Chang2018Colloquium}%
  \BibitemOpen
  \bibfield  {author} {\bibinfo {author} {\bibfnamefont {D.~E.}\ \bibnamefont {Chang}}, \bibinfo {author} {\bibfnamefont {J.~S.}\ \bibnamefont {Douglas}}, \bibinfo {author} {\bibfnamefont {A.}~\bibnamefont {{Gonz{\'a}lez-Tudela}}}, \bibinfo {author} {\bibfnamefont {C.-L.}\ \bibnamefont {Hung}},\ and\ \bibinfo {author} {\bibfnamefont {H.~J.}\ \bibnamefont {Kimble}},\ }\bibfield  {title} {\bibinfo {title} {{Colloquium: {{Quantum}} Matter Built from Nanoscopic Lattices of Atoms and Photons}},\ }\href {https://doi.org/10.1103/RevModPhys.90.031002} {\bibfield  {journal} {\bibinfo  {journal} {Rev. Mod. Phys.}\ }\textbf {\bibinfo {volume} {90}},\ \bibinfo {pages} {031002} (\bibinfo {year} {2018})}\BibitemShut {NoStop}%
\bibitem [{\citenamefont {Zhang}\ \emph {et~al.}(2021)\citenamefont {Zhang}, \citenamefont {Esteban}, \citenamefont {Boto}, \citenamefont {Urbieta}, \citenamefont {Arrieta}, \citenamefont {Shan}, \citenamefont {Li}, \citenamefont {Baumberg},\ and\ \citenamefont {Aizpurua}}]{Zhang2021Addressing}%
  \BibitemOpen
  \bibfield  {author} {\bibinfo {author} {\bibfnamefont {Y.}~\bibnamefont {Zhang}}, \bibinfo {author} {\bibfnamefont {R.}~\bibnamefont {Esteban}}, \bibinfo {author} {\bibfnamefont {R.~A.}\ \bibnamefont {Boto}}, \bibinfo {author} {\bibfnamefont {M.}~\bibnamefont {Urbieta}}, \bibinfo {author} {\bibfnamefont {X.}~\bibnamefont {Arrieta}}, \bibinfo {author} {\bibfnamefont {C.}~\bibnamefont {Shan}}, \bibinfo {author} {\bibfnamefont {S.}~\bibnamefont {Li}}, \bibinfo {author} {\bibfnamefont {J.~J.}\ \bibnamefont {Baumberg}},\ and\ \bibinfo {author} {\bibfnamefont {J.}~\bibnamefont {Aizpurua}},\ }\bibfield  {title} {\bibinfo {title} {{Addressing Molecular Optomechanical Effects in Nanocavity-Enhanced {{Raman}} Scattering beyond the Single Plasmonic Mode}},\ }\href {https://doi.org/10.1039/D0NR06649D} {\bibfield  {journal} {\bibinfo  {journal} {Nanoscale}\ }\textbf {\bibinfo {volume} {13}},\ \bibinfo {pages} {1938} (\bibinfo {year} {2021})}\BibitemShut {NoStop}%
\bibitem [{\citenamefont {Esteban}\ \emph {et~al.}(2022)\citenamefont {Esteban}, \citenamefont {Baumberg},\ and\ \citenamefont {Aizpurua}}]{Esteban2022}%
  \BibitemOpen
  \bibfield  {author} {\bibinfo {author} {\bibfnamefont {R.}~\bibnamefont {Esteban}}, \bibinfo {author} {\bibfnamefont {J.~J.}\ \bibnamefont {Baumberg}},\ and\ \bibinfo {author} {\bibfnamefont {J.}~\bibnamefont {Aizpurua}},\ }\bibfield  {title} {\bibinfo {title} {{Molecular {{Optomechanics Approach}} to {{Surface-Enhanced Raman Scattering}}}},\ }\href {https://doi.org/10.1021/acs.accounts.1c00759} {\bibfield  {journal} {\bibinfo  {journal} {Acc. Chem. Res.}\ }\textbf {\bibinfo {volume} {55}},\ \bibinfo {pages} {1889} (\bibinfo {year} {2022})}\BibitemShut {NoStop}%
\bibitem [{\citenamefont {Sheremet}\ \emph {et~al.}(2023)\citenamefont {Sheremet}, \citenamefont {Petrov}, \citenamefont {Iorsh}, \citenamefont {Poshakinskiy},\ and\ \citenamefont {Poddubny}}]{Sheremet2023}%
  \BibitemOpen
  \bibfield  {author} {\bibinfo {author} {\bibfnamefont {A.~S.}\ \bibnamefont {Sheremet}}, \bibinfo {author} {\bibfnamefont {M.~I.}\ \bibnamefont {Petrov}}, \bibinfo {author} {\bibfnamefont {I.~V.}\ \bibnamefont {Iorsh}}, \bibinfo {author} {\bibfnamefont {A.~V.}\ \bibnamefont {Poshakinskiy}},\ and\ \bibinfo {author} {\bibfnamefont {A.~N.}\ \bibnamefont {Poddubny}},\ }\bibfield  {title} {\bibinfo {title} {{Waveguide Quantum Electrodynamics: {{Collective}} Radiance and Photon-Photon Correlations}},\ }\href {https://doi.org/10.1103/RevModPhys.95.015002} {\bibfield  {journal} {\bibinfo  {journal} {Rev. Mod. Phys.}\ }\textbf {\bibinfo {volume} {95}},\ \bibinfo {pages} {015002} (\bibinfo {year} {2023})}\BibitemShut {NoStop}%
\bibitem [{\citenamefont {{Dur{\'a}-Azor{\'i}n}}\ \emph {et~al.}(2024)\citenamefont {{Dur{\'a}-Azor{\'i}n}}, \citenamefont {Manjavacas},\ and\ \citenamefont {{Fern{\'a}ndez-Dom{\'i}nguez}}}]{Dura-Azorin2024}%
  \BibitemOpen
  \bibfield  {author} {\bibinfo {author} {\bibfnamefont {B.}~\bibnamefont {{Dur{\'a}-Azor{\'i}n}}}, \bibinfo {author} {\bibfnamefont {A.}~\bibnamefont {Manjavacas}},\ and\ \bibinfo {author} {\bibfnamefont {A.~I.}\ \bibnamefont {{Fern{\'a}ndez-Dom{\'i}nguez}}},\ }\bibfield  {title} {\bibinfo {title} {{Geometric {{Antibunching}} and {{Directional Shaping}} of {{Photon Anticorrelations}}}},\ }\Eprint {https://arxiv.org/abs/2410.17911} {arXiv:2410.17911}  (\bibinfo {year} {2024})\BibitemShut {NoStop}%
\bibitem [{\citenamefont {{Gonzalez-Tudela}}\ \emph {et~al.}(2013)\citenamefont {{Gonzalez-Tudela}}, \citenamefont {Laussy}, \citenamefont {Tejedor}, \citenamefont {Hartmann},\ and\ \citenamefont {del Valle}}]{Gonzalez-Tudela2013Two-photon}%
  \BibitemOpen
  \bibfield  {author} {\bibinfo {author} {\bibfnamefont {A.}~\bibnamefont {{Gonzalez-Tudela}}}, \bibinfo {author} {\bibfnamefont {F.~P.}\ \bibnamefont {Laussy}}, \bibinfo {author} {\bibfnamefont {C.}~\bibnamefont {Tejedor}}, \bibinfo {author} {\bibfnamefont {M.~J.}\ \bibnamefont {Hartmann}},\ and\ \bibinfo {author} {\bibfnamefont {E.}~\bibnamefont {del Valle}},\ }\bibfield  {title} {\bibinfo {title} {{Two-Photon Spectra of Quantum Emitters}},\ }\href {https://doi.org/10.1088/1367-2630/15/3/033036} {\bibfield  {journal} {\bibinfo  {journal} {New J. Phys.}\ }\textbf {\bibinfo {volume} {15}},\ \bibinfo {pages} {033036} (\bibinfo {year} {2013})}\BibitemShut {NoStop}%
\bibitem [{\citenamefont {S{\'a}nchez~Mu{\~n}oz}\ \emph {et~al.}(2014)\citenamefont {S{\'a}nchez~Mu{\~n}oz}, \citenamefont {{del Valle}}, \citenamefont {Tejedor},\ and\ \citenamefont {Laussy}}]{SanchezMunoz2014Violation}%
  \BibitemOpen
  \bibfield  {author} {\bibinfo {author} {\bibfnamefont {C.}~\bibnamefont {S{\'a}nchez~Mu{\~n}oz}}, \bibinfo {author} {\bibfnamefont {E.}~\bibnamefont {{del Valle}}}, \bibinfo {author} {\bibfnamefont {C.}~\bibnamefont {Tejedor}},\ and\ \bibinfo {author} {\bibfnamefont {F.~P.}\ \bibnamefont {Laussy}},\ }\bibfield  {title} {\bibinfo {title} {{Violation of Classical Inequalities by Photon Frequency Filtering}},\ }\href {https://doi.org/10.1103/PhysRevA.90.052111} {\bibfield  {journal} {\bibinfo  {journal} {Phys. Rev. A}\ }\textbf {\bibinfo {volume} {90}},\ \bibinfo {pages} {052111} (\bibinfo {year} {2014})}\BibitemShut {NoStop}%
\bibitem [{\citenamefont {S{\'a}nchez~Mu{\~n}oz}\ \emph {et~al.}(2018)\citenamefont {S{\'a}nchez~Mu{\~n}oz}, \citenamefont {Laussy}, \citenamefont {del Valle}, \citenamefont {Tejedor},\ and\ \citenamefont {{Gonz{\'a}lez-Tudela}}}]{SanchezMunoz2018Filtering}%
  \BibitemOpen
  \bibfield  {author} {\bibinfo {author} {\bibfnamefont {C.}~\bibnamefont {S{\'a}nchez~Mu{\~n}oz}}, \bibinfo {author} {\bibfnamefont {F.~P.}\ \bibnamefont {Laussy}}, \bibinfo {author} {\bibfnamefont {E.}~\bibnamefont {del Valle}}, \bibinfo {author} {\bibfnamefont {C.}~\bibnamefont {Tejedor}},\ and\ \bibinfo {author} {\bibfnamefont {A.}~\bibnamefont {{Gonz{\'a}lez-Tudela}}},\ }\bibfield  {title} {\bibinfo {title} {{Filtering Multiphoton Emission from State-of-the-Art Cavity Quantum Electrodynamics}},\ }\href {https://doi.org/10.1364/OPTICA.5.000014} {\bibfield  {journal} {\bibinfo  {journal} {Optica}\ }\textbf {\bibinfo {volume} {5}},\ \bibinfo {pages} {14} (\bibinfo {year} {2018})}\BibitemShut {NoStop}%
\bibitem [{\citenamefont {Schmidt}\ \emph {et~al.}(2021)\citenamefont {Schmidt}, \citenamefont {Esteban}, \citenamefont {Giedke}, \citenamefont {Aizpurua},\ and\ \citenamefont {{Gonz{\'a}lez-Tudela}}}]{Schmidt2021}%
  \BibitemOpen
  \bibfield  {author} {\bibinfo {author} {\bibfnamefont {M.~K.}\ \bibnamefont {Schmidt}}, \bibinfo {author} {\bibfnamefont {R.}~\bibnamefont {Esteban}}, \bibinfo {author} {\bibfnamefont {G.}~\bibnamefont {Giedke}}, \bibinfo {author} {\bibfnamefont {J.}~\bibnamefont {Aizpurua}},\ and\ \bibinfo {author} {\bibfnamefont {A.}~\bibnamefont {{Gonz{\'a}lez-Tudela}}},\ }\bibfield  {title} {\bibinfo {title} {{Frequency-Resolved Photon Correlations in Cavity Optomechanics}},\ }\href {https://doi.org/10.1088/2058-9565/abe569} {\bibfield  {journal} {\bibinfo  {journal} {Quantum Sci. Technol.}\ }\textbf {\bibinfo {volume} {6}},\ \bibinfo {pages} {034005} (\bibinfo {year} {2021})}\BibitemShut {NoStop}%
\bibitem [{\citenamefont {{Mart{\'i}nez-Garc{\'i}a}}\ and\ \citenamefont {{Mart{\'i}n-Cano}}(2024)}]{Martinez-Garcia2024}%
  \BibitemOpen
  \bibfield  {author} {\bibinfo {author} {\bibfnamefont {M.~{\'A}.}\ \bibnamefont {{Mart{\'i}nez-Garc{\'i}a}}}\ and\ \bibinfo {author} {\bibfnamefont {D.}~\bibnamefont {{Mart{\'i}n-Cano}}},\ }\bibfield  {title} {\bibinfo {title} {{Coherent {{Electron-Vibron Interactions}} in {{Surface-Enhanced Raman Scattering}} ({{SERS}})}},\ }\href {https://doi.org/10.1103/PhysRevLett.132.093601} {\bibfield  {journal} {\bibinfo  {journal} {Phys. Rev. Lett.}\ }\textbf {\bibinfo {volume} {132}},\ \bibinfo {pages} {093601} (\bibinfo {year} {2024})}\BibitemShut {NoStop}%
\bibitem [{\citenamefont {{Juan-Delgado}}\ \emph {et~al.}(2024)\citenamefont {{Juan-Delgado}}, \citenamefont {Esteban}, \citenamefont {Nodar}, \citenamefont {Trebbia}, \citenamefont {Lounis},\ and\ \citenamefont {Aizpurua}}]{Juan-Delgado2024}%
  \BibitemOpen
  \bibfield  {author} {\bibinfo {author} {\bibfnamefont {A.}~\bibnamefont {{Juan-Delgado}}}, \bibinfo {author} {\bibfnamefont {R.}~\bibnamefont {Esteban}}, \bibinfo {author} {\bibfnamefont {{\'A}.}~\bibnamefont {Nodar}}, \bibinfo {author} {\bibfnamefont {J.-B.}\ \bibnamefont {Trebbia}}, \bibinfo {author} {\bibfnamefont {B.}~\bibnamefont {Lounis}},\ and\ \bibinfo {author} {\bibfnamefont {J.}~\bibnamefont {Aizpurua}},\ }\bibfield  {title} {\bibinfo {title} {{Tailoring the Statistics of Light Emitted from Two Interacting Quantum Emitters}},\ }\href {https://doi.org/10.1103/PhysRevResearch.6.023207} {\bibfield  {journal} {\bibinfo  {journal} {Phys. Rev. Res.}\ }\textbf {\bibinfo {volume} {6}},\ \bibinfo {pages} {023207} (\bibinfo {year} {2024})}\BibitemShut {NoStop}%
\bibitem [{\citenamefont {Mollow}(1969)}]{Mollow1969}%
  \BibitemOpen
  \bibfield  {author} {\bibinfo {author} {\bibfnamefont {B.~R.}\ \bibnamefont {Mollow}},\ }\bibfield  {title} {\bibinfo {title} {{Power {{Spectrum}} of {{Light Scattered}} by {{Two-Level Systems}}}},\ }\href {https://doi.org/10.1103/PhysRev.188.1969} {\bibfield  {journal} {\bibinfo  {journal} {Phys. Rev.}\ }\textbf {\bibinfo {volume} {188}},\ \bibinfo {pages} {1969} (\bibinfo {year} {1969})}\BibitemShut {NoStop}%
\bibitem [{\citenamefont {Carmichael}\ and\ \citenamefont {Walls}(1976)}]{Carmichael1976}%
  \BibitemOpen
  \bibfield  {author} {\bibinfo {author} {\bibfnamefont {H.~J.}\ \bibnamefont {Carmichael}}\ and\ \bibinfo {author} {\bibfnamefont {D.~F.}\ \bibnamefont {Walls}},\ }\bibfield  {title} {\bibinfo {title} {{Proposal for the Measurement of the Resonant {{Stark}} Effect by Photon Correlation Techniques}},\ }\href {https://doi.org/10.1088/0022-3700/9/4/001} {\bibfield  {journal} {\bibinfo  {journal} {J. Phys. B}\ }\textbf {\bibinfo {volume} {9}},\ \bibinfo {pages} {L43} (\bibinfo {year} {1976})}\BibitemShut {NoStop}%
\bibitem [{\citenamefont {Kimble}(1977)}]{Kimble1977}%
  \BibitemOpen
  \bibfield  {author} {\bibinfo {author} {\bibfnamefont {H.~J.}\ \bibnamefont {Kimble}},\ }\bibfield  {title} {\bibinfo {title} {{Photon {{Antibunching}} in {{Resonance Fluorescence}}}},\ }\href {https://doi.org/10.1103/PhysRevLett.39.691} {\bibfield  {journal} {\bibinfo  {journal} {Phys. Rev. Lett.}\ }\textbf {\bibinfo {volume} {39}},\ \bibinfo {pages} {691} (\bibinfo {year} {1977})}\BibitemShut {NoStop}%
\bibitem [{\citenamefont {Hanschke}\ \emph {et~al.}(2020)\citenamefont {Hanschke}, \citenamefont {Schweickert}, \citenamefont {Carre{\~n}o}, \citenamefont {Sch{\"o}ll}, \citenamefont {Zeuner}, \citenamefont {Lettner}, \citenamefont {Casalengua}, \citenamefont {Reindl}, \citenamefont {{da Silva}}, \citenamefont {Trotta}, \citenamefont {Finley}, \citenamefont {Rastelli}, \citenamefont {{del Valle}}, \citenamefont {Laussy}, \citenamefont {Zwiller}, \citenamefont {M{\"u}ller},\ and\ \citenamefont {J{\"o}ns}}]{Hanschke2020}%
  \BibitemOpen
  \bibfield  {author} {\bibinfo {author} {\bibfnamefont {L.}~\bibnamefont {Hanschke}}, \bibinfo {author} {\bibfnamefont {L.}~\bibnamefont {Schweickert}}, \bibinfo {author} {\bibfnamefont {J.~C.~L.}\ \bibnamefont {Carre{\~n}o}}, \bibinfo {author} {\bibfnamefont {E.}~\bibnamefont {Sch{\"o}ll}}, \bibinfo {author} {\bibfnamefont {K.~D.}\ \bibnamefont {Zeuner}}, \bibinfo {author} {\bibfnamefont {T.}~\bibnamefont {Lettner}}, \bibinfo {author} {\bibfnamefont {E.~Z.}\ \bibnamefont {Casalengua}}, \bibinfo {author} {\bibfnamefont {M.}~\bibnamefont {Reindl}}, \bibinfo {author} {\bibfnamefont {S.~F.~C.}\ \bibnamefont {{da Silva}}}, \bibinfo {author} {\bibfnamefont {R.}~\bibnamefont {Trotta}}, \bibinfo {author} {\bibfnamefont {J.~J.}\ \bibnamefont {Finley}}, \bibinfo {author} {\bibfnamefont {A.}~\bibnamefont {Rastelli}}, \bibinfo {author} {\bibfnamefont {E.}~\bibnamefont {{del Valle}}}, \bibinfo {author} {\bibfnamefont {F.~P.}\ \bibnamefont {Laussy}}, \bibinfo {author} {\bibfnamefont {V.}~\bibnamefont {Zwiller}}, \bibinfo {author} {\bibfnamefont {K.}~\bibnamefont {M{\"u}ller}},\ and\ \bibinfo {author} {\bibfnamefont {K.~D.}\ \bibnamefont {J{\"o}ns}},\ }\bibfield  {title} {\bibinfo {title} {{Origin of {{Antibunching}} in {{Resonance Fluorescence}}}},\ }\href {https://doi.org/10.1103/PhysRevLett.125.170402} {\bibfield  {journal} {\bibinfo  {journal} {Phys. Rev. Lett.}\ }\textbf {\bibinfo {volume} {125}},\ \bibinfo {pages} {170402} (\bibinfo {year} {2020})}\BibitemShut {NoStop}%
\bibitem [{\citenamefont {Phillips}\ \emph {et~al.}(2020)\citenamefont {Phillips}, \citenamefont {Brash}, \citenamefont {McCutcheon}, \citenamefont {{Iles-Smith}}, \citenamefont {Clarke}, \citenamefont {Royall}, \citenamefont {Skolnick}, \citenamefont {Fox},\ and\ \citenamefont {Nazir}}]{Phillips2020}%
  \BibitemOpen
  \bibfield  {author} {\bibinfo {author} {\bibfnamefont {C.~L.}\ \bibnamefont {Phillips}}, \bibinfo {author} {\bibfnamefont {A.~J.}\ \bibnamefont {Brash}}, \bibinfo {author} {\bibfnamefont {D.~P.~S.}\ \bibnamefont {McCutcheon}}, \bibinfo {author} {\bibfnamefont {J.}~\bibnamefont {{Iles-Smith}}}, \bibinfo {author} {\bibfnamefont {E.}~\bibnamefont {Clarke}}, \bibinfo {author} {\bibfnamefont {B.}~\bibnamefont {Royall}}, \bibinfo {author} {\bibfnamefont {M.~S.}\ \bibnamefont {Skolnick}}, \bibinfo {author} {\bibfnamefont {A.~M.}\ \bibnamefont {Fox}},\ and\ \bibinfo {author} {\bibfnamefont {A.}~\bibnamefont {Nazir}},\ }\bibfield  {title} {\bibinfo {title} {{Photon {{Statistics}} of {{Filtered Resonance Fluorescence}}}},\ }\href {https://doi.org/10.1103/PhysRevLett.125.043603} {\bibfield  {journal} {\bibinfo  {journal} {Phys. Rev. Lett.}\ }\textbf {\bibinfo {volume} {125}},\ \bibinfo {pages} {043603} (\bibinfo {year} {2020})}\BibitemShut {NoStop}%
\bibitem [{\citenamefont {Zubizarreta~Casalengua}\ \emph {et~al.}(2020)\citenamefont {Zubizarreta~Casalengua}, \citenamefont {L{\'o}pez~Carre{\~n}o}, \citenamefont {Laussy},\ and\ \citenamefont {{del Valle}}}]{ZubizarretaCasalengua2020Tuning}%
  \BibitemOpen
  \bibfield  {author} {\bibinfo {author} {\bibfnamefont {E.}~\bibnamefont {Zubizarreta~Casalengua}}, \bibinfo {author} {\bibfnamefont {J.~C.}\ \bibnamefont {L{\'o}pez~Carre{\~n}o}}, \bibinfo {author} {\bibfnamefont {F.~P.}\ \bibnamefont {Laussy}},\ and\ \bibinfo {author} {\bibfnamefont {E.}~\bibnamefont {{del Valle}}},\ }\bibfield  {title} {\bibinfo {title} {{Tuning Photon Statistics with Coherent Fields}},\ }\href {https://doi.org/10.1103/PhysRevA.101.063824} {\bibfield  {journal} {\bibinfo  {journal} {Phys. Rev. A}\ }\textbf {\bibinfo {volume} {101}},\ \bibinfo {pages} {063824} (\bibinfo {year} {2020})}\BibitemShut {NoStop}%
\bibitem [{\citenamefont {{S{\'a}nchez-Barquilla}}\ \emph {et~al.}(2020)\citenamefont {{S{\'a}nchez-Barquilla}}, \citenamefont {Silva},\ and\ \citenamefont {Feist}}]{Sanchez-Barquilla2020}%
  \BibitemOpen
  \bibfield  {author} {\bibinfo {author} {\bibfnamefont {M.}~\bibnamefont {{S{\'a}nchez-Barquilla}}}, \bibinfo {author} {\bibfnamefont {R.~E.~F.}\ \bibnamefont {Silva}},\ and\ \bibinfo {author} {\bibfnamefont {J.}~\bibnamefont {Feist}},\ }\bibfield  {title} {\bibinfo {title} {{Cumulant Expansion for the Treatment of Light-Matter Interactions in Arbitrary Material Structures}},\ }\href {https://doi.org/10.1063/1.5138937} {\bibfield  {journal} {\bibinfo  {journal} {J. Chem. Phys.}\ }\textbf {\bibinfo {volume} {152}},\ \bibinfo {pages} {034108} (\bibinfo {year} {2020})}\BibitemShut {NoStop}%
\bibitem [{\citenamefont {Feist}\ \emph {et~al.}(2021)\citenamefont {Feist}, \citenamefont {{Fern{\'a}ndez-Dom{\'i}nguez}},\ and\ \citenamefont {{Garc{\'i}a-Vidal}}}]{Feist2021}%
  \BibitemOpen
  \bibfield  {author} {\bibinfo {author} {\bibfnamefont {J.}~\bibnamefont {Feist}}, \bibinfo {author} {\bibfnamefont {A.~I.}\ \bibnamefont {{Fern{\'a}ndez-Dom{\'i}nguez}}},\ and\ \bibinfo {author} {\bibfnamefont {F.~J.}\ \bibnamefont {{Garc{\'i}a-Vidal}}},\ }\bibfield  {title} {\bibinfo {title} {{Macroscopic {{QED}} for Quantum Nanophotonics: Emitter-Centered Modes as a Minimal Basis for Multiemitter Problems}},\ }\href {https://doi.org/10.1515/nanoph-2020-0451} {\bibfield  {journal} {\bibinfo  {journal} {Nanophotonics}\ }\textbf {\bibinfo {volume} {10}},\ \bibinfo {pages} {477} (\bibinfo {year} {2021})}\BibitemShut {NoStop}%
\bibitem [{\citenamefont {Gro{\ss}}\ \emph {et~al.}(2018)\citenamefont {Gro{\ss}}, \citenamefont {Hamm}, \citenamefont {Tufarelli}, \citenamefont {Hess},\ and\ \citenamefont {Hecht}}]{Gross2018}%
  \BibitemOpen
  \bibfield  {author} {\bibinfo {author} {\bibfnamefont {H.}~\bibnamefont {Gro{\ss}}}, \bibinfo {author} {\bibfnamefont {J.~M.}\ \bibnamefont {Hamm}}, \bibinfo {author} {\bibfnamefont {T.}~\bibnamefont {Tufarelli}}, \bibinfo {author} {\bibfnamefont {O.}~\bibnamefont {Hess}},\ and\ \bibinfo {author} {\bibfnamefont {B.}~\bibnamefont {Hecht}},\ }\bibfield  {title} {\bibinfo {title} {{Near-Field Strong Coupling of Single Quantum Dots}},\ }\href {https://doi.org/10.1126/sciadv.aar4906} {\bibfield  {journal} {\bibinfo  {journal} {Science Adv.}\ }\textbf {\bibinfo {volume} {4}},\ \bibinfo {pages} {eaar4906} (\bibinfo {year} {2018})}\BibitemShut {NoStop}%
\bibitem [{\citenamefont {Li}\ \emph {et~al.}(2022)\citenamefont {Li}, \citenamefont {Li}, \citenamefont {Liu}, \citenamefont {Zhong}, \citenamefont {Liu}, \citenamefont {Chen},\ and\ \citenamefont {Wang}}]{Li2022Room}%
  \BibitemOpen
  \bibfield  {author} {\bibinfo {author} {\bibfnamefont {J.-Y.}\ \bibnamefont {Li}}, \bibinfo {author} {\bibfnamefont {W.}~\bibnamefont {Li}}, \bibinfo {author} {\bibfnamefont {J.}~\bibnamefont {Liu}}, \bibinfo {author} {\bibfnamefont {J.}~\bibnamefont {Zhong}}, \bibinfo {author} {\bibfnamefont {R.}~\bibnamefont {Liu}}, \bibinfo {author} {\bibfnamefont {H.}~\bibnamefont {Chen}},\ and\ \bibinfo {author} {\bibfnamefont {X.-H.}\ \bibnamefont {Wang}},\ }\bibfield  {title} {\bibinfo {title} {{Room-{{Temperature Strong Coupling Between}} a {{Single Quantum Dot}} and a {{Single Plasmonic Nanoparticle}}}},\ }\href {https://doi.org/10.1021/acs.nanolett.2c00606} {\bibfield  {journal} {\bibinfo  {journal} {Nano Lett.}\ }\textbf {\bibinfo {volume} {22}},\ \bibinfo {pages} {4686} (\bibinfo {year} {2022})}\BibitemShut {NoStop}%
\bibitem [{\citenamefont {Thu Ha~Do}\ \emph {et~al.}(2024)\citenamefont {Thu Ha~Do}, \citenamefont {Nonahal}, \citenamefont {Li}, \citenamefont {Valuckas}, \citenamefont {Tan}, \citenamefont {Kuznetsov}, \citenamefont {Nguyen}, \citenamefont {Aharonovich},\ and\ \citenamefont {Ha}}]{ThuHaDo2024}%
  \BibitemOpen
  \bibfield  {author} {\bibinfo {author} {\bibfnamefont {T.}~\bibnamefont {Thu Ha~Do}}, \bibinfo {author} {\bibfnamefont {M.}~\bibnamefont {Nonahal}}, \bibinfo {author} {\bibfnamefont {C.}~\bibnamefont {Li}}, \bibinfo {author} {\bibfnamefont {V.}~\bibnamefont {Valuckas}}, \bibinfo {author} {\bibfnamefont {H.~H.}\ \bibnamefont {Tan}}, \bibinfo {author} {\bibfnamefont {A.~I.}\ \bibnamefont {Kuznetsov}}, \bibinfo {author} {\bibfnamefont {H.~S.}\ \bibnamefont {Nguyen}}, \bibinfo {author} {\bibfnamefont {I.}~\bibnamefont {Aharonovich}},\ and\ \bibinfo {author} {\bibfnamefont {S.~T.}\ \bibnamefont {Ha}},\ }\bibfield  {title} {\bibinfo {title} {{Room-Temperature Strong Coupling in a Single-Photon Emitter-Metasurface System}},\ }\href {https://doi.org/10.1038/s41467-024-46544-w} {\bibfield  {journal} {\bibinfo  {journal} {Nat. Commun.}\ }\textbf {\bibinfo {volume} {15}},\ \bibinfo {pages} {2281} (\bibinfo {year} {2024})}\BibitemShut {NoStop}%
\bibitem [{\citenamefont {Hu}\ \emph {et~al.}(2024)\citenamefont {Hu}, \citenamefont {Huang}, \citenamefont {Arul}, \citenamefont {{S{\'a}nchez-Iglesias}}, \citenamefont {Xiong}, \citenamefont {{Liz-Marz{\'a}n}},\ and\ \citenamefont {Baumberg}}]{Hu2024}%
  \BibitemOpen
  \bibfield  {author} {\bibinfo {author} {\bibfnamefont {S.}~\bibnamefont {Hu}}, \bibinfo {author} {\bibfnamefont {J.}~\bibnamefont {Huang}}, \bibinfo {author} {\bibfnamefont {R.}~\bibnamefont {Arul}}, \bibinfo {author} {\bibfnamefont {A.}~\bibnamefont {{S{\'a}nchez-Iglesias}}}, \bibinfo {author} {\bibfnamefont {Y.}~\bibnamefont {Xiong}}, \bibinfo {author} {\bibfnamefont {L.~M.}\ \bibnamefont {{Liz-Marz{\'a}n}}},\ and\ \bibinfo {author} {\bibfnamefont {J.~J.}\ \bibnamefont {Baumberg}},\ }\bibfield  {title} {\bibinfo {title} {{Robust Consistent Single Quantum Dot Strong Coupling in Plasmonic Nanocavities}},\ }\href {https://doi.org/10.1038/s41467-024-51170-7} {\bibfield  {journal} {\bibinfo  {journal} {Nat. Commun.}\ }\textbf {\bibinfo {volume} {15}},\ \bibinfo {pages} {6835} (\bibinfo {year} {2024})}\BibitemShut {NoStop}%
\bibitem [{\citenamefont {Liu}\ \emph {et~al.}(2024)\citenamefont {Liu}, \citenamefont {Geng}, \citenamefont {Ai}, \citenamefont {Fan}, \citenamefont {Liu}, \citenamefont {Lu}, \citenamefont {Kuang}, \citenamefont {Liu}, \citenamefont {Guo},\ and\ \citenamefont {Wu}}]{Liu2024Deterministic}%
  \BibitemOpen
  \bibfield  {author} {\bibinfo {author} {\bibfnamefont {R.}~\bibnamefont {Liu}}, \bibinfo {author} {\bibfnamefont {M.}~\bibnamefont {Geng}}, \bibinfo {author} {\bibfnamefont {J.}~\bibnamefont {Ai}}, \bibinfo {author} {\bibfnamefont {X.}~\bibnamefont {Fan}}, \bibinfo {author} {\bibfnamefont {Z.}~\bibnamefont {Liu}}, \bibinfo {author} {\bibfnamefont {Y.-W.}\ \bibnamefont {Lu}}, \bibinfo {author} {\bibfnamefont {Y.}~\bibnamefont {Kuang}}, \bibinfo {author} {\bibfnamefont {J.-F.}\ \bibnamefont {Liu}}, \bibinfo {author} {\bibfnamefont {L.}~\bibnamefont {Guo}},\ and\ \bibinfo {author} {\bibfnamefont {L.}~\bibnamefont {Wu}},\ }\bibfield  {title} {\bibinfo {title} {{Deterministic Positioning and Alignment of a Single-Molecule Exciton in Plasmonic Nanodimer for Strong Coupling}},\ }\href {https://doi.org/10.1038/s41467-024-46831-6} {\bibfield  {journal} {\bibinfo  {journal} {Nat. Commun.}\ }\textbf {\bibinfo {volume} {15}},\ \bibinfo {pages} {4103} (\bibinfo {year} {2024})}\BibitemShut {NoStop}%
\bibitem [{\citenamefont {Li}\ \emph {et~al.}(2016)\citenamefont {Li}, \citenamefont {{Hern{\'a}ngomez-P{\'e}rez}}, \citenamefont {{Garc{\'i}a-Vidal}},\ and\ \citenamefont {{Fern{\'a}ndez-Dom{\'i}nguez}}}]{Li2016Transformation}%
  \BibitemOpen
  \bibfield  {author} {\bibinfo {author} {\bibfnamefont {R.-Q.}\ \bibnamefont {Li}}, \bibinfo {author} {\bibfnamefont {D.}~\bibnamefont {{Hern{\'a}ngomez-P{\'e}rez}}}, \bibinfo {author} {\bibfnamefont {F.~J.}\ \bibnamefont {{Garc{\'i}a-Vidal}}},\ and\ \bibinfo {author} {\bibfnamefont {A.~I.}\ \bibnamefont {{Fern{\'a}ndez-Dom{\'i}nguez}}},\ }\bibfield  {title} {\bibinfo {title} {{Transformation {{Optics Approach}} to {{Plasmon-Exciton Strong Coupling}} in {{Nanocavities}}}},\ }\href {https://doi.org/10.1103/PhysRevLett.117.107401} {\bibfield  {journal} {\bibinfo  {journal} {Phys. Rev. Lett.}\ }\textbf {\bibinfo {volume} {117}},\ \bibinfo {pages} {107401} (\bibinfo {year} {2016})}\BibitemShut {NoStop}%
\bibitem [{\citenamefont {Novotny}\ and\ \citenamefont {Hecht}(2012)}]{Novotny2012}%
  \BibitemOpen
  \bibfield  {author} {\bibinfo {author} {\bibfnamefont {L.}~\bibnamefont {Novotny}}\ and\ \bibinfo {author} {\bibfnamefont {B.}~\bibnamefont {Hecht}},\ }\href {https://doi.org/10.1017/CBO9780511794193} {\emph {\bibinfo {title} {{Principles of {{Nano-Optics}}}}}},\ \bibinfo {edition} {2nd}\ ed.\ (\bibinfo  {publisher} {Cambridge University Press},\ \bibinfo {address} {Cambridge},\ \bibinfo {year} {2012})\BibitemShut {NoStop}%
\bibitem [{\citenamefont {L{\'o}pez~Carre{\~n}o}\ \emph {et~al.}(2018)\citenamefont {L{\'o}pez~Carre{\~n}o}, \citenamefont {{del Valle}},\ and\ \citenamefont {Laussy}}]{LopezCarreno2018}%
  \BibitemOpen
  \bibfield  {author} {\bibinfo {author} {\bibfnamefont {J.~C.}\ \bibnamefont {L{\'o}pez~Carre{\~n}o}}, \bibinfo {author} {\bibfnamefont {E.}~\bibnamefont {{del Valle}}},\ and\ \bibinfo {author} {\bibfnamefont {F.~P.}\ \bibnamefont {Laussy}},\ }\bibfield  {title} {\bibinfo {title} {{Frequency-Resolved {{Monte Carlo}}}},\ }\href {https://doi.org/10.1038/s41598-018-24975-y} {\bibfield  {journal} {\bibinfo  {journal} {Sci. Rep.}\ }\textbf {\bibinfo {volume} {8}},\ \bibinfo {pages} {6975} (\bibinfo {year} {2018})}\BibitemShut {NoStop}%
\end{thebibliography}%

\end{document}